\documentclass[a4paper,11pt]{article}
\usepackage[english]{babel}
\usepackage[utf8]{inputenc}
\pdfoutput=1
\usepackage{braket}
\usepackage{physics}
\usepackage{color}
\usepackage{amssymb}
\usepackage{amsmath}
\usepackage{graphicx}
\usepackage{epstopdf}
\usepackage{subcaption}
\usepackage{comment}
\usepackage{mathtools}
\usepackage{ragged2e}
\usepackage{hyperref}
\hypersetup{
	colorlinks=true,
	linkcolor=blue,
	filecolor=cyan, 
	urlcolor=blue,
	citecolor=red,
} 
\usepackage[titletoc,toc,title]{appendix}
\numberwithin{equation}{section}
\usepackage{cleveref}
\usepackage[margin=2.5cm]{geometry}
\usepackage{cite}
\usepackage{epsfig}
\usepackage{float}
\usepackage{cancel}
\usepackage{amsfonts}
\usepackage{enumitem}
\usepackage[font={footnotesize,it}]{caption}
\usepackage{authblk}

\begin{document}
	
	\title{Defect extremal surfaces for entanglement negativity}

	\author[1]{Debarshi Basu\thanks{\noindent E-mail:~ {\tt \href{mailto:debarshi@iitk.ac.in} {debarshi@iitk.ac.in}}}}

	\author[1]{Himanshu Parihar\thanks{\noindent E-mail:~ {\tt \href{mailto:himansp@iitk.ac.in} {himansp@iitk.ac.in}}}}

	\author[1]{Vinayak Raj\thanks{\noindent E-mail:~ {\tt \href{mailto:vraj@iitk.ac.in} {vraj@iitk.ac.in}}}}

	\author[1]{Gautam Sengupta\thanks{\noindent E-mail:~ {\tt \href{mailto:sengupta@iitk.ac.in} {sengupta@iitk.ac.in}}}}

	\affil[1]{
		Department of Physics\\
		
		Indian Institute of Technology\\ 
		
		Kanpur 208 016, India
	}
	
	\date{}
	
	\maketitle
	
	\thispagestyle{empty}
	
	\begin{abstract}
		
		\noindent
		We propose a doubly holographic version of the semi-classical island formula for the entanglement negativity in the framework of the defect AdS/BCFT correspondence where the AdS bulk contains a defect conformal matter theory. In this context, we propose a defect extremal surface (DES) formula for computing the entanglement negativity modified by the contribution from the defect matter theory on the end-of-the-world brane. The equivalence of the DES proposal and the semi-classical island formula for the entanglement negativity is demonstrated in AdS$_3$/BCFT$_2$ framework. Furthermore, in the time-dependent AdS$_3$/BCFT$_2$ scenarios involving eternal black holes in the lower dimensional effective description, we investigate the time evolution of the entanglement negativity through the DES and the island formulae and obtain the analogues of the Page curves.
		
		\justify

	\end{abstract}
	
	\clearpage
	
	\tableofcontents
	
	\clearpage

\section{Introduction} \label{sec:Intro}
From the past few decades, the study of the black hole information loss paradox has led to several key insights about semi-classical and quantum gravity. Recently, tremendous progress has been made towards a possible resolution of this paradox which involves the appearance of regions termed {\it ``islands"} in the black hole geometry at late times \cite{Almheiri:2019hni, Almheiri:2019psf, Almheiri:2019qdq, Almheiri:2019psy, Almheiri:2019yqk, Almheiri:2020cfm}. This leads to the Page curve \cite{Page:1993wv, Page:1993df, Page:2013dx}, which indicates that the process of black hole formation and evaporation follows a unitary evolution. The appearance of the islands stems from the late time dominance of the \textit{replica wormhole} saddles in the gravitational path integral for the R\'enyi entanglement entropy. The resultant island formula was inspired by the advent of quantum extremal surfaces (QES) introduced earlier, to compute the quantum corrections to the holographic entanglement entropy \cite{Ryu:2006bv, Hubeny:2007xt, Faulkner:2013ana, Engelhardt:2014gca}. In this connection, in \cite{Almheiri:2019hni, Almheiri:2019yqk, Almheiri:2019qdq, Almheiri:2020cfm} a quantum dot (e.g. SYK model) coupled to a CFT$_2$ on a half-line was regarded as the holographic dual to a $2$-dimensional conformal field theory coupled to semi-classical gravity on a hybrid manifold\footnote{In such holographic dual theories, the hybrid manifold on which the CFT is defined consists of a flat bath along with a curved geometry with dynamical gravity.}. For such $2$-dimensional conformal field theories coupled to semi-classical gravity, the island formula involves the fine-grained entropy of the Hawking radiation in a region $R$, obtained through the extremization over the entanglement entropy island region $I(R)$ and is expressed as follows
\begin{equation}
S [R] =  \text{min} ~\underset{I (R)}{\text{ext}} \left[ \frac{\text{Area} [\partial I (R)]}{4 G_N} + S_\text{eff} \big(R \cup I (R)\big) \right] ,
\end{equation}
where $G_N$ is the Newton's constant and $S_\text{eff} (X)$ corresponds to the effective semi-classical entanglement entropy of quantum matter fields located on $X$. For recent related works, see \cite{Anderson:2020vwi, Chen:2019iro, Balasubramanian:2020hfs, Chen:2020wiq, Gautason:2020tmk, Bhattacharya:2020ymw, Anegawa:2020ezn, Hashimoto:2020cas, Hartman:2020swn, Krishnan:2020oun, Alishahiha:2020qza, Geng:2020qvw, Li:2020ceg, Chandrasekaran:2020qtn, Bak:2020enw, Krishnan:2020fer, Karlsson:2020uga, Hartman:2020khs, Balasubramanian:2020coy, Balasubramanian:2020xqf, Sybesma:2020fxg, Chen:2020hmv, Ling:2020laa, Hernandez:2020nem, Marolf:2020rpm, Matsuo:2020ypv, Akal:2020twv, Caceres:2020jcn, Raju:2020smc, Deng:2020ent, Anous:2022wqh, Bousso:2022gth, Hu:2022ymx, Grimaldi:2022suv, Akers:2022max, Yu:2021rfg, Geng:2021mic, Chou:2021boq, Hollowood:2021lsw, He:2021mst, Arefeva:2021kfx, Ling:2021vxe, Bhattacharya:2021dnd, Azarnia:2021uch, Saha:2021ohr, Hollowood:2021wkw, Sun:2021dfl, Li:2021dmf, Aguilar-Gutierrez:2021bns, Ahn:2021chg, Yu:2021cgi, Lu:2021gmv, Caceres:2021fuw, Akal:2021foz, Arefeva:2022cam, Arefeva:2022guf, Bousso:2022ntt, Krishnan:2021ffb, Zeng:2021kyb, Teresi:2021qff, Okuyama:2021bqg, Chen:2021jzx, Pedraza:2021ssc, Guo:2021blh, Kibe:2021gtw, Renner:2021qbe, Dong:2021oad, Raju:2021lwh, Nam:2021bml, Kames-King:2021etp, Chen:2021lnq, Sato:2021ftf, Kudler-Flam:2021alo, Wang:2021afl, Ageev:2021ipd, Buoninfante:2021ijy, Cadoni:2021ypx, Marolf:2021ghr, Chu:2021gdb, Urbach:2021zil, Li:2021lfo, Neuenfeld:2021bsb, Aalsma:2021bit, Ghosh:2021axl, Bhattacharya:2021jrn, Geng:2021wcq, Krishnan:2021faa, Verheijden:2021yrb, Bousso:2021sji, Karananas:2020fwx, Goto:2020wnk, Bhattacharya:2020uun, Chen:2020jvn, Agon:2020fqs, Laddha:2020kvp, Akers:2019nfi, Chen:2019uhq, Basak:2022acg,Geng:2020fxl,Geng:2021iyq, Afrasiar:2022}.

A natural description for the island formulation was provided through a \textit{double holographic} framework \cite{Almheiri:2019hni} where the $d$-dimensional conformal field theory coupled to semi-classical gravity may be interpreted as a lower dimensional effective description of a bulk $(d+1)$-dimensional theory of gravity. In this scenario, the $d$-dimensional conformal field theory is considered to possess a dual bulk $(d+1)$-dimensional gravitational theory in the AdS$_{d+1}$/CFT$_d$ framework. In the double holographic picture the computation of the entanglement entropy through the island formula in the lower dimensional theory reduces to its holographic characterization through the (H)RT formula \cite{Ryu:2006bv, Hubeny:2007xt} in the bulk dual AdS$_{d+1}$ geometry. This may be understood as a realization of the ER=EPR proposal \cite{Maldacena:2013xja} where the island region in the black hole interior is contained within the entanglement wedge of the radiation bath through the double holographic perspective.

On a separate note, CFT$_2$s on a manifold with a boundary, termed as boundary conformal field theories (BCFT$_2$s) \cite{Cardy:2004hm} have received considerable attention in the recent past. The holographic dual of such BCFT$_2$s \cite{Takayanagi:2011zk, Fujita:2011fp, Rozali:2019day, Sully:2020pza,Kastikainen:2021ybu} involves an asymptotically AdS$_3$ spacetime truncated by an end-of-the-world (EOW) brane $\mathbb{Q}$ with Neumann boundary condition.
An extension of this AdS$_3$/BCFT$_2$ duality studied in \cite{Deng:2020ent}, involved additional \textit{defect} conformal matter on the EOW brane $\mathbb{Q}$ which resulted in the modification of the Neumann boundary condition. The entanglement entropy of an interval in this defect BCFT$_2$ was also computed in \cite{Deng:2020ent, Chu:2021gdb} through a modification of the quantum corrected RT formula. This was termed as the defect extremal surface (DES) formula as it involved contributions from the defect conformal matter fields. Interestingly, this DES formula has been proposed to be the doubly holographic counterpart of the island formula in the context of the defect AdS$_3$/BCFT$_2$ scenario \cite{Deng:2020ent}. The authors in \cite{Deng:2020ent} compared the entanglement entropy computed through the DES formula in the $3d$ bulk geometry with that computed through the island formula in the effective $2d$ description and found an exact agreement. Subsequently, the time dependent AdS$_3$/BCFT$_2$ scenario was studied in \cite{Chu:2021gdb}, where in the effective $2d$ description, an eternal black hole emerges on the EOW brane. The entanglement entropy for the Hawking radiation from the eternal black hole, obtained through the DES formula reproduced the Page curve and was consistent with the island proposal. 

The fine grained entanglement entropy is a viable measure of entanglement for bipartite pure states. For configurations involving bipartite pure states in black hole geometries, the island proposal in the effective picture or the DES formula in the doubly holographic scenario correctly encode the entanglement structure of the Hawking radiation. However, entanglement entropy fails to characterize the structure of entanglement for bipartite mixed states as it receives contributions from irrelevant classical and quantum correlations. For such cases involving bipartite mixed states, it is required to consider alternative mixed state correlation or entanglement measures. Several of such correlation and entanglement measures like the reflected entropy \cite{Dutta:2019gen, Akers:2021pvd}, the entanglement negativity \cite{Vidal:2002zz, Plenio:2005cwa}, the entanglement of purification \cite{Takayanagi:2017knl, Nguyen:2017yqw} and the balanced partial entanglement entropy \cite{Wen:2021qgx, Camargo:2022mme} have been studied in the literature.

In this context, the crucial issue of characterization of the entanglement structure of bipartite mixed states was addressed in \cite{Li:2021dmf} through the computation of the reflected entropy in the time dependent framework involving an eternal black hole in the AdS$_3$/BCFT$_2$ scenario. The authors proposed a $3d$ bulk DES formula for the reflected entropy and compared their results with the $2d$ effective field theory computations involving islands. They obtained the analogues of the Page curves for the reflected entropy and demonstrated the appearance of islands at late times.


The above developments bring into sharp focus the crucial issue of the characterization of the mixed state entanglement structure of the Hawking radiation from black holes. In this context, the non-convex entanglement monotone termed the \textit{entanglement negativity} \cite{Vidal:2002zz, Plenio:2005cwa} serves as a natural candidate to investigate the entanglement structure of such mixed states. The entanglement negativity has been explored in conformal field theories \cite{Calabrese:2012ew, Calabrese:2012nk, Calabrese:2014yza} through appropriate replica techniques\footnote{For an extension of this replica technique in the Galilean conformal field theories, see \cite{Malvimat:2018izs}.}. Subsequently several holographic constructions for computing the entanglement negativity in the context of the AdS/CFT correspondence was advanced in a series of interesting works\footnote{For analogues of these proposals in the context of flat holography, see \cite{Basu:2021axf, Setare:2021ryi}.}  in \cite{Rangamani:2014ywa, Chaturvedi:2016rft, Chaturvedi:2016rcn, Jain:2017aqk, Jain:2017xsu, Chaturvedi:2017znc, Jain:2017uhe, Jain:2018bai, Malvimat:2018txq, Malvimat:2018ood, KumarBasak:2020viv, Mondal:2021kzj, Afrasiar:2021hld, Basu:2022nds} which reproduced the field theoretic results in the large central charge limit \cite{Kulaxizi:2014nma,Malvimat:2017yaj, Basu:2021axf}. Interestingly, in \cite{Kudler-Flam:2018qjo, Kusuki:2019zsp, KumarBasak:2020eia, KumarBasak:2021lwm, Basu:2021awn}, an alternative holographic proposal based on the bulk entanglement wedge cross-section (EWCS) was also investigated. In this connection, an island formulation for the entanglement negativity was recently established in \cite{KumarBasak:2020ams} following a similar island construction for the reflected entropy developed in \cite{Chandrasekaran:2020qtn, Li:2020ceg}. Furthermore, a geometric construction based on the double holographic framework was discussed qualitatively in \cite{KumarBasak:2020ams} and subsequently investigated in \cite{KumarBasak:2021rrx} through a partial dimensional reduction \cite{Verheijden:2021yrb} of the $3d$ bulk space time. In this article, we generalize these doubly holographic scenarios to the framework of AdS/BCFT with defect conformal matter on the EOW brane. We propose a DES formula for computing the bulk entanglement negativity in asymptotically AdS$_3$ geometries truncated by an EOW brane. Furthermore, we demonstrate the equivalence of the DES results with the corresponding island computations for the entanglement negativity of bipartite mixed states in both static and time-dependent configurations involving black hole/bath systems in the effective lower dimensional theory.

The rest of the article is organized as follows. In \cref{sec:review}, we recollect various aspects of the DES formula for the entanglement entropy and the corresponding effective lower dimensional picture involving the entanglement islands. In \cref{sec:EN-DES}, we provide the island construction for the entanglement negativity \cite{KumarBasak:2020ams}, and propose the DES formulas for computing the bulk entanglement negativity for disjoint and adjacent subsystems on the conformal boundary of asymptotically AdS$_3$ geometries with defect conformal matter on the EOW brane. In \cref{sec:EN-static}, we compute the entanglement negativity for disjoint and adjacent intervals in a static time slice of the conformal boundary. Beginning with a brief review of the eternal black hole configuration in the $2d$ effective semi-classical picture, we describe DES and island computations for the entanglement negativity between interior regions of the black hole, between the black hole and radiation in the bath region, and between radiation segments, and demonstrate the equivalence of the two formulations in \cref{sec:Time-dependent-EN}. Finally, in \cref{sec:summary}, we summarize our results and comment on possible future directions.

\section{Review of earlier literature}\label{sec:review}
In this section, we will briefly recall the salient features of the holographic model under consideration. We first review the AdS/BCFT scenario \cite{Takayanagi:2011zk} modified through the inclusion of conformal matter on the end-of-the-world (EOW) brane which was proposed in \cite{Deng:2020ent, Chu:2021gdb}. Following this, we describe the defect extremal surface (DES) formula \cite{Deng:2020ent} for computing the entanglement entropy in the bulk AdS geometry truncated by the EOW brane. We will also briefly elucidate the effective $2d$ description of the model and the semi-classical island formula for computing the entanglement entropy of a subsystem in the effective description. 

\subsection{AdS$_3$/BCFT$_2$}
As described in \cite{Takayanagi:2011zk, Fujita:2011fp} the bulk dual of a BCFT$_2$ defined on the half line $x\geq 0$ is given by an AdS$_3$ geometry truncated by an EOW brane $\mathbb{Q}$ with Neumann boundary conditions. The gravitational action of the bulk manifold $\mathcal{N}$ is given by
\begin{align}
	I=\int_{\mathcal{N}}\sqrt{-g}\,(R-2\Lambda)+2\int_{\mathbb{Q}}\sqrt{-h}\,(K-T)\,,
\end{align}
where $h_{ab}$ is the induced metric, $K$ is the trace of the extrinsic curvature $K_{ab}$ on the EOW brane $\mathbb{Q}$ with a tension $T$. The Neumann boundary condition on the EOW brane is given as $K_{ab}=(K-T)h_{ab}$. The $3d$ bulk geometry may be described by two sets of relevant coordinate charts, $(t,x,z)$ and $(t,\rho,y)$, which are related through
\begin{align}\label{x-y-coordinates}
	x=y \tanh\left(\frac{\rho}{\ell}\right)~~,~~z=-y \sech\left(\frac{\rho}{\ell}\right)\,.
\end{align}
The bulk metric in these coordinates is given by the standard Poincar\'e slicing, as follows
\begin{align}
	ds^2&=d\rho^2+\cosh^2\left(\frac{\rho}{\ell}\right)\frac{-dt^2+dy^2}{y^2}\notag\\
	&=\frac{\ell^2}{z^2}(-dt^2+dx^2+dz^2)\,,
\end{align}
where $\ell$ is the AdS$_3$ radius. In the Poincar\'e slicing\footnote{A convenient choice for a polar coordinate is $\theta=\text{arccos}\left[\sech\left(\frac{\rho}{\ell}\right)\right]$, which determines the angular position of the brane from the vertical as shown in \cref{fig:DES-EE}.} described by the $(t,\rho,y)$ coordinate chart the EOW brane is situated at a constant $\rho=\rho_0$ slice and the induced metric on the brane is given by that of  an AdS$_2$ geometry \cite{Takayanagi:2011zk}.

An extension to this usual AdS$_3$/BCFT$_2$ framework was proposed in \cite{Deng:2020ent} where one essentially begins with an orthogonal brane with zero tension and through the addition of conformal matter onto it, turns on a finite tension. The Neumann boundary condition on the EOW brane $\mathbb{Q}$ is modified by the stress tensor of this defect CFT$_2$. The EOW brane $\mathbb{Q}$ is then treated as a defect in the bulk geometry.

\subsection{Defect extremal surface}
\begin{figure}[h!]
	\centering
	\includegraphics[scale=0.6]{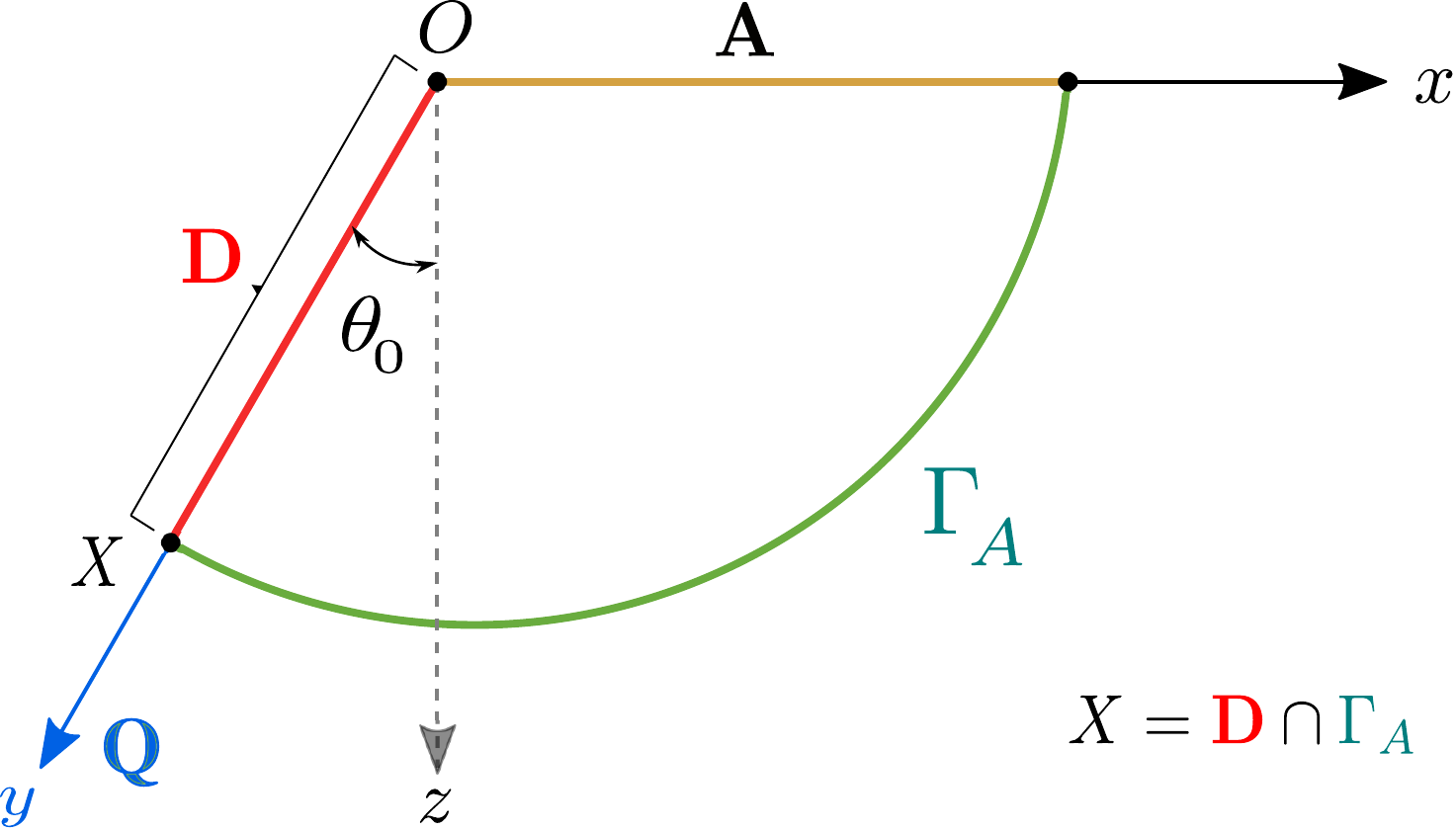}
	\caption{Schematics of the defect extremal surface for the entanglement entropy of a subsystem $A$. Figure modified from \cite{Deng:2020ent}.}
	\label{fig:DES-EE}
\end{figure}

For the modified bulk picture with defect conformal matter on $\mathbb{Q}$, the entanglement entropy of an interval $A$ in the original BCFT$_2$ involves contributions from the defect matter, and the usual RT formula \cite{Ryu:2006bv} is modified to the defect extremal surface (DES) formula \cite{Deng:2020ent, Chu:2021gdb} given as
\begin{align}
	S_{\text{DES}}(A)=\underset{\Gamma_A,X}{\text{min}~\text{ext}}\left[\frac{\mathcal{A}\left(\Gamma_A\right)}{4G_N}+S_{\text{defect}}(D)\right]~~,~~X=\Gamma_A\cap D\, ,
\end{align}
where $\Gamma_A$ is a co-dimension two extremal surface homologous to the subsystem $A$ and $D$ is the defect region along the EOW brane $\mathbb{Q}$ as depicted in \cref{fig:DES-EE}.

For an interval $A=[0,L]$ in the BCFT$_2$, the generalized entanglement entropy corresponding to a defect $D=[- a,0]$ on the brane CFT$_2$ may be computed through the DES formula as follows\footnote{Note that we are using the standard geodesic length formula for Poincar\'e AdS$_3$ instead of the AdS/BCFT techniques employed in \cite{Deng:2020ent,Chu:2021gdb} as both the procedures lead to the same answer and are therefore complementary.} \cite{Deng:2020ent}
\begin{align}
	S_{\text{gen}}(a)&=\frac{\mathcal{A}\left(\Gamma_A\right)}{4G_N}+S_{\text{defect}}([-a,0])\notag\\
	&=\frac{\ell}{4G_N}\cosh^{-1}\left[\frac{(L+a\sin\theta_0)^2+(a\cos\theta_0)^2}{2 \epsilon \,a \cos\theta_0}\right]+\frac{c}{6}\log\left(\frac{2\ell}{\epsilon_y\cos\theta_0}\right),
\end{align}
Note that the defect contribution to the generalized entropy is a constant which implies that the defect extremal surface is same as the RT surface for the subsystem $A$. Extremization with respect to the position $a$ of the defect leads to the entanglement entropy of the subsystem $A$ as follows
\begin{align}
	S_{\text{DES}}([0,L])=\frac{c}{6}\left[\log\left(\frac{2L}{\epsilon}\right)+\tanh^{-1}(\sin\theta_0)+\log\left(\frac{2\ell}{\epsilon_y\cos\theta_0}\right)\right]\,.\label{DES_EE-example}
\end{align}
where both the central charges of the original BCFT and the defect CFT$_2$ are taken to be equal\footnote{Note that, the equality of the two central charges is essential in order to relate the present bulk description to the effective $2d$ island scenario which involves a CFT$_2$ on the complete hybrid manifold.} to $c$.
\subsection{Effective description and boundary island formula}
The lower dimensional effective semi-classical theory for the bulk configuration described above may be obtained through a combination of a partial Randall-Sundrum reduction \cite{Karch:2000ct,Randall:1999vf} and the usual AdS/BCFT duality \cite{Maldacena:1997re}. As described in \cite{Deng:2020ent, Chu:2021gdb, Li:2021dmf}, this is implemented by dividing the AdS$_3$ bulk into two parts through the insertion of an imaginary co-dimension one surface $\mathbb{Q}'$ orthogonal to the asymptotic boundary, with transparent boundary conditions. The portion of the bulk enclosed between $\mathbb{Q}$ and $\mathbb{Q}'$ is dimensionally reduced along the $\rho$ direction using a partial Randall-Sundrum reduction thereby obtaining a effective $2d$ gravitational theory coupled with the matter CFT$_2$ on $\mathbb{Q}$. On the other hand, the rest of the bulk is dual to the original BCFT$_2$ on the half line $x\geq 0$ from the usual AdS/BCFT duality. The transparent boundary conditions along $\mathbb{Q}'$ naturally glues the gravity theory on $\mathbb{Q}$ and the BCFT$_2$ on the half line $x\geq 0$, leading to an effective $2d$ semi-classical theory on a hybrid manifold, similar to that considered in \cite{Almheiri:2019hni, Almheiri:2019qdq}.

In the effective semi-classical description described above, one may utilize the island formula \cite{Almheiri:2019hni, Almheiri:2019qdq} to compute the entanglement entropy. For a subsystem $A=[0,L]$ in the flat CFT$_2$ on the asymptotic boundary, an island region $I_A=[-a,0]$ appears in the gravitational sector on the EOW brane $\mathbb{Q}$. The entanglement entropy is obtained by extremizing the generalized entropy functional as
\begin{align}
	S_{\text{bdy}}=\underset{a}{\text{ext}}\,S_{\text{gen}}(a)&=\underset{a}{\text{ext}}\left[\mathcal{A}(y=-a)+S_{\text{matter}}([-a,L])\right]\notag\\
	&=\frac{c}{6}\tanh^{-1}(\sin\theta_0)+\frac{c}{6}\log\left(\frac{4L\ell}{\epsilon\,\epsilon_y\cos\theta_0}\right)\,.
\end{align}
The first term in the above expression is due to the constant area of the quantum extremal surface in the AdS$_3$/BCFT$_2$ framework, given as \cite{Deng:2020ent}
\begin{align}
	\mathcal{A}(\partial I_A)\equiv\frac{\rho_0}{4G_N}=\frac{\ell}{4G_N}\tanh^{-1}(\sin\theta_0)\,,\label{area-term}
\end{align}
where $\theta_0$ is the angle of the EOW brane with the vertical. It is observed from the above that the island formula leads to the same expression for the entanglement entropy as the DES result in \cref{DES_EE-example}. In other words, the DES formula may be considered as the doubly holographic counterpart of the island formula in the defect AdS/BCFT framework.

\subsection{Entanglement negativity} \label{sec:EN-review}
In this subsection, we will briefly review the salient features of the mixed state entanglement measure termed the entanglement negativity and its holographic characterization in the context of AdS$_3$/CFT$_2$ scenario. In a seminal work \cite{Vidal:2002zz}, Vidal and Werner introduced the computable mixed state entanglement measure, the entanglement negativity which is defined as the trace norm of the density matrix partially transposed with respect to one of the subsystems. 
%
%
%
%
%
In \cite{Calabrese:2012ew, Calabrese:2012nk, Calabrese:2014yza}, replica techniques were developed to obtain the entanglement negativity for subsystems in CFT$_2$s which involved the even parity $n_e$ of the replica index. The entanglement negativity was obtained through the analytic continuation of the replica index $n_e \to 1$ as follows
\begin{equation}
	\mathcal{E} = \lim_{n_e \to 1} \log \text{Tr} (\rho_{AB}^{T_B})^{n_e} ,
\end{equation}
where the superscript $T_B$ denotes partial transposition with respect to the subsystem $B$. The trace $\text{Tr} (\rho_{AB}^{T_B})^{n_e}$ may be expressed as a twist field correlator in the CFT$_2$, corresponding to the bipartite state under consideration. As an example, we consider the generic bipartite mixed state of two disjoint intervals $A = [u_1, v_1]$ and $B = [u_2, v_2]$ in a CFT$_2$. The trace $\text{Tr} (\rho_{AB}^{T_B})^{n_e}$ is then given by the following four-point correlator of \textit{twist fields},
\begin{equation}
	\text{Tr} (\rho_{AB}^{T_B})^{n_e} = \langle \mathcal{T}_{n_e} (u_1) \bar{\mathcal{T}}_{n_e} (v_1) \bar{\mathcal{T}}_{n_e} (u_2) \mathcal{T}_{n_e} (v_2) \rangle_{\text{CFT}^{\bigotimes n_e}}\,,
\end{equation}
where the twist fields $\mathcal{T}_{n_e}$ and $\bar{\mathcal{T}}_{n_e}$ are primary fields with conformal dimensions
\begin{align}
	\Delta_{n_e}=\frac{c}{12}\left(n_e-\frac{1}{n_e}\right)\,.
\end{align}
Subsequently, in a series of works \cite{Chaturvedi:2016rft, Chaturvedi:2016rcn, Jain:2017aqk, Malvimat:2018txq}, several holographic proposals for the entanglement negativity were proposed for specific bipartite mixed states. These proposals involved appropriate algebraic sums of the lengths of codimension two bulk static minimal surfaces homologous to various subsystems describing the mixed state. In particular, for two disjoint intervals $A$ and $B$ sandwiching another interval $C$ in a CFT$_2$, the holographic entanglement negativity may be obtained geometrically in the context of the AdS$_3$/CFT$_2$ correspondence as follows \cite{Malvimat:2018txq}
\begin{align}
	\mathcal{E}(A:B)=\frac{3}{16 G_N}\left(\mathcal{L}_{AC}+\mathcal{L}_{BC}-\mathcal{L}_{C}-\mathcal{L}_{ABC}\right)\,,\label{HEN-disj-basic}
\end{align}
where $\mathcal{L}_X$ denotes the length of the extremal curve homologous to subsystem $X$. The configuration of two adjacent intervals $A$ and $B$ may be obtained through the limit $C\to\emptyset$ of the above, and the holographic entanglement negativity is given as \cite{Jain:2017aqk}
\begin{align}
	\mathcal{E}(A:B)=\frac{3}{16 G_N}\left(\mathcal{L}_{A}+\mathcal{L}_{B}-\mathcal{L}_{AB}\right)\,.\label{HEN-adj-basic}
\end{align}
Note that, these proposals have further been extended to various other holographic frameworks including flat holography \cite{Basu:2021axf}, anomalous AdS/CFT \cite{Basu:2022nds} as well as higher dimensional scenarios \cite{Jain:2017xsu, KumarBasak:2020viv, Mondal:2021kzj, Afrasiar:2021hld}.

\section{Defect extremal surface for entanglement negativity}\label{sec:EN-DES}
\begin{figure}[h!]
	\centering
	\includegraphics[scale=0.65]{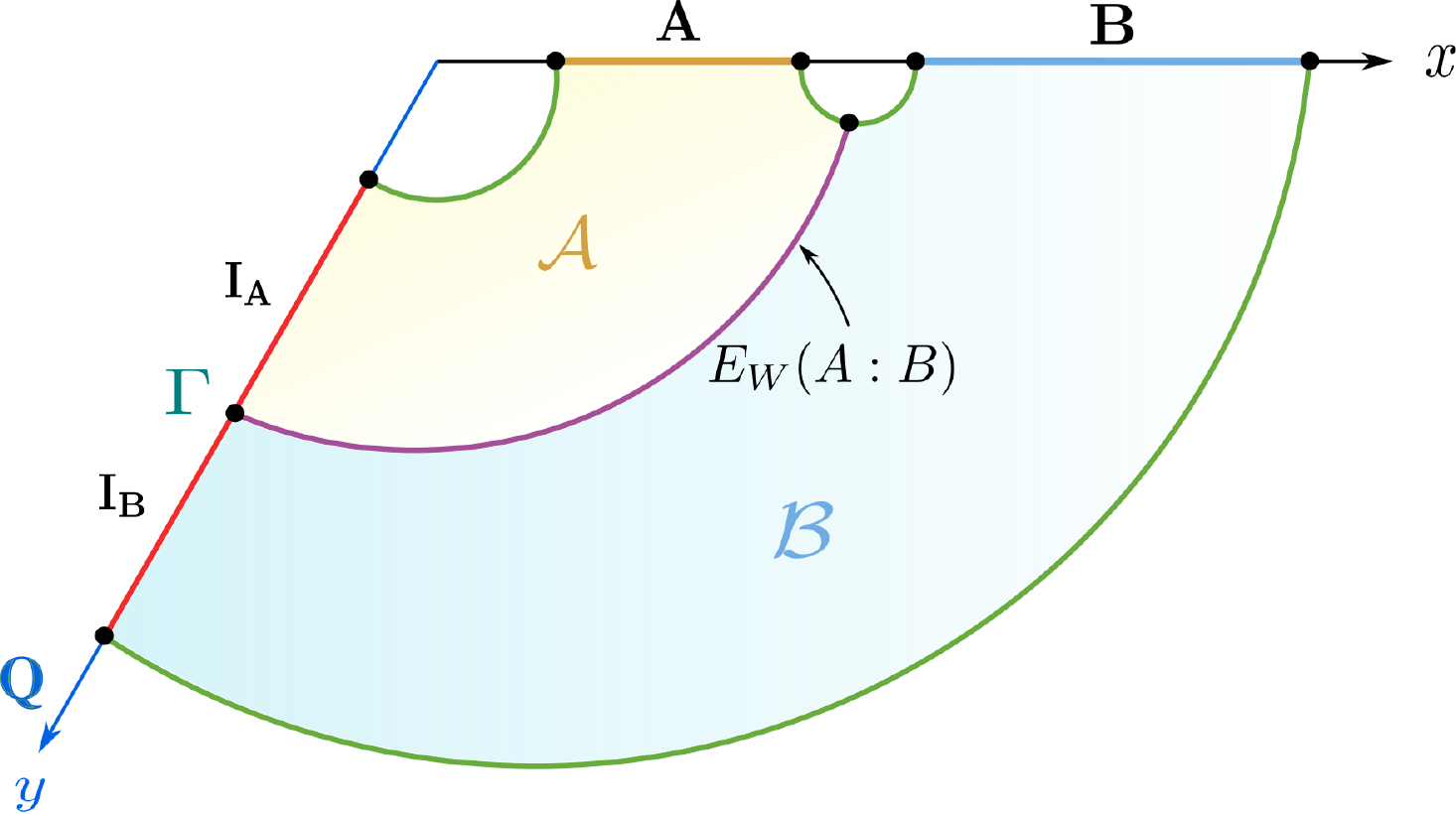}
	\caption{Schematics of the quantum extremal surface for the entanglement negativity between two disjoint intervals $A$ and $B$, where $I_A$ and $I_B$ are the entanglement negativity islands satisfying the constraint $I_A\cup I_B = I(A\cup B)$. The island cross-section is given by $\Gamma\equiv\partial I_A\cap \partial I_B$. In the double holographic picture described in \cite{KumarBasak:2020ams}, the $3d$ bulk entanglement wedge cross section ending at the point $\Gamma$ on the EOW brane $\mathbb{Q}$ splits the entanglement wedge corresponding to $A\cup B$ into two parts $\mathcal{A}$ and $\mathcal{B}$. Figure modified from \cite{Li:2021dmf}.}
	\label{fig:EN-QES}
\end{figure}

In this section, we propose the defect extremal surface (DES) formula for the entanglement negativity in the AdS/BCFT models which include defect conformal matter on the EOW brane \cite{Deng:2020ent, Chu:2021gdb, Li:2021dmf}. To begin with, we recall the semi-classical QES formula for the entanglement negativity involving entanglement islands in the lower dimensional effective picture discussed earlier. As described in \cite{KumarBasak:2020ams, KumarBasak:2021rrx}, the QES proposal for the entanglement negativity between two disjoint intervals in the effective boundary description\footnote{Note that, in this article, we use the nomenclature \textit{boundary description} and \textit{lower dimensional effective description} interchangeably.} is given by
\begin{align}
\mathcal{E}^{\text{bdy}}(A:B)=\text{min}~\underset{\Gamma=\partial I_A\cap \partial I_B}{\text{ext}}\left[\frac{3}{16 G_N}\Big(\mathcal{A}(\partial I_A)+\mathcal{A}(\partial I_B)-\mathcal{A}(\partial I_{AB})\Big)+\mathcal{E}^{\text{eff}}\left(A\cup I_A:B\cup I_B\right)\right],\label{QES-island}
\end{align}
where $I_A$ and $I_B$ are the entanglement negativity islands corresponding to subsystems $A$ and $B$, respectively. The entanglement negativity islands obeys the condition $I_A\cup I_B = I(A\cup B)$, where $I(A\cup B)$ denotes the entanglement entropy island for $A\cup B$, as illustrated in \cref{fig:EN-QES}. Furthermore, the extremization in the QES formula is performed over the location of the \textit{island cross-section} $\Gamma\equiv\partial I_A\cap \partial I_B$.
In this context, utilizing the constraint $I_A\cup I_B = I(A\cup B)$, the algebraic sum of the area contributions in \cref{QES-island} may be reduced to that corresponding to the island cross-section $\Gamma$. Hence, the QES formula may be expressed as \cite{KumarBasak:2020ams}
\begin{align}
\mathcal{E}^{\text{bdy}}(A:B)=\text{min}~\underset{\Gamma}{\text{Ext}}\left[\frac{3}{8 G_N}\,\mathcal{A}\left(\Gamma=\partial I_A\cap \partial I_B\right)+\mathcal{E}^{\text{eff}}\left(A\cup I_A:B\cup I_B\right)\right]\,.\label{EN_QES}
\end{align}

Inspired by the holographic characterizations for the entanglement negativity described earlier, we now propose DES formulae to obtain the entanglement negativity in the doubly holographic framework of the defect AdS$_3$/BCFT$_2$ scenario. In the presence of the bulk defect theory, the entanglement negativity for a bipartite mixed state $\rho_{AB}$ in the dual BCFT$_2$ involves corrections from the bulk matter fields. Following \cite{Faulkner:2013ana, Engelhardt:2014gca}, the effective matter contribution is given by the bulk entanglement negativity between the regions $\mathcal{A}$ and $\mathcal{B}$ which are obtained by splitting the codimension one region dual to $\rho_{AB}$ via the entanglement wedge cross section\footnote{Note that a defect extremal surface formula for the reflected entropy was developed in \cite{Li:2021dmf} utilizing a similar construction. Furthermore, the authors in \cite{Li:2021dmf} demonstrated the equivalence of the DES and QES formulae for the reflected entropy in the framework of defect AdS$_3$/BCFT$_2$.}. For the bipartite mixed state configuration described by two disjoint intervals $A$ and $B$ in the dual CFT$_2$, the $3d$ bulk dual DES formula for the entanglement negativity is therefore given by  
\begin{align}
\mathcal{E}^{\text{bulk}}(\mathcal{A}:\mathcal{B})=\text{min}~\underset{\Gamma}{\text{Ext}}\left[\frac{3}{16 G_N}\Big(\mathcal{L}(\gamma_{AC})+\mathcal{L}(\gamma_{BC})-\mathcal{L}(\gamma_{C})-\mathcal{L}(\gamma_{ABC})\Big)+\mathcal{E}^{\text{eff}}\left(\mathcal{A}:\mathcal{B}\right)\right],\label{DES-disj}
\end{align}
%
%
%
%
%
where $\mathcal{L}(\gamma_X)$ is the length of the bulk extremal curve homologous to the interval $X$ on the boundary CFT$_2$ and $\mathcal{E}^{\text{eff}}\left(\mathcal{A}:\mathcal{B}\right)$ denotes the effective entanglement negativity between the quantum matter fields inside the bulk regions $\mathcal{A}$ and $\mathcal{B}$. The bulk effective term in \cref{DES-disj} reduces to the effective entanglement negativity between the entanglement negativity islands $I_A$ and $I_B$ on the EOW brane as the conformal matter is present only on the EOW brane. Note that if the intervals are far away such that their entanglement wedges are disconnected, the contributions coming from the combination of bulk extremal curves vanishes identically due to phase transitions to other entropy saddles \cite{Malvimat:2018ood}.
\begin{figure}[h!]
	\centering
	\includegraphics[scale=0.65]{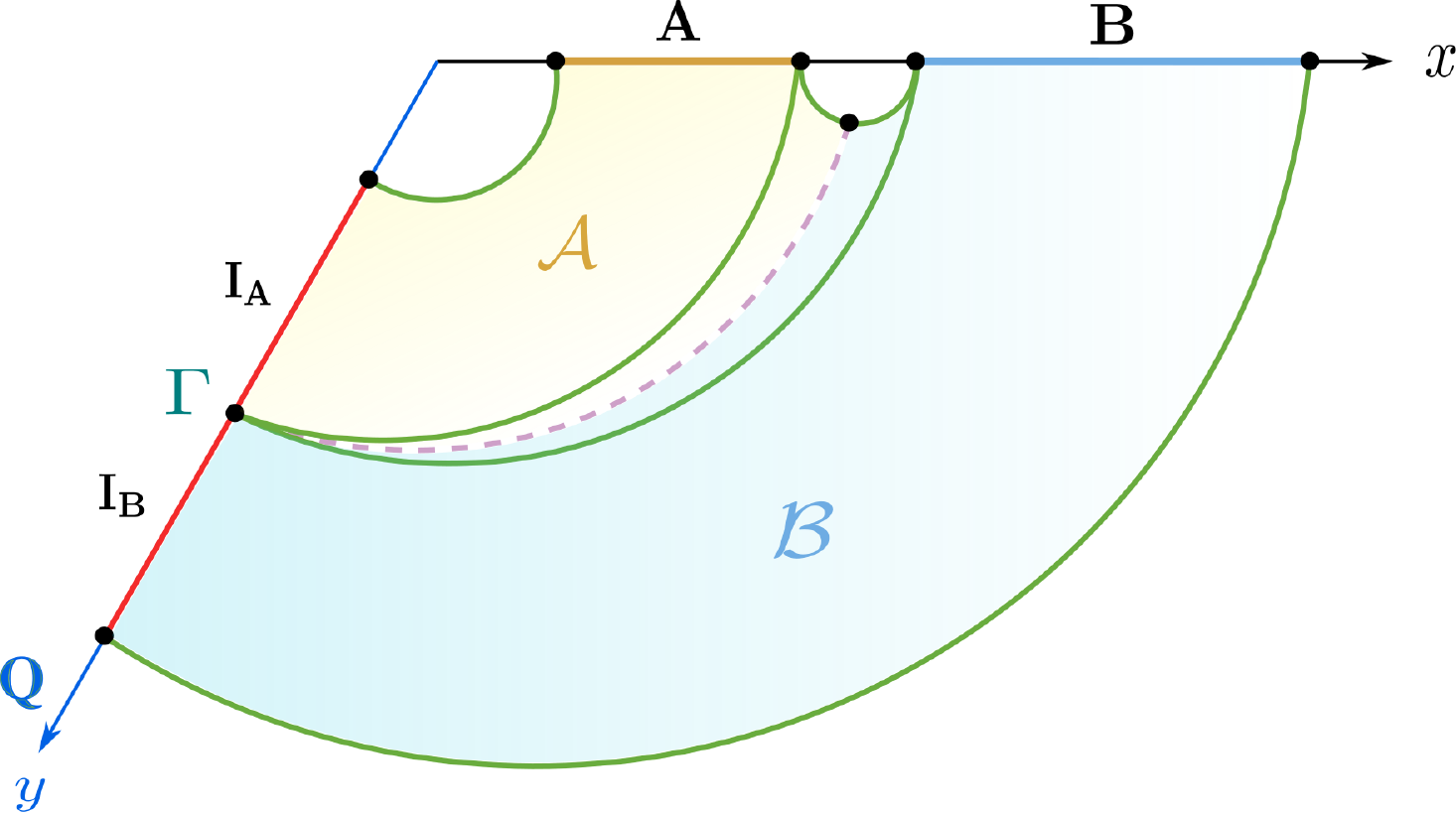}
	\caption{Schematics of the defect extremal surfaces for the entanglement negativity between two disjoint intervals $A$ and $B$. $I_A$ and $I_B$ denote the entanglement negativity islands corresponding to $A$ and $B$, respectively. The interval $C$ sandwiched between $A$ and $B$ does not have an island.}
	\label{fig:EN-DES}
\end{figure}

The DES formula for two adjacent intervals $A$ and $B$ in the bulk description may be obtained from \cref{DES-disj} through the limit $C\to\emptyset$ as follows
\begin{align}
\mathcal{E}^{\text{bulk}}(\mathcal{A}:\mathcal{B})=\text{min}~\underset{\Gamma}{\text{Ext}}\left[\frac{3}{16 G_N}\Big(\mathcal{L}(\gamma_{A})+\mathcal{L}(\gamma_{B})-\mathcal{L}(\gamma_{AB})\Big)+\mathcal{E}^{\text{eff}}\left(\mathcal{A}:\mathcal{B}\right)\right].\label{DES-adj}
\end{align}
In the following we will compute the entanglement negativity for various bipartite mixed states in a defect BCFT$_2$ through the island and the DES formulae and find exact agreement between the bulk and the boundary results.

\section{Entanglement negativity on a fixed time slice}\label{sec:EN-static}

\subsection{Two disjoint intervals}\label{sec:disj}
In this subsection we focus on the computation of the entanglement negativity for the bipartite mixed state of two disjoint intervals $A=[b_1,b_2]$ and $B=[b_3,\infty]$ on a static time-slice in the defect AdS$_3$/BCFT$_2$ framework. There are three possible phases for the entanglement negativity for this mixed state configuration based on the subsystem sizes, which we investigate below.

\subsubsection{Phase-I}\label{sec:disj-1}
\subsubsection*{Boundary description}
\begin{figure}[H]
	\centering
	\begin{subfigure}[b]{0.51\textwidth}
		\centering
		\includegraphics[width=\textwidth]{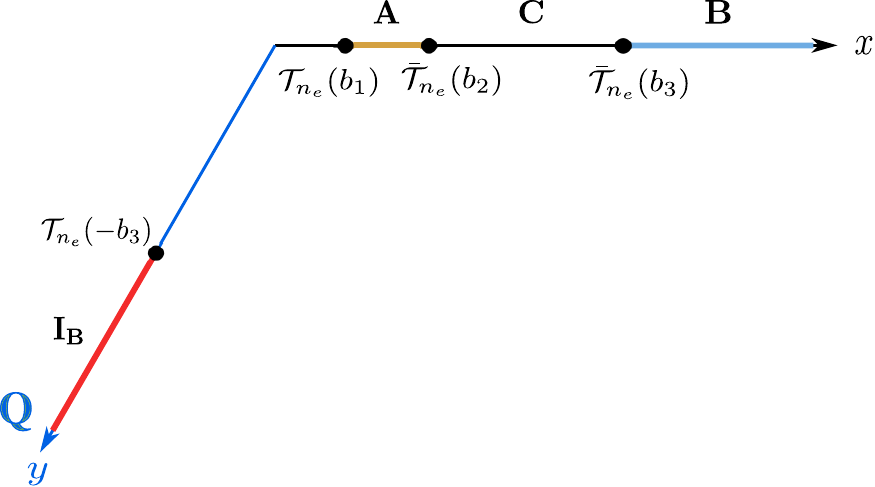}
		\caption{QES}
		\label{fig:Disj-phase1-qes}
	\end{subfigure}
	\hfill
	\begin{subfigure}[b]{0.48\textwidth}
		\centering
		\includegraphics[width=\textwidth]{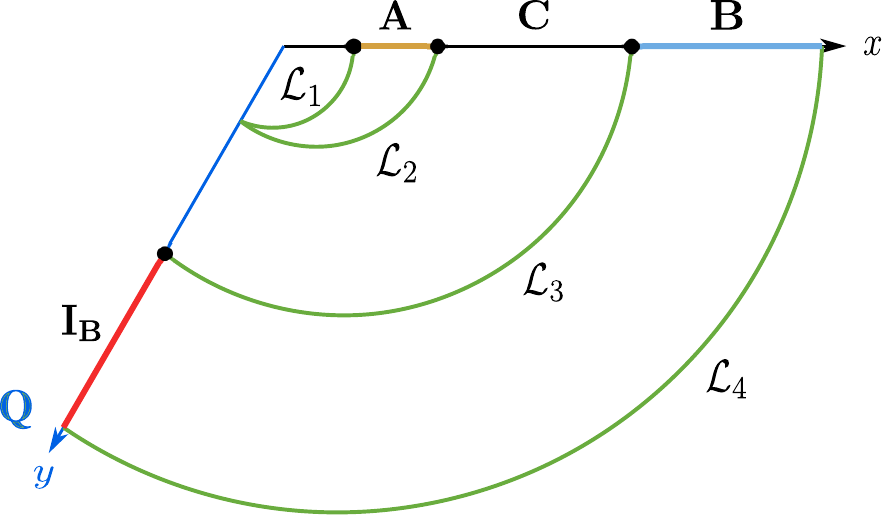}
		\caption{DES}
		\label{fig:Disj-phase1-des}
	\end{subfigure}
	\caption{Schematics of the defect extremal surface for the entanglement negativity between two disjoint intervals $A$ and $B$ in phase-I.}
	\label{fig:Disj-phase1}
\end{figure}

In this phase, the interval $C$ separating the two disjoint intervals $A$ and $B$ is large\footnote{Note that, in this phase the interval $C$ has an entanglement island. In the bulk description, this corresponds to a disconnected entanglement wedge for $A\cup B$.\label{footnote-Cisland}} and the interval $A$ is small enough such that it does not possess an entanglement entropy island. Consequently, there is no non-trivial island cross-section on the EOW brane as shown in \cref{fig:Disj-phase1}. Hence $\Gamma=\emptyset$, and the area term in the QES formula \cref{EN_QES} vanishes, namely $\mathcal{A}(\Gamma)=0$. 

The effective semi-classical entanglement negativity in this phase may be obtained through a correlation function of twist operators located at the endpoints of the intervals as follows
\begin{align}
	\mathcal{E}^{\text{eff}}\left(A:B\cup I_B\right)&=\lim_{n_e\to 1}\log\Big[\left(\epsilon_y \,\Omega(-b_3)\right)^{\Delta_{n_e}}\left<\mathcal{T}_{n_e}(b_1)\bar{\mathcal{T}}_{n_e}(b_2)\bar{\mathcal{T}}_{n_e}(b_3)\mathcal{T}_{n_e}(-b_3) \right>_{\mathrm{CFT}^{\bigotimes n_e}}\Big]\notag\\
	&\approx \lim_{n_e\to 1}\log \left[\left(\epsilon_y \,\Omega(-b_3)\right)^{\Delta_{n_e}}\left<\mathcal{T}_{n_e}(b_1)\bar{\mathcal{T}}_{n_e}(b_2)\right>_{n_e}\left<\bar{\mathcal{T}}_{n_e}(b_3)\mathcal{T}_{n_e}(-b_3)\right>_{n_e}\right]\notag\\
	&=0\,,\label{Disj-1-bdy}
\end{align}
where $\epsilon_y$ is the UV cut-off on the EOW brane $\mathbb{Q}$ and the warp factor $\Omega$ is given by \cite{Deng:2020ent}
\begin{align}
ds^2_{\text{brane}}=\Omega^{-2}(y)ds^2_{\text{flat}}~~,~~\Omega(-b_3)=\left|\frac{b_3\cos\theta_0}{\ell}\right|\,.\label{warp-factor}
\end{align}
In the second equality of \cref{Disj-1-bdy}, we have factorized the given four-point function utilizing the corresponding OPE channels. Consequently, in this phase the total entanglement negativity for the two disjoint intervals in the boundary description is vanishing.

\subsubsection*{Bulk description}
The dual bulk description for this phase has a disconnected entanglement wedge and hence we have $\Gamma=\emptyset$ similar to the boundary description. Furthermore, as the bulk matter fields are only localized on the EOW brane $\mathbb{Q}$ and $A$ has no corresponding island, the effective entanglement negativity between bulk quantum matter fields also vanishes as follows
\begin{align}
	\mathcal{E}^{\text{eff}}\left(\mathcal{A}:\mathcal{B}\right)= \mathcal{E}^{\text{eff}}\left(\emptyset:I_{B}\right)\equiv 0\,.\label{eff-vanish}
\end{align}
Hence, in the bulk description the holographic entanglement negativity is entirely given by the contribution from the areas of the defect extremal surfaces. The lengths of the bulk DES homologous to various subsystems are given by
\begin{align}
	&\mathcal{L}_{AC}=\mathcal{L}_1+\mathcal{L}_3~~~~,~~~~\mathcal{L}_{BC}=\mathcal{L}_2+\mathcal{L}_4\notag\\
	&\mathcal{L}_{C}=\mathcal{L}_2+\mathcal{L}_3~~~~~\,\,,~~~~\mathcal{L}_{ABC}=\mathcal{L}_1+\mathcal{L}_4\,.
\end{align}
Now utilizing the bulk DES formula for the entanglement negativity for two disjoint intervals in \cref{DES-disj}, we obtain
\begin{align}
	\mathcal{E}^{\text{bulk}}(\mathcal{A}:\mathcal{B})&=\frac{3}{16G_N}\left(\mathcal{L}_{AC}+\mathcal{L}_{BC}-\mathcal{L}_{C}-\mathcal{L}_{ABC}\right)=0\,.
\end{align}
Therefore, the boundary and bulk description match trivially, leading to a vanishing entanglement negativity in this phase.

\subsubsection{Phase-II}\label{sec:disj-2}
\subsubsection*{Boundary description}
Next we turn our attention towards the phase where the interval $A$ still does not possess an island, but the interval $C$ sandwiched between $A$ and $B$ is small and is therefore does not lead to an entanglement entropy island as well (cf. \cref{footnote-Cisland}). In this phase, there is no non-trivial island cross-section as depicted in \cref{fig:Disj-phase2} and hence the area term in \cref{EN_QES} vanishes identically. On the other hand, the effective semi-classical entanglement negativity is given by
\begin{align}
	\mathcal{E}^{\text{eff}}\left(A:B\cup I_B\right)&=\lim_{n_e\to 1}\log\Big[\left(\epsilon_y \,\Omega(-b_1)\right)^{\Delta_{n_e}}\left<\mathcal{T}_{n_e}(-b_1)\mathcal{T}_{n_e}(b_1)\bar{\mathcal{T}}_{n_e}(b_2)\bar{\mathcal{T}}_{n_e}(b_3) \right>_{\mathrm{CFT}^{\bigotimes n_e}} \Big].\label{en-qes-phase2}
\end{align}
As described in \cite{Kulaxizi:2014nma, Malvimat:2018txq, Malvimat:2018ood}, in the large-central charge limit the above four-point correlation function of the twist operators has the following form
\begin{align}
	\left<\mathcal{T}_{n_e}(-b_1)\mathcal{T}_{n_e}(b_1)\bar{\mathcal{T}}_{n_e}(b_2)\bar{\mathcal{T}}_{n_e}(b_3) \right>_{\mathrm{CFT}^{\bigotimes n_e}}=(1-x)^{\hat{\Delta}}\label{large-c}
\end{align}
where the conformal dimension $\hat{\Delta}$ corresponding to the dominant Virasoro conformal block, and the cross-ratio $x$ are given as
\begin{align}
	\hat{\Delta}=\frac{c}{6}\left(\frac{n_e}{2}-\frac{2}{n_e}\right)~~,~~x=\frac{(b_2-b_1)(b_3+b_1)}{(b_2+b_1)(b_3-b_1)}.\label{h-x}
\end{align}
We may now obtain the the entanglement negativity for this phase in the boundary description by substituting \cref{large-c,h-x} in \cref{en-qes-phase2} to be
\begin{align}
	\mathcal{E}^{\text{bdy}}(A:B)=\frac{c}{4}\log\left[\frac{(b_1+b_2)(b_3-b_1)}{2b_1 (b_3-b_2)}\right]\,.\label{disj2-QES-fin}
\end{align}

\subsubsection*{Bulk description}
From the bulk perspective, in this phase the entanglement wedge corresponding to $A\cup B$ is connected. However, as the interval $A$ does not have an island, the minimal entanglement wedge cross-section does not meet the EOW brane $\mathbb{Q}$ resulting in a trivial island cross section $\Gamma=\emptyset$. Hence, the effective entanglement negativity between the bulk quantum matter fields vanishes similar to \cref{eff-vanish}.
\begin{figure}[H]
	\centering
	\begin{subfigure}[b]{0.49\textwidth}
		\centering
		\includegraphics[width=\textwidth]{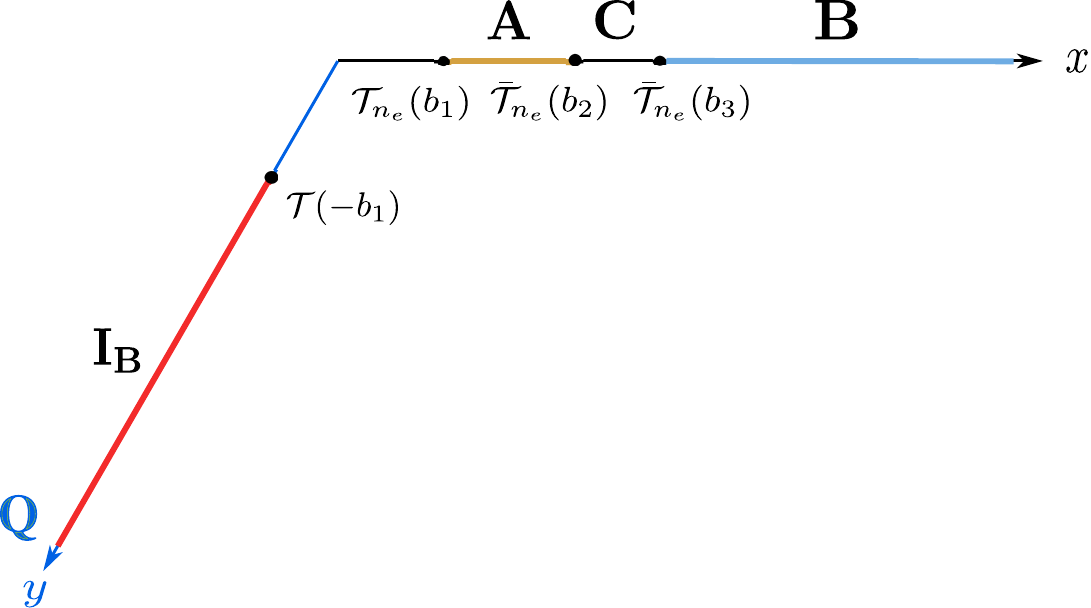}
		\caption{QES}
		\label{fig:Disj-phase2-qes}
	\end{subfigure}
	\hfill
	\begin{subfigure}[b]{0.49\textwidth}
		\centering
		\includegraphics[width=\textwidth]{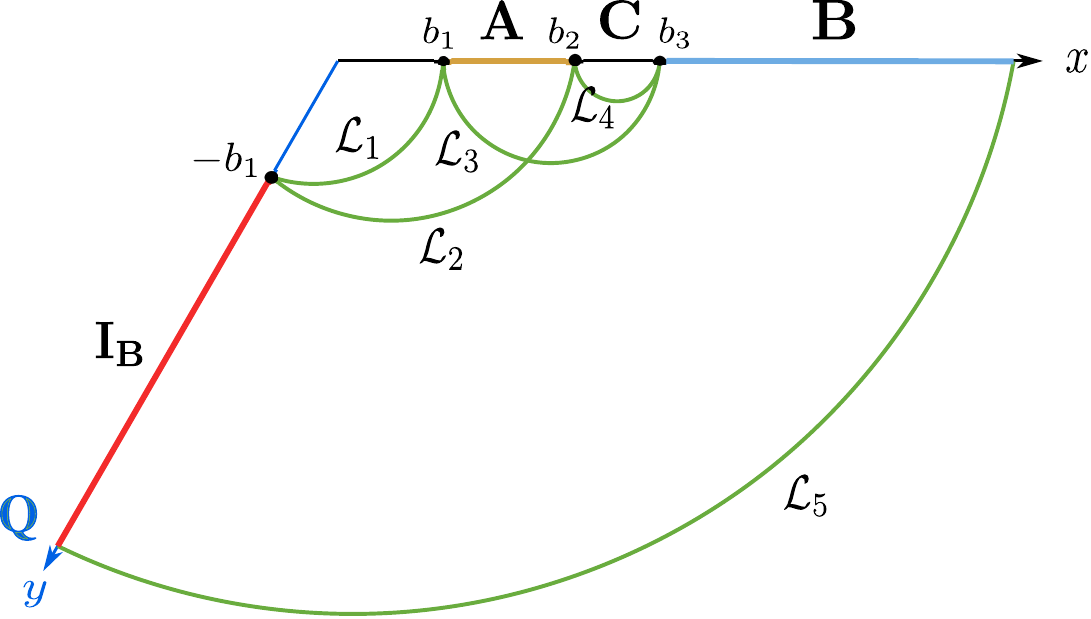}
		\caption{DES}
		\label{fig:Disj-phase2-des}
	\end{subfigure}
	\caption{Schematics of the defect extremal surface for the entanglement negativity between two disjoint intervals $A$ and $B$ in phase II.}
	\label{fig:Disj-phase2}
\end{figure}
The bulk entanglement negativity consists of the contributions from the combination of the defect extremal surfaces as depicted in \cref{fig:Disj-phase2-des}. Now utilizing \cref{DES-disj}, we may obtain the entanglement negativity between $A$ and $B$ in this phase as follows
\begin{align}
	\mathcal{E}^{\text{bulk}}(\mathcal{A}:\mathcal{B})&=\frac{3}{16G_N}\left[\mathcal{L}_3+(\mathcal{L}_2+\mathcal{L}_5)-\mathcal{L}_4-(\mathcal{L}_1+\mathcal{L}_5)\right]\notag\\
	&=\frac{3}{16G_N}\left(\mathcal{L}_2+\mathcal{L}_3-\mathcal{L}_1-\mathcal{L}_4\right)\label{disj-ph2-bulk}
\end{align}
In the framework of defect AdS$_3$/BCFT$_2$ \cite{Deng:2020ent, Chu:2021gdb, Li:2021dmf}, it was observed that the defect extremal surfaces have the same structure as the corresponding RT surfaces since the contribution from the defect matter fields turned out to be constant. The lengths of the defect extremal surfaces $\mathcal{L}_3$ and $\mathcal{L}_4$ in \cref{disj-ph2-bulk} are given by \cite{Ryu:2006bv,Takayanagi:2011zk}
\begin{align}
	\mathcal{L}_3=2 \ell\,\log\left(\frac{b_3-b_1}{\epsilon}\right)~~,~~\mathcal{L}_4=2\ell\,\log\left(\frac{b_3-b_2}{\epsilon}\right)\,,\label{L3-L4}
\end{align}
where $\epsilon$ is a UV cut-off in the dual BCFT$_2$. As described in \cite{Deng:2020ent, Chu:2021gdb}, the length of the defect extremal surface $\mathcal{L}_1$ ending on the brane $\mathbb{Q}$ is given by
\begin{align}
	\mathcal{L}_1=\ell\,\log\left(\frac{2b_1}{\epsilon}\right)+\ell\,\tanh^{-1}(\sin\theta_0)\,.\label{L1-disj2}
\end{align}
Furthermore, the length of the defect extremal surface $\mathcal{L}_2$ may be obtained as follows \cite{Ryu:2006bv}
\begin{align}
	\mathcal{L}_2=\ell\,\cosh^{-1}\left[\frac{(b_2+b_1\sin\theta_0)^2+(b_1\cos\theta_0)^2}{2 \epsilon (b_1\cos\theta_0)}\right]\,.
\end{align}
Note that, in this phase the interval $A$ is very small and therefore we may approximate the above length in the following way
\begin{align}
	\mathcal{L}_2=\ell\,\cosh^{-1}\left(\frac{b_2^2+b_1^2+2b_1b_2\sin\theta_0}{b_2^2-b_1^2}\right)+\ell\,\log\left(\frac{b_2^2-b_1^2}{\epsilon b_1}\right)\,.
\end{align}
Now utilizing the identity $\cosh^{-1}\,x+\cosh^{-1}\,y=\cosh^{-1}\left(xy+\sqrt{(x^2-1)(y^2-1)}\right)$ we finally obtain
\begin{align}
	\mathcal{L}_2&=\ell\left[\cosh^{-1}\left(\frac{b_2^2-b_1^2}{b_2^2-b_1^2}\right)+\log\left(\frac{b_2^2-b_1^2}{\epsilon b_1}\right)+\cosh^{-1}\left(\frac{1}{\cos\theta_0}\right)\right]\notag\\
	&=\ell\,\log\left(\frac{(b_1+b_2)^2}{\epsilon b_1}\right)+\ell\,\tanh^{-1}(\sin\theta_0)\,.\label{L2-final-ph2}
\end{align}
Substituting \cref{L2-final-ph2,L2-final-ph2,L3-L4} in \cref{disj-ph2-bulk} we may now obtain the entanglement negativity between $A$ and $B$ in the bulk description as follows
\begin{align}
		\mathcal{E}^{\text{bulk}}(\mathcal{A}:\mathcal{B})=\frac{3\ell}{4G_N}\log\left[\frac{(b_1+b_2)(b_3-b_1)}{2b_1 (b_3-b_2)}\right]\,.
\end{align}
Upon employing the Brown-Henneaux formula \cite{Brown:1986nw}, we observe an exact matching with the island result in \cref{disj2-QES-fin}.

\subsubsection{Phase-III}\label{sec:disj-3}
\subsubsection*{Boundary description}
\begin{figure}[ht]
	\centering
	\includegraphics[scale=0.6]{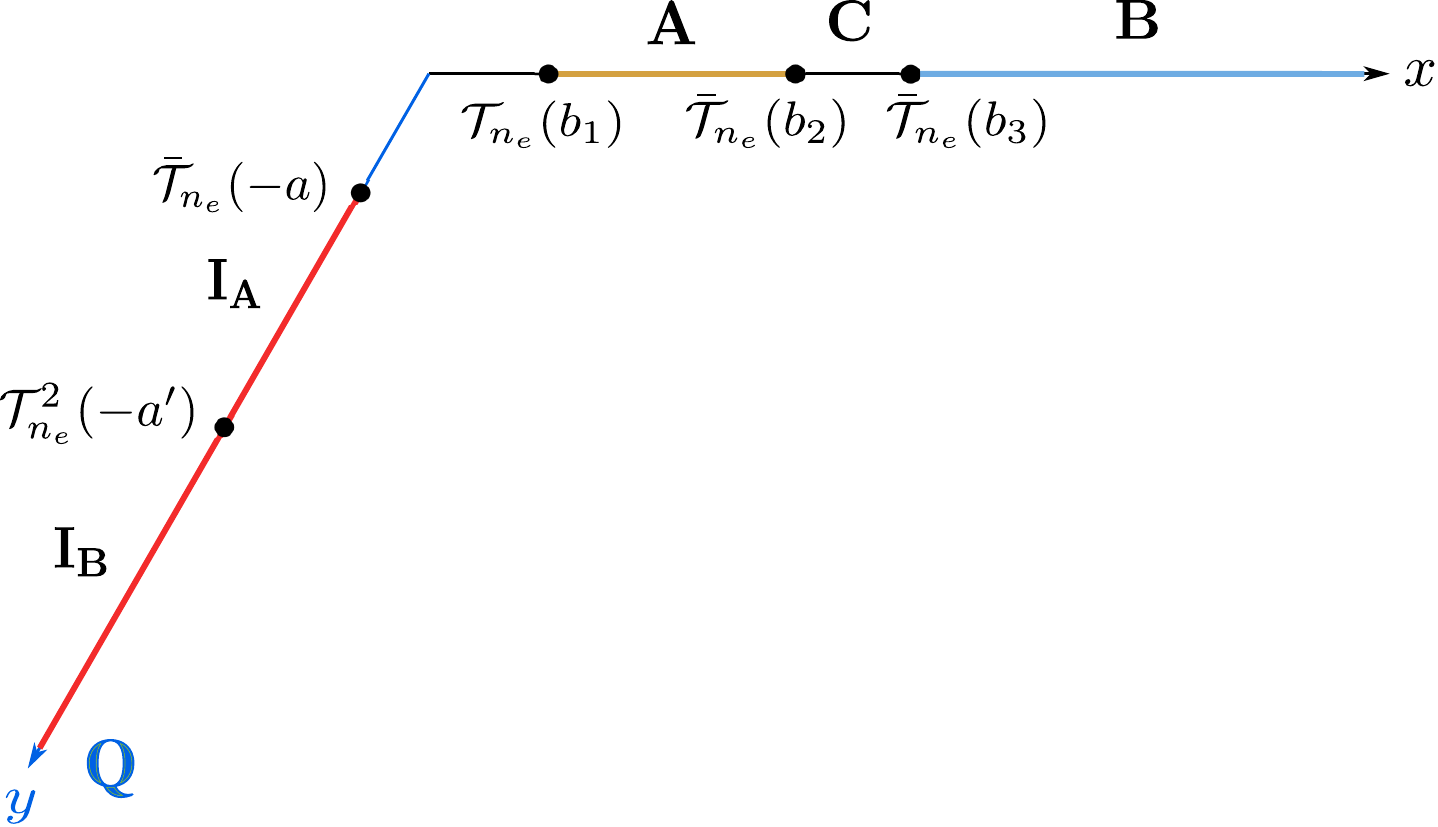}
	\caption{Schematics of the quantum extremal surface for the entanglement negativity between two disjoint intervals $A$ and $B$ in phase-III. In this phase, we have a non-trivial island cross-section on the brane at coordinate $a'$.}
	\label{fig:Disj-phase3-QES}
\end{figure}

In the final phase both the intervals $A$ and $B$ are large enough to posses entanglement islands $I_A\equiv[-a,-a']$ and $I_B\equiv[-a',-\infty]$ respectively. They are also considered to be in proximity such that they have a connected entanglement wedge as shown in \cref{fig:Disj-phase3-QES}. The area term in \cref{EN_QES} for the island cross-section $\Gamma \equiv \partial I_A \cap \partial I_B$ is then given as \cite{Deng:2020ent, Chu:2021gdb, Li:2021dmf}
\begin{align}\label{area-disj3}
	\mathcal{A}(\Gamma)=\frac{\ell}{4G_N}\tanh^{-1}(\sin\theta_0)\,,
\end{align}
The semi-classical effective entanglement negativity may be obtained in terms of the following five-point twist correlator
\begin{align}
	\mathcal{E}^{\text{eff}}\left(A\cup I_A:B\cup I_B\right)=\lim_{n_e\to 1}\log\Big[&\left(\epsilon_y\Omega(-a)\right)^{\Delta_{n_e}}\left(\epsilon_y\Omega(-a')\right)^{\Delta_{n_e}^{(2)}}\notag\\
	&\times\left<\mathcal{T}_{n_e}(b_1)\bar{\mathcal{T}}_{n_e}(-a)\bar{\mathcal{T}}_{n_e}(b_2)\mathcal{T}_{n_e}^2(-a')\bar{\mathcal{T}}_{n_e}(b_3)\right>_{\mathrm{CFT}^{\bigotimes n_e}}\Big],\label{eff-neg-disj3}
\end{align}
where $\epsilon_y$ is a UV regulator on the AdS$_2$ brane $\mathbb{Q}$ and the warp factor $\Omega(-a')$ is given in \cref{warp-factor}. The five-point twist correlator in \cref{eff-neg-disj3} have the following factorization \cite{KumarBasak:2020ams} in the corresponding OPE channel
\begin{align}
	&\left<\mathcal{T}_{n_e}(b_1)\bar{\mathcal{T}}_{n_e}(-a)\bar{\mathcal{T}}_{n_e}(b_2)\mathcal{T}_{n_e}^2(-a')\bar{\mathcal{T}}_{n_e}(b_3)\right> \approx \left<\mathcal{T}_{n_e}(b_1)\bar{\mathcal{T}}_{n_e}(-a)\right>\left<\bar{\mathcal{T}}_{n_e}(b_2)\mathcal{T}_{n_e}^2(-a')\bar{\mathcal{T}}_{n_e}(b_3)\right>\notag\\
	&\qquad \qquad \qquad \qquad \qquad \qquad \qquad =\frac{1}{(a+b_1)^{2\Delta_{n_e}}}\frac{C_{\bar{\mathcal{T}}_{n_e}\mathcal{T}^2_{n_e}\bar{\mathcal{T}}_{n_e}}}{(b_3+a')^{\Delta_{n_e}^{(2)}}(b_2+a')^{\Delta_{n_e}^{(2)}}(b_3-b_2)^{2\Delta_{n_e}-\Delta_{n_e}^{(2)}}},\label{factorization}
\end{align}
where $\Delta_{n_e}$ and $\Delta_{n_e}^{(2)}$ are the conformal dimensions of the twist operators $\mathcal{T}_{n_e}$ and $\mathcal{T}_{n_e}^2$ respectively and are given as \cite{Calabrese:2012ew, Calabrese:2012nk}
\begin{align}
	\Delta_{n_e}=\frac{c}{12}\left(1-\frac{1}{n_e}\right)~~,~~\Delta_{n_e}^{(2)}=\frac{c}{6}\left(\frac{n_e}{2}-\frac{2}{n_e}\right)\,.\label{Conformal-dim}
\end{align}
Note that the point $a$ on the brane is determined by the DES for the subsystem $A$ to be $a=b_1$ \cite{Deng:2020ent}. Therefore, by utilizing the contractions in \eqref{factorization} along with the areas term in \cref{area-disj3}, we may obtain the generalized negativity in the boundary description from \cref{QES-island} to be
\begin{align}
	\mathcal{E}^{\text{bdy}}_{\text{gen}}(A:B)=\frac{c}{4}\left[\tanh^{-1}(\sin\theta_0)+\log\frac{\ell(b_2+a')(b_3+a')}{a'(b_3-b_2)\epsilon_y\cos\theta_0}\right].\label{disj3-QES-N-ext}
\end{align}
The extremization with respect to the island cross-section $\Gamma$ with the coordinate $a'$ on the brane leads to
\begin{align}
	\partial_{a'}\,\mathcal{E}^{\text{bdy}}_{\text{gen}}=0~~\implies~~ a'=\sqrt{b_2b_3} \, ~.
\end{align} 
Substituting this into \cref{disj3-QES-N-ext}, we may obtain the total entanglement negativity between $A$ and $B$ in phase-III from the boundary description to be
\begin{align}
	\mathcal{E}^{\text{bdy}}(A:B)=\frac{c}{4}\left[\tanh^{-1}(\sin\theta_0)+\log\left(\frac{\sqrt{b_3}+\sqrt{b_2}}{\sqrt{b_3}-\sqrt{b_2}}\right)+\log\left(\frac{\ell}{\epsilon_y\cos\theta_0}\right)\right].\label{disj3-QES-fin}
\end{align}

\subsubsection*{Bulk description}
\begin{figure}[ht]
	\centering
	\includegraphics[scale=0.65]{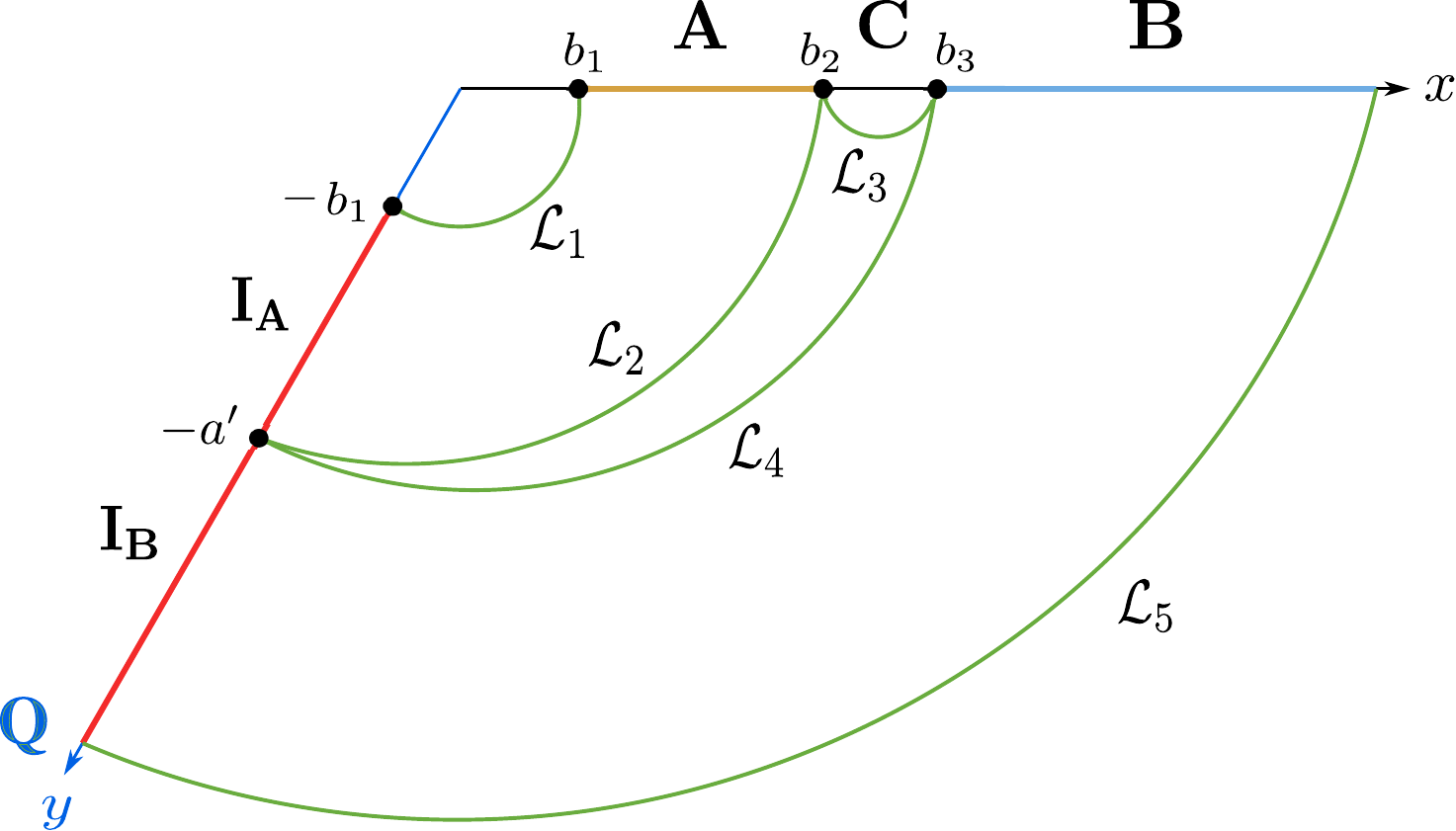}
	\caption{Schematics of the defect extremal surface for the entanglement negativity between two disjoint intervals $A$ and $B$ in phase-III. In this phase, the EWCS ends on the island cross-section $\Gamma$ on the EOW brane.}
	\label{fig:Disj-phase3-DES}
\end{figure}

The bulk description in phase-III consists of a connected entanglement wedge and the minimal cross-section ends on the EOW brane. 
The configuration is sketched in \cref{fig:Disj-phase3-DES}. Since the bulk quantum matter is entirely situated on the EOW brane, the effective entanglement negativity between the bulk quantum matter fields in the bulk regions $\mathcal{A}$ and $\mathcal{B}$ reduces to the effective matter negativity between the corresponding island regions $I_A$ and $I_B$,
\begin{align}
	\mathcal{E}^{\text{eff}}\left(\mathcal{A}:\mathcal{B}\right)&\equiv\mathcal{E}^{\text{eff}}(I_A:I_B)=\lim_{n_e\to 1}\log\left[\left(\epsilon_y\Omega(-a')\right)^{\Delta_{n_e}^{(2)}}\left<\mathcal{T}_{n_e}(-b_1)\bar{\mathcal{T}}^2_{n_e}(-a')\right>_{\mathrm{BCFT}^{\bigotimes n_e}}\right],
\end{align}
where $\epsilon_y$ is the UV cut-off on the EOW brane and $\Omega$ is the conformal factor as given in \cref{warp-factor}. Utilizing the doubling trick \cite{Cardy:2004hm, Sully:2020pza} the above two-point function in the defect BCFT$_2$ may be reduced to a four-point correlator of chiral twist fields in a CFT$_2$ defined on the whole complex plane. As described in \cite{Sully:2020pza}, the four-point correlator in the chiral CFT$_2$ has two dominant channels depending on the cross-ratio as follows.
\paragraph{I. BOE channel:} In this channel the two point correlator factorizes into two one-point functions in the BCFT$_2$ as follows
\begin{align}
	\left<\mathcal{T}_{n_e}(b_1)\bar{\mathcal{T}}^2_{n_e}(-a')\right>_{\mathrm{BCFT}^{\bigotimes n_e}}&=\left<\mathcal{T}_{n_e}(-b_1)\right>_{\mathrm{BCFT}^{\bigotimes n_e}}\left<\bar{\mathcal{T}}^2_{n_e}(-a')\right>_{\mathrm{BCFT}^{\bigotimes n_e}}\notag\\
	&=\frac{\epsilon_y^{\Delta_{n_e}+\Delta_{n_e}^{(2)}}}{(2b_1)^{\Delta_{n_e}}(2a')^{\Delta_{n_e}^{(2)}}}
\end{align}
%
Therefore, the effective bulk entanglement negativity in this phase is given by
\begin{align}
	\mathcal{E}^{\text{eff}}(I_A:I_B)=\frac{c}{4}\log\frac{2\ell}{\epsilon_y\cos\theta_0}\,.\label{BOE-neg}
\end{align}
Note that this effective entanglement negativity is equal to the R\'enyi entropy of order half for the interval $I_A$ (or $I_B$) which is consistent with the expectations from quantum information theory.

\paragraph{II. OPE channel:} In this channel, the two-point correlator of twist fields on the BCFT$_2$ reduces to a three-point correlator of chiral twist fields on the full complex plane as follows \cite{Cardy:2004hm, Sully:2020pza, Li:2021dmf}
\begin{align}
	\left<\mathcal{T}_{n_e}(-b_1)\bar{\mathcal{T}}^2_{n_e}(-a')\right>_{\mathrm{BCFT}^{\bigotimes n_e}}&=	\left<\bar{\mathcal{T}}_{n_e}(-b_1)\mathcal{T}_{n_e}(b_1)\bar{\mathcal{T}}^2_{n_e}(-a')\right>_{\mathrm{CFT}^{\bigotimes n_e}}\notag\\
	&=\frac{C_{\bar{\mathcal{T}}_{n_e}\mathcal{T}^2_{n_e}\bar{\mathcal{T}}_{n_e}}}{(a'^2-b_1^2)^{\Delta_{n_e}^{(2)}}(2b_1)^{\Delta_{n_e}^{(2)}-2\Delta_{n_e}}}\,.
\end{align}
Therefore, the effective bulk entanglement negativity  in this channel is given by
\begin{align}
	\mathcal{E}^{\text{eff}}(I_A:I_B)=\frac{c}{4}\log\left[\frac{\ell (a'^2-b_1^2)}{a' b_1\epsilon_y\cos\theta_0}\right]\,.
\end{align}

As shown in \cref{fig:Disj-phase3-DES}, the contribution to the bulk entanglement negativity from the defect extremal surfaces homologous  to different combinations of subsystems is given by
\begin{align}
	&\frac{3}{16G_N}\left(\mathcal{L}_2+\mathcal{L}_4-\mathcal{L}_3\right)\notag\\ 
&=\frac{3\ell}{16G_N}\Bigg[\cosh^{-1}\left\{\frac{(b_2+a'\sin\theta_0)^2+(a'\cos\theta_0)^2}{2\epsilon (a'\cos\theta_0)}\right\}+\cosh^{-1}\left\{\frac{(b_3+a'\sin\theta_0)^2+(a'\cos\theta_0)^2}{2\epsilon (a'\cos\theta_0)}\right\}\notag\\&\quad\quad\quad\quad\quad\quad\quad\quad\quad\quad\quad\quad\quad\quad\quad\quad\quad\quad\quad\quad\quad\quad-2\log\left(\frac{b_3-b_2}{\epsilon}\right)\Bigg]
\end{align}
The entanglement negativity between the disjoint intervals $A$ and $B$ is obtained by extremizing the generalized negativity over the position of the island cross-section $\Gamma$. For the OPE channel of the effective bulk entanglement negativity there is no extremal solution while for the BOE channel we obtain
\begin{align}
	\partial_{a'}\,\mathcal{E}^{\text{bulk}}_{\text{gen}}=0~~\implies~~ a'=\sqrt{b_2b_3}
\end{align}
Substituting this and utilizing the proximity limit $b_3\to b_2$ in the intermediate step, we obtain the entanglement negativity between $A$ and $B$ in the bulk description as follows
\begin{align}
	\mathcal{E}^{\text{bulk}}&=\frac{3\ell}{16G_N}\left[\cosh^{-1}\left(\frac{b_2+b_3+2\sqrt{b_2b_3}\sin\theta_0}{(b_3-b_2)\cos\theta_0}\right)\right]+\frac{c}{4}\log \left( \frac{2\ell}{\epsilon_y \cos \theta_0} \right)\notag\\
	&=\frac{3\ell}{16G_N}\left[\log\left(\frac{\sqrt{b_3}+\sqrt{b_2}}{\sqrt{b_3}-\sqrt{b_2}}\right)+\cosh^{-1}\left(\frac{1}{\cos\theta_0}\right)\right]+\frac{c}{4}\log \left( \frac{2\ell}{\epsilon_y \cos \theta_0} \right)
\end{align}
The above expression for the holographic entanglement negativity matches exactly with the QES result in \cref{disj3-QES-fin} obtained through the island formula \cref{QES-island}. This provides yet another consistency check of our holographic construction for the entanglement negativity in the defect AdS$_3$/BCFT$_2$ scenario.

\subsection{Two adjacent intervals} \label{sec:adj}
Having computed the entanglement negativity for configurations involving two disjoint intervals, we now turn our attention to the mixed state of two adjacent intervals $A = [0,b_1]$ and $B = [b_1,b_2]$ on a fixed time-slice in the AdS$_3$/BCFT$_2$ model. The interval $A$ in this case always possess an entanglement island as it starts from the interface between the EOW brane and the asymptotic boundary. We however, have two possible phases for this case based on the size of the interval $B$ which are described below.

\subsubsection{Phase-I}\label{sec:adj-1}

\subsubsection*{Boundary description}
\begin{figure}[ht]
	\centering
	\includegraphics[scale=0.65]{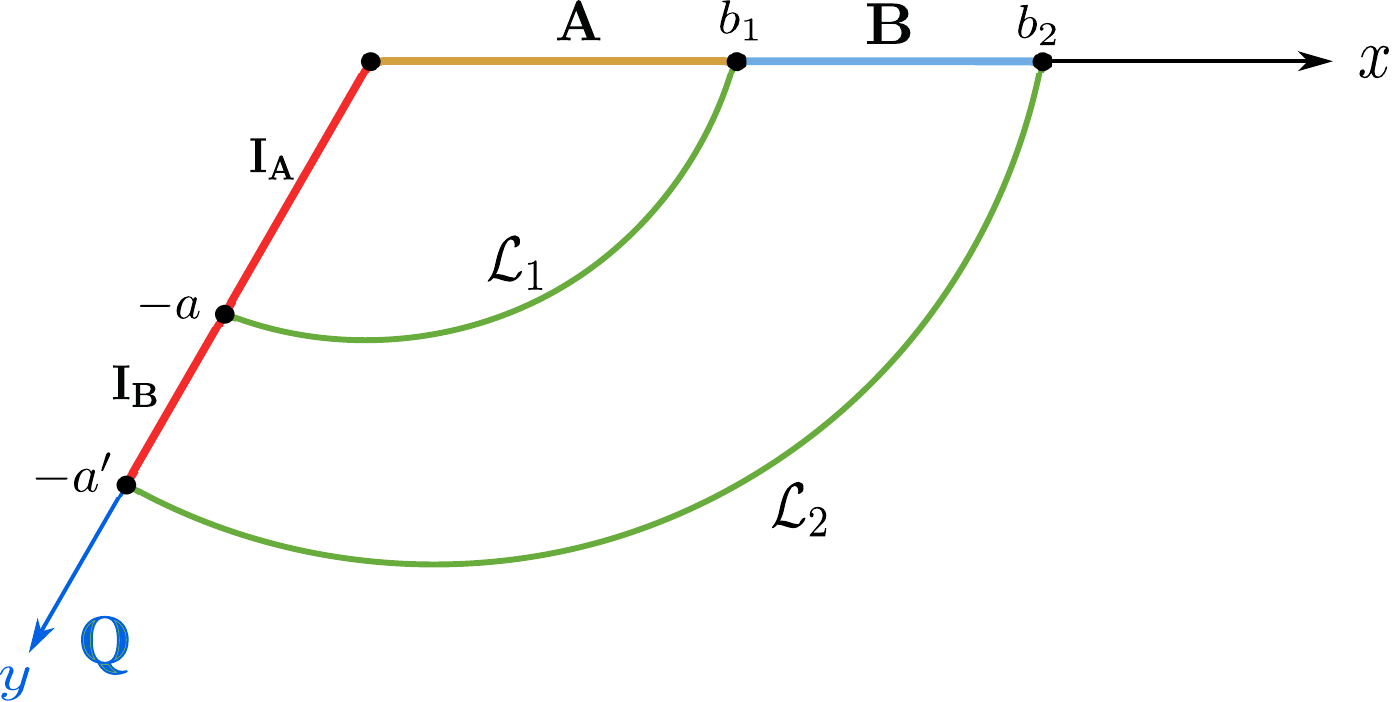}
	\caption{Schematics of the defect extremal surface for the entanglement negativity between two adjacent intervals $A$ and $B$ in phase-I. $I_A$ and $I_B$ on the EOW brane describe the entanglement island corresponding to intervals $A$ and $B$ respectively.}
	\label{fig:Adj-phase-1}
\end{figure}

For this phase, we consider that the interval $B$ is large enough to posses an entanglement island described as $I_B$ in \cref{fig:Adj-phase-1}. The area term in \cref{EN_QES} for the point $\Gamma = \partial I_A \cap \partial I_B$ is as given in \cref{area-term}. The effective semi-classical entanglement negativity is given by the following two point twist correlator
\begin{equation}\label{qes-eff-adj-1}
\begin{aligned}
\mathcal{E}^\text{eff}(A \cup I_A : B \cup I_B)& = \lim_{n_e \to 1} \log \left[\left(\epsilon_y \, \Omega(-a)\right)^{\Delta_{n_e}^{(2)}}\left<\mathcal{T}^2_{n_e}(b_1)\bar{\mathcal{T}}^2_{n_e}(-a)\right>_{\mathrm{CFT}^{\bigotimes n_e}}\right] \\
& = \frac{c}{4} \log \left[ \frac{\ell (b_1 + a)^2}{\epsilon \, \epsilon_y \, a \cos \theta_0}\right],
\end{aligned}
\end{equation}
where $\epsilon$ and $\epsilon_y$ are the UV cut-offs on the asymptotic boundary and the EOW brane $\mathbb{Q}$ respectively, and the warp factor $\Omega(a)$ is as given in \cref{warp-factor}. The point $a = b_1$ on the brane is determined through the entanglement entropy computation of the interval $A$. Using this in \cref{qes-eff-adj-1} along with the area term, we may obtain the total entanglement negativity in the boundary description to be
\begin{equation} \label{qes-adj-1}
\mathcal{E}^\text{bdy}(A:B) = \frac{c}{4} \left[\log \left( \frac{2 b_1}{\epsilon} \right) + \log \left( \frac{2\ell}{\epsilon_y \cos \theta_0} \right) + \tanh^{-1}(\sin \theta_0) \right],
\end{equation}
where we have used the Brown-Henneaux formula in the area term \cite{Brown:1986nw}.
\subsubsection*{Bulk description}
In the double holographic description, the entanglement wedge corresponding to the subsystem $A \cup B$ is connected in the bulk. The contribution to the effective entanglement negativity between the bulk matter fields in regions $\mathcal{A}$ and $\mathcal{B}$ arises solely from the quantum matter fields situated on the EOW brane as follows
\begin{equation}\label{des-eff-adj-1}
\begin{aligned}
\mathcal{E}^\text{eff} (\mathcal{A} : \mathcal{B}) = \mathcal{E}^\text{eff} (I_A : I_B) &= \lim_{n_e \to 1} \log \left[\left(\epsilon_y \, \Omega(-a)\right)^{\Delta_{n_e}^{(2)}}\left<\mathcal{T}_{n_e}(-a')\bar{\mathcal{T}}^2_{n_e}(-a)\right>_{\mathrm{BCFT}^{\bigotimes n_e}}\right]\\
& = \frac{c}{4} \log \left( \frac{2\ell}{\epsilon_y \cos \theta_0} \right) \, .
\end{aligned}
\end{equation}
Utilizing \cref{DES-adj}, the total entanglement negativity for this case including the contribution from the combinations of the bulk extremal curves is obtained to be
\begin{equation} \label{des-adj-1}
\begin{aligned}
\mathcal{E}^\text{bulk}(A : B) & = \mathcal{E}^\text{eff} (\mathcal{A} : \mathcal{B}) + \frac{3}{16 G_N} 2 \mathcal{L}_1\\
& = \frac{c}{4} \log \left( \frac{2\ell}{\epsilon_y \cos \theta_0} \right) + \frac{3\ell}{8 G_N} \left[\log \left( \frac{2 b_1}{\epsilon} \right) + \tanh^{-1}(\sin \theta_0) \right],
\end{aligned}
\end{equation}
where we have used the fact that the entanglement entropy computation for the interval $A$ fixes $a = b_1$. On utilization of the Brown-Henneaux formula \cite{Brown:1986nw}, the above expression matches exactly with the result obtained from the boundary perspective in \cref{qes-adj-1}.

\subsubsection{Phase-II}\label{sec:adj-2}

\subsubsection*{Boundary description}
\begin{figure}[ht]
	\centering
	\includegraphics[scale=0.65]{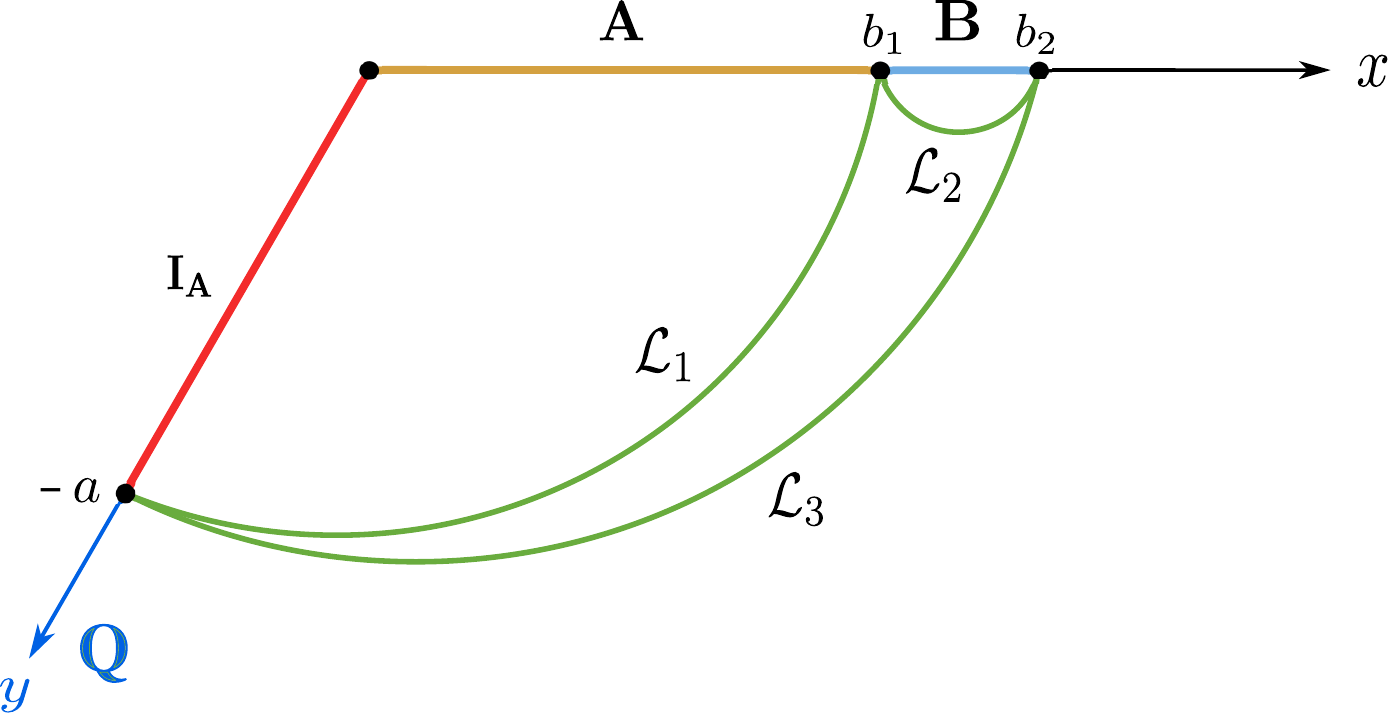}
	\caption{Schematics of the defect extremal surface for the entanglement negativity between two adjacent intervals $A$ and $B$ in phase-II. In this phase, the island of the interval $B$ is an empty set.}
	\label{fig:Adj-phase-2}
\end{figure}

For this phase, we now consider the case where the interval $B$ is small such that it lacks an entanglement entropy island as shown in \cref{fig:Adj-phase-2}. This implies that the island cross-section $\Gamma$ is a null set. The remaining effective semi-classical entanglement negativity is obtained through the following three point twist correlator
\begin{equation} \label{qes-eff-adj-2}
\begin{aligned}
\mathcal{E}^\text{eff}(A \cup I_A : B \cup I_B)& = \lim_{n_e \to 1} \log \left[\left(\epsilon_y \, \Omega(-a)\right)^{\Delta_{n_e}}\left<\mathcal{T}_{n_e}(-a) \bar{\mathcal{T}}^2_{n_e}(b_1) \mathcal{T}_{n_e}(b_2)\right>_{\mathrm{CFT}^{\bigotimes n_e}}\right] \\
& = \frac{c}{4} \log \left[ \frac{(b_1 + a) (b_2 - b_1)}{(b_2 + a) \epsilon} \right].
\end{aligned}
\end{equation}
Again, the point on the AdS$_2$ brane $\mathbb{Q}$ is fixed to be $a = b_1$ through the entanglement entropy computation of the interval $A$. Utilizing this value of $a$, we may obtain the total entanglement negativity for this phase in the boundary description to be
\begin{equation} \label{qes-adj-2}
\mathcal{E}^\text{bdy}(A : B) = \frac{c}{4} \log \left[ \frac{2 b_1 (b_2 - b_1)}{(b_2 + b_1) \epsilon} \right].
\end{equation}

\subsubsection*{Bulk description}
For the bulk description of this phase, we observe in \cref{fig:Adj-phase-2} that the entanglement wedge for the subsystem $A \cup B$ is connected. However, since the interval $B$ does not have an entanglement island, the effective entanglement negativity term in \cref{DES-adj} vanishes. The only contribution to the total entanglement negativity comes from the lengths of the extremal curves labelled as $\mathcal{L}_i$ ($i=1,2,3$) in \cref{fig:Adj-phase-2}. To this end, we note that the lengths of the extremal curves $\mathcal{L}_1$ and $\mathcal{L}_2$ have the same form as given in \cref{L1-disj2,L3-L4} respectively. Using similar approximations as were employed for the bulk description in subsection \ref{sec:disj-2}, the length of the extremal curve $\mathcal{L}_3$ may be computed to be
\begin{equation}
\begin{aligned}
\mathcal{L}_3 = \ell\,\log \left( \frac{(b_1+b_2)^2}{\epsilon b_1} \right) + \ell\,\tanh^{-1}(\sin \theta_0)\,,
\end{aligned}
\end{equation}
where we have used $a = b_1$. We may now obtain the total entanglement negativity for this phase using \cref{DES-adj} to be
\begin{equation} \label{des-adj-2}
\mathcal{E}^\text{bdy}(A : B) = \frac{3\ell}{8 G_N} \log \left[ \frac{2 b_1 (b_2 - b_1)}{(b_2 + b_1) \epsilon} \right],
\end{equation}
which on utilization of the usual Brown-Henneaux formula \cite{Brown:1986nw} matches exactly with the result obtained through the boundary description in \cref{qes-adj-2}.

\section{Time dependent entanglement negativity in black holes}\label{sec:Time-dependent-EN}
In this section we investigate the nature of mixed state entanglement through the entanglement negativity in a time-dependent defect AdS$_3$/BCFT$_2$ scenario involving an eternal black hole in the effective two-dimensional description \cite{Li:2021dmf, Chu:2021gdb}. The lower dimensional effective model involves the appearance of entanglement islands during the emission of the Hawking radiation from the eternal black hole.

\subsection{Review of the eternal black hole in AdS/BCFT}
\label{sec:review-BH}
As described in \cite{Li:2021dmf,Chu:2021gdb}, we consider a BCFT$_2$ defined on the half-plane $(x,\tau \geq 0)$. The corresponding bulk dual is described by the Poincar\'e AdS$_3$ geometry truncated by an end-of-the-world (EOW) brane located at the hypersurface $\tau=-z \tan\theta_0$. Here $\theta_0$ is the angle made by the EOW brane with the vertical, and $\tau$ and $z$ are the timelike\footnote{Note that in the Euclidean signature, there is no essential difference between the timelike and spacelike coordinates and the present parametrization is a convenient choice adapted in \cite{Li:2021dmf, Chu:2021gdb}.} and holographic coordinates respectively.

\begin{figure}[H]
	\centering
\begin{subfigure}[b]{0.58\textwidth}
	\centering
	\includegraphics[width=\textwidth]{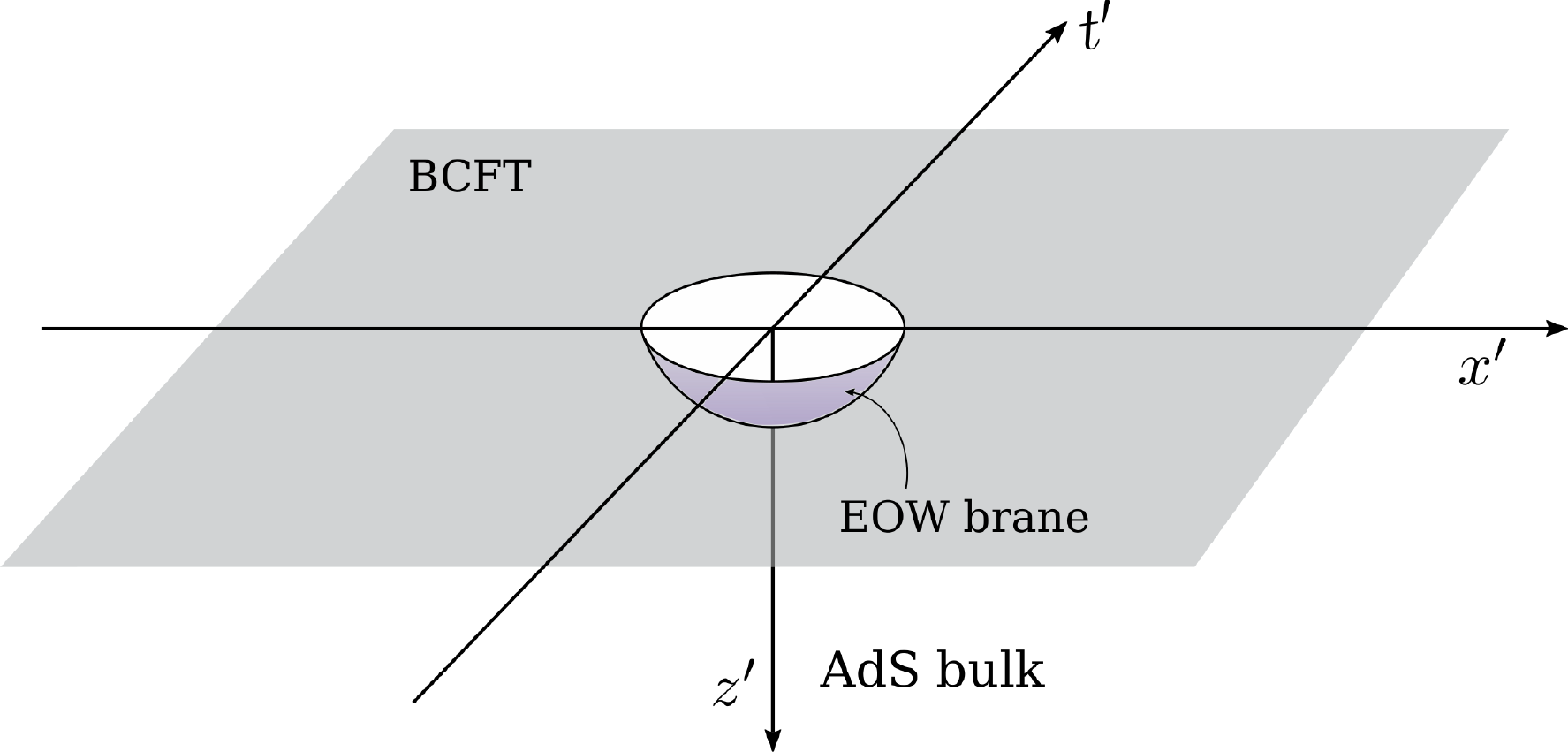}
	\caption{Euclidean AdS/BCFT with the BCFT defined outside the circle.}
	\label{fig:BH-model1}
\end{subfigure}
\hfill
\begin{subfigure}[b]{0.41\textwidth}
	\centering
	\includegraphics[width=\textwidth]{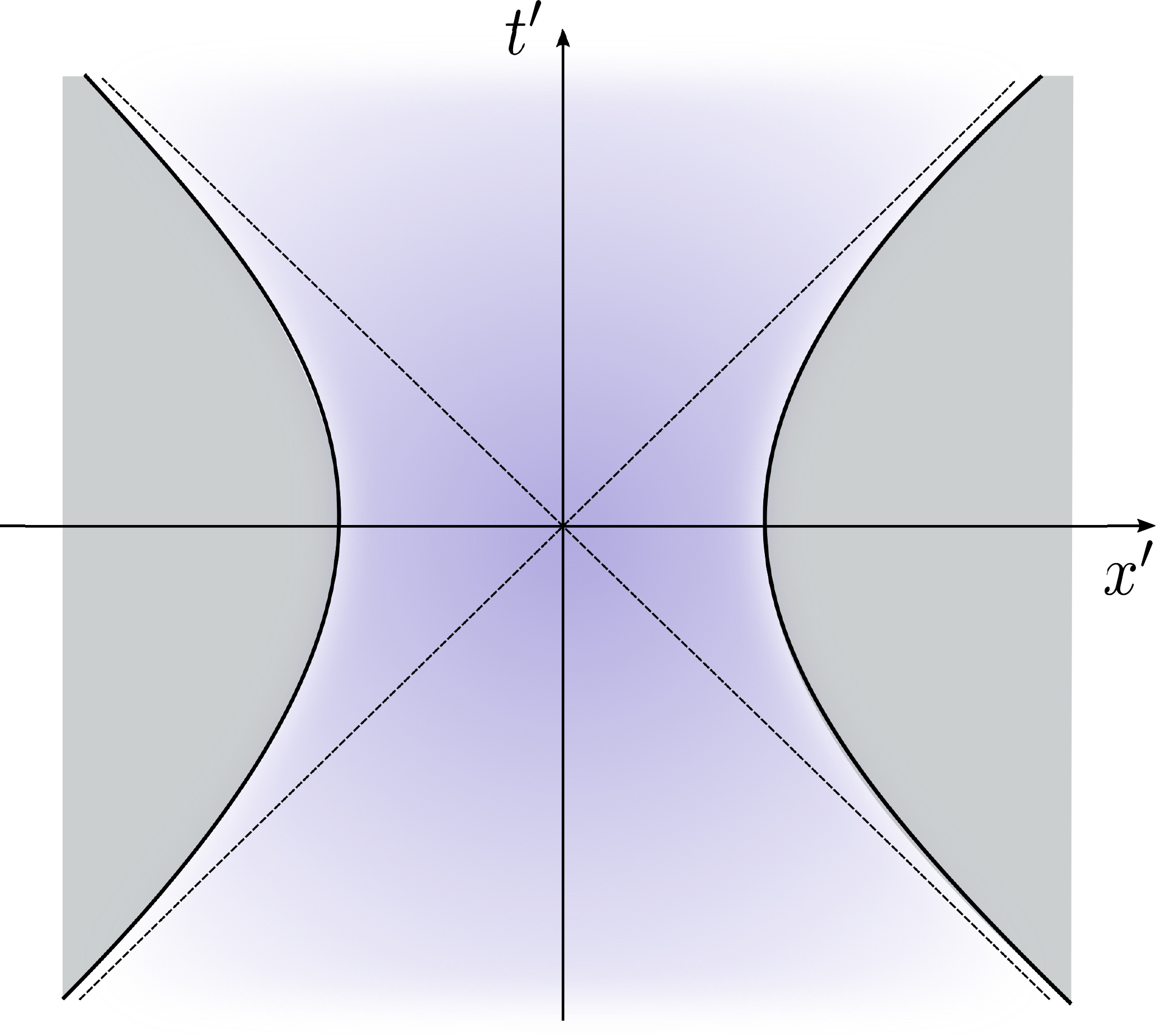}
	\caption{2d eternal black hole in Lorentzian signature.}
	\label{fig:BH-model-Lorentzian}
\end{subfigure}
\caption{ }
\label{fig:BH-model2}
\end{figure}

Utilizing a global conformal map, the boundary of the BCFT$_2$ is then mapped to a circle 
\begin{align}
	x'^2+\tau'^2=1.\label{circle}
\end{align}
The bulk dual of such a global conformal transformation is given by the following Banados map \cite{Takayanagi:2011zk,Li:2021dmf,Chu:2021gdb}
\begin{align}
	&\tau'=1+\frac{\tau-\frac{1}{2} \left(\tau^2+x^2+z^2\right)}{1-\tau+\frac{1}{4} \left(\tau^2+x^2+z^2\right)},\notag\\
	&x'=\frac{x}{1-\tau+\frac{1}{4} \left(\tau^2+x^2+z^2\right)},\label{Banados}\\
	&z'=\frac{z}{1-\tau+\frac{1}{4} \left(\tau^2+x^2+z^2\right)}\,.\notag 
\end{align}
The EOW brane is mapped to a portion of a sphere under these bulk transformations,
\begin{align}
	x'^2+\tau'^2+(z'+\tan\theta_0)^2=\sec^2\theta_0.
\end{align}
Note that, as the above transformation is a global conformal map, the metric in the bulk dual spacetime as well as the metric induced on the EOW brane are preserved under the Banados map \cref{Banados}. The schematics of this time-dependent AdS/BCFT scenario is depicted in \cref{fig:BH-model1}.

Finally employing the partial Randall-Sundrum reduction combined with the AdS$_3$/BCFT$_2$ correspondence discussed in \cite{Li:2021dmf, Chu:2021gdb, Deng:2020ent}, one obtains a two-sided $(1+1)$-dimensional eternal black hole on the EOW brane which is coupled to the BCFT$_2$ outside the circle \cref{circle} in the $2d$ effective description. The schematics of the configuration is depicted in \cref{fig:BH-model-Lorentzian}. The hybrid manifold consisting of a $2d$ eternal black hole with a fluctuating geometry coupled to the flat BCFT$_2$ may conveniently be described in terms of the Rindler coordinates $(X,T)$ defined through
\begin{align}
	x'=e^X \cosh T~~,~~t'\equiv-i\tau'=e^X \sinh T. \label{Rindler}
\end{align}
These Rindler coordinates naturally capture the near-horizon geometry of the $2d$ black hole \cite{Chu:2021gdb}.

In the following, we will compute the entanglement negativity for various bipartite states involving two disjoint and two adjacent intervals in the time-dependent defect AdS/BCFT scenario discussed above. In this regard, we will employ the semi-classical island formula \cref{EN_QES} in the lower dimensional effective description as well as the doubly holographic defect extremal surface proposals in \cref{DES-disj,DES-adj} and find exact agreement between the two.

\subsection{Entanglement negativity between black hole interiors}
In this subsection, we compute the time-dependent entanglement negativity between different regions of the black hole interior. As described in \cite{Li:2021dmf,Chu:2021gdb} the black hole region $B$ is defined as the space-like interval from $Q \equiv (t'_0,-x'_0)$ to $P \equiv (t'_0,x'_0)$ as shown in \cref{fig:BH-BH-conn}. We perform the computations in the Euclidean signature with $\tau'_0=it'_0$ and subsequently obtain the final result in Lorentzian signature through an analytic continuation. Depending on the configuration of the extremal surface for the entanglement entropy of $B$, there are two possible phases for the extremal surfaces corresponding to the entanglement negativity between the black hole subsystems $B_L$ and $B_R$.

\subsubsection{Connected phase}
\begin{figure}[ht]
	\centering
	\includegraphics[scale=.7]{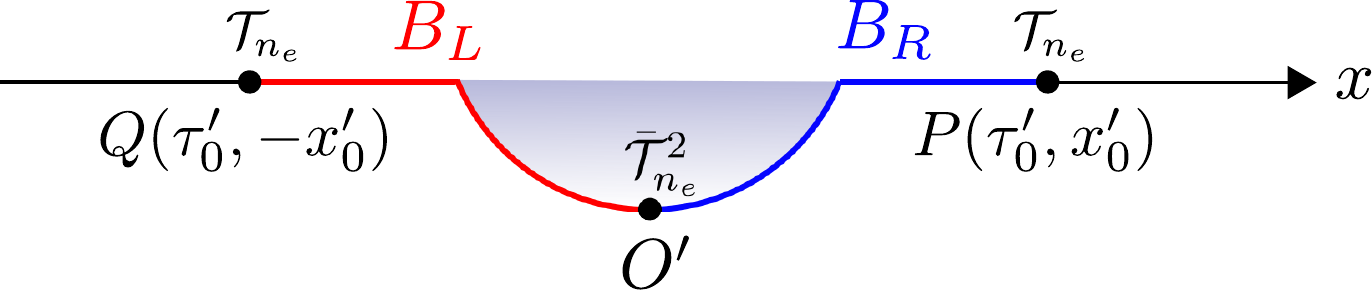}
	\caption{Schematics of the quantum extremal surface for the entanglement negativity between black hole interiors in the connected phase at a constant time slice.}
	\label{fig:BH-BH-conn}
\end{figure}

The connected phase corresponds to the scenario where there is no entanglement island for the radiation bath in the effective boundary description as shown in \cref{fig:BH-BH-conn}. From the bulk perspective, this corresponds to a connected extremal surface for $B_L \cup B_R$. In this phase, it is required to compute the entanglement negativity between the two adjacent intervals $B_L\equiv|O'Q|$ and $B_R\equiv |O'P|$, where the point $O'$ is dynamical as it resides on the EOW brane with a gravitational theory. 

\subsubsection*{Boundary description}
In the $2d$ boundary description, the effective semi-classical entanglement negativity between $B_L$ and $B_R$ may be computed through the three-point correlation function of twist operators as follows
\begin{align}
	\mathcal{E}^{\text{eff}}\left(B_L:B_R\right)=\lim_{n_e\to 1}\log \left[\left(\epsilon_y\Omega_{O'}\right)^{\Delta_{n_e}^{(2)}}\left<\mathcal{T}_{n_e}(Q)\bar{\mathcal{T}}^2_{n_e}(O')\mathcal{T}_{n_e}(P)\right>\right]\label{BH_BH_1}.
\end{align}
It is convenient to perform the computations in the un-primed coordinates $(y,x)$ given in \cref{x-y-coordinates}, where $y$ measures the distance along the EOW brane and $x$ is the spatial coordinate describing the BCFT$_2$. In these coordinates, the conformal factor associated with the dynamical point $O'$ on the EOW brane $\mathbb{Q}$ is given by \cite{Deng:2020ent,Chu:2021gdb,Li:2021dmf}
\begin{align}
	\Omega_{O'}(y)=\left|\frac{y\cos\theta_0}{\ell}\right|\,.\label{warp-factor-BH}
\end{align}
where $\ell$ is the AdS$_3$ radius inherited from the bulk geometry.
The form of the CFT$_2$ three-point function in \cref{BH_BH_1} is given by
\begin{align}
	\left<\mathcal{T}_{n_e}(Q)\bar{\mathcal{T}}^2_{n_e}(O')\mathcal{T}_{n_e}(P)\right>=C_{\mathcal{T}^{}_{n_e}\mathcal{T}_{n_e}^2\mathcal{T}^{}_{n_e}}|O'P|^{-\Delta_{n_e}^{(2)}}|O'Q|^{-\Delta_{n_e}^{(2)}}|PQ|^{-2\Delta^{}_{n_e}+\Delta_{n_e}^{(2)}}\,.\label{3-point}
\end{align}
where $C_{\mathcal{T}^{}_{n_e}\mathcal{T}_{n_e}^2\mathcal{T}_{n_e}^{}}$ is the constant OPE coefficient which is neglected henceforth. Substituting \cref{warp-factor-BH,3-point} in \cref{BH_BH_1}, we may obtain the following expression for the generalized entanglement negativity between $B_L$ and $B_R$ in the boundary description
\begin{align}
\mathcal{E}^{\text{bdy}}_{\text{gen}}\left(B_L:B_R\right)=\frac{c}{4}\log\left[\frac{(\tau_0+y)^2+x_0^2}{2x_0}\right]+\frac{c}{4}\log\left(\frac{\ell}{\epsilon_y\,y\cos\theta_0}\right)+\frac{c}{4}\tanh^{-1}(\sin\theta_0)\,,\label{BH-BH-gen}
\end{align}
where we have added the area term \cref{area-term} corresponding to the point $O'$ on the EOW brane in the QES formula. The above expression is extremized over the position $y$ of the dynamical point $O'$ to obtain
\begin{align}
	y_0=\sqrt{\tau_0^2+x_0^2}\,.
\end{align}
Substituting the above expression in \cref{BH-BH-gen}, the semi-classical entanglement negativity in the $2d$ effective boundary description is obtained as follows
\begin{align}
	\mathcal{E}^{\text{bdy}}\left(B_L:B_R\right)=\frac{c}{4}\log\left[\frac{\tau_0+\sqrt{\tau_0^2+x_0^2}}{x_0}\right]+\frac{c}{4}\log\left(\frac{\ell}{\epsilon_y\cos\theta_0}\right)+\frac{c}{4}\tanh^{-1}(\sin\theta_0)\,.
\end{align}
Now transforming back to the primed coordinates using \cref{Banados} and analytically continuing to the Lorentzian signature, the above expression reduces to
\begin{align}
	\mathcal{E}^{\text{bdy}}\left(B_L:B_R\right)=\frac{c}{4}\log\left[\frac{x^{\prime \,2}_0-t^{\prime \,2}_0-1+\sqrt{4x^{\prime \,2}_0+\left(x^{\prime \,2}_0-t^{\prime \,2}_0-1\right)^2}}{2x'_0}\right]&+\frac{c}{4}\log\left(\frac{\ell}{\epsilon_y\cos\theta_0}\right)\notag\\
	&+\frac{c}{4}\tanh^{-1}(\sin\theta_0)\,.
\end{align}
In terms of the Rindler coordinates $(X,T)$, the final result for the entanglement negativity between the black hole interiors becomes
\begin{align}
	\mathcal{E}^{\text{bdy}}\left(B_L:B_R\right)=\frac{c}{4}\log\left[\frac{e^{2X_0}-1+\sqrt{4 e^{2X_0}\cosh^2 T+\left(e^{2X_0}-1\right)^2}}{2 e^{X_0}\cosh T}\right]&+\frac{c}{4}\log\left(\frac{\ell}{\epsilon_y\cos\theta_0}\right)\notag\\
	&+\frac{c}{4}\log\left(\frac{\cos\theta_0}{1-\sin\theta_0}\right)\,,\label{BH-BH-bdy-fin}
\end{align}
where $X_0$ describes the boundary of the black hole region at a fixed Rindler time $T$. Note that the above expression for the entanglement negativity between $B_L$ and $B_R$ is a decreasing function of the Rindler time $T$ in this phase.

\subsubsection*{Bulk description}
\begin{figure}[ht]
	\centering
	\includegraphics[scale=0.7]{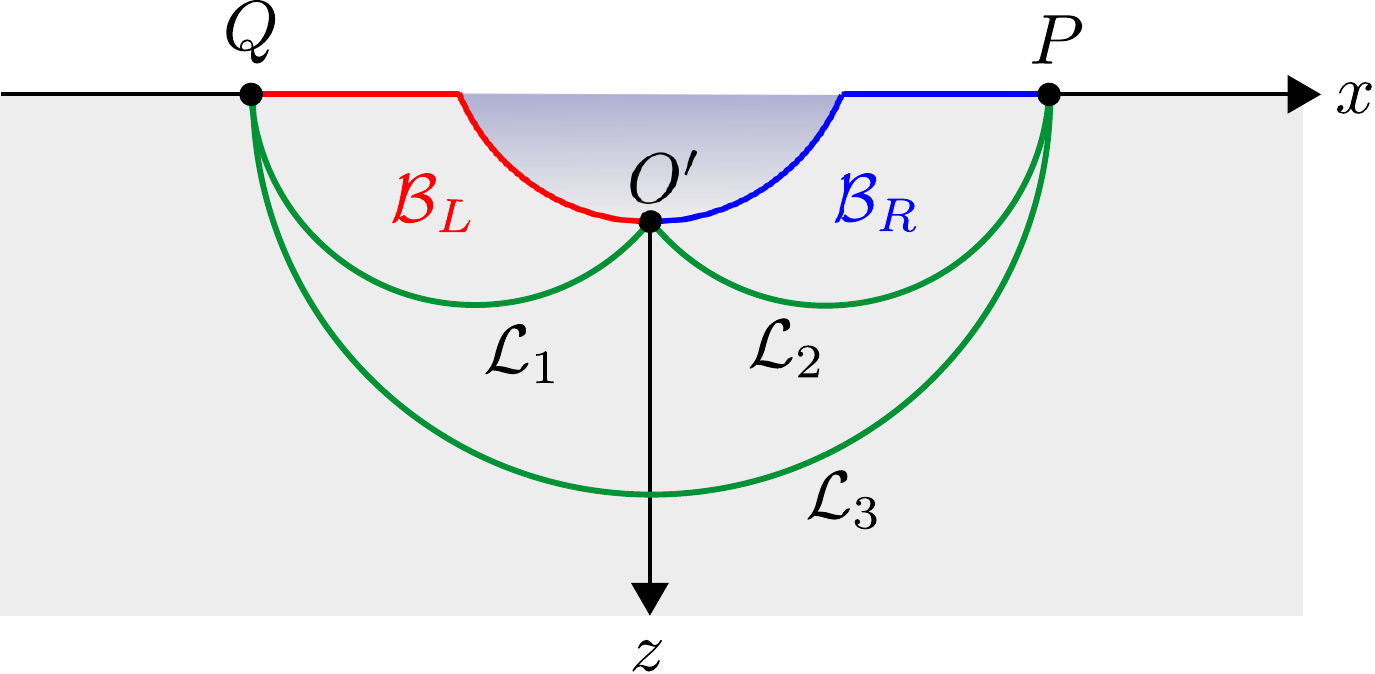}
	\caption{Schematics of the defect extremal surface for the entanglement negativity between black hole interiors in the connected phase. The bulk extremal curves homologous to $B_L$, $B_R$ and $B_L\cup B_R$ are given by $\mathcal{L}_1$, $\mathcal{L}_2$ and $\mathcal{L}_3$ respectively.}
	\label{fig:BH-BH-conn-DES}
\end{figure}

Next we focus on the three-dimensional bulk description for the connected phase  of the entanglement negativity between the black hole interiors. To compute the holographic entanglement negativity, we note that the mixed state configuration described by $B_L$ and $B_R$ corresponds to the case of two adjacent intervals $|O'P|$ and $|O'Q|$. The configuration of the bulk extremal curves homologous to various subsystems under consideration is depicted in \cref{fig:BH-BH-conn-DES}. Now employing the DES formula given in \cref{DES-adj}, we may obtain
\begin{align}
	\mathcal{E}^{\text{bulk}}_{\text{gen}}\left(\mathcal{B}_L:\mathcal{B}_R\right)=\frac{3}{16G_N}\left(\mathcal{L}_1+\mathcal{L}_2-\mathcal{L}_3\right)+\mathcal{E}^{\text{eff}}\left(\mathcal{B}_L:\mathcal{B}_R\right)\,,\label{DES-BH-BH}
\end{align}
where $\mathcal{L}_1$, $\mathcal{L}_2$ and $\mathcal{L}_3$ are the lengths of the bulk extremal curves homologous to $|O'P|$, $|O'Q|$ and $|PQ|$ respectively and $\mathcal{E}^{\text{eff}}\left(\mathcal{B}_L:\mathcal{B}_R\right)$ denotes the effective entanglement negativity between bulk quantum matter fields residing on the EOW brane.

In the un-primed coordinates, the Cauchy slice on the EOW brane is in a pure state as described in \cite{Li:2021dmf}. Hence, the effective entanglement negativity in \cref{DES-BH-BH} may be obtained through the R\'enyi entropy of order half for a part of matter fields on the EOW brane. Consequently, similar to \cref{BOE-neg}, the effective entanglement negativity is a constant given by
\begin{align}
	\mathcal{E}^{\text{eff}}\left(\mathcal{B}_L:\mathcal{B}_R\right)=\frac{c}{4}\log\frac{2\ell}{\epsilon_y\cos\theta_0}\,.\label{eff-BH-BH}
\end{align}

As the effective entanglement negativity turns out to be a constant, the entanglement negativity in this phase is determined entirely through the algebraic sum of the lengths of the extremal curves in \cref{DES-BH-BH}. To obtain the lengths of these extremal curve, we employ the un-primed coordinate system with the Poincar\'e AdS$_3$ metric \cite{Li:2021dmf}. Under the bulk map in \cref{Banados}, the coordinates of $P$ and $Q$ may be mapped to $(\tau_0,x_0,0)$ and $(\tau_0,-x_0,0)$ where
\begin{align}
	\tau_0=\frac{2(x^{\prime 2}_0+\tau^{\prime 2}_0-1)}{(\tau^{\prime }_0+1)^2+x^{\prime 2}_0}~~,~~x_0=\frac{4x'_0}{(\tau'_0+1)^2+x^{\prime 2}_0}\,.
\end{align}
Utilizing the left-right $\mathbb{Z}_2$ symmetry of the configuration, we may set the coordinates of the dynamical point $O'$ on the brane as $O': (-z\tan\theta_0,0,z)$ where $z$ is determined through the extremization of the generalized negativity functional in \cref{DES-BH-BH}. The lengths of the extremal curves may now be obtained in the un-primed coordinates through the standard Poincar\'e AdS$_3$ result as follows \cite{Ryu:2006bv, Hubeny:2007xt}
\begin{align}
	&\mathcal{L}_1=\ell\,\cosh^{-1}\left[\frac{(\tau_0+z\tan\theta_0)^2+x_0^2+z^2}{2\,z}\right]-\ell\,\log\left[\frac{4\epsilon}{(\tau'_0+1)^2+x^{\prime 2}_0}\right]=\mathcal{L}_2\,,\notag\\
	&\mathcal{L}_3=2\ell\,\log(2x_0)-2\ell\,\log\left[\frac{4\epsilon}{(\tau'_0+1)^2+x^{\prime 2}_0}\right]\,. \label{L-BH-BH-conn}
\end{align}
In the above expression, $\epsilon$ is the UV cut-off for the original BCFT$_2$ in the primed coordinates and the second logarithmic term arises due to the cut-off in the un-primed coordinates (cf. the Banados map in \cref{Banados}). Now extremizing the generalized negativity with respect to $z$ we may obtain the position of $O'$ to be
\begin{align}
	\partial_z\mathcal{E}^{\text{bulk}}_{\text{gen}}=0\implies z=\sqrt{x_0^2+\tau_0^2}\cos\theta_0\,.
\end{align}
Substituting the above value of $z$ in \cref{DES-BH-BH}, we may obtain the bulk entanglement negativity between $\mathcal{B}_L$ and $\mathcal{B}_R$ as follows
\begin{align}
	\mathcal{E}^{\text{bulk}}\left(\mathcal{B}_L:\mathcal{B}_R\right)=\frac{c}{4}\left[\cosh^{-1}\left(\frac{\sqrt{x_0^2+\tau_0^2}+\tau_0\sin\theta_0}{ x_0\cos\theta_0}\right)+\log\frac{\ell}{\epsilon_y\cos\theta_0}\right]\,,\label{DES-BH-BH1}
\end{align}
where the effective contribution from the quantum matter fields given in \cref{eff-BH-BH} has been included. Now utilizing the hyperbolic identity
\begin{align}
	\cosh^{-1}\left(\frac{\sqrt{x_0^2+\tau_0^2}+\tau_0\sin\theta_0}{ x_0\cos\theta_0}\right)=\log\left(\frac{\tau_0+\sqrt{x_0^2+\tau_0^2}}{x_0}\right)+\cosh^{-1}(\sec\theta_0)\,,
\end{align}
\cref{DES-BH-BH1} may be expressed as
\begin{align}
	\mathcal{E}^{\text{bulk}}\left(\mathcal{B}_L:\mathcal{B}_R\right)=\frac{c}{4}\left[\log\left(\frac{\tau_0+\sqrt{x_0^2+\tau_0^2}}{ x_0}\right)+\log\frac{\ell}{\epsilon_y\cos\theta_0}+\log\frac{\cos\theta_0}{1-\sin\theta_0}\right]
\end{align}
Transforming to the primed coordinates using \cref{Banados} and subsequently to the Rindler coordinates \cref{Rindler} via the analytic continuation $\tau' = i t'$, we may obtain the bulk entanglement negativity between $\mathcal{B}_L$ and $\mathcal{B}_R$ to be
\begin{align}
	\mathcal{E}^{\text{bulk}}\left(\mathcal{B}_L:\mathcal{B}_R\right)=\frac{c}{4}\log\left[\frac{e^{2X_0}-1+\sqrt{4 e^{2X_0}\cosh^2 T+\left(e^{2X_0}-1\right)^2}}{2 e^{X_0}\cosh T}\right]&+\frac{c}{4}\log\left(\frac{\ell}{\epsilon_y\cos\theta_0}\right)\notag\\
	&+\frac{c}{4}\log\left(\frac{\cos\theta_0}{1-\sin\theta_0}\right)\,.
\end{align}
The above expression matches identically with the boundary QES result in \cref{BH-BH-bdy-fin} which provides a strong consistency check of our holographic construction.


\subsubsection{Disconnected phase}
\begin{figure}[ht]
	\centering
	\includegraphics[scale=.7]{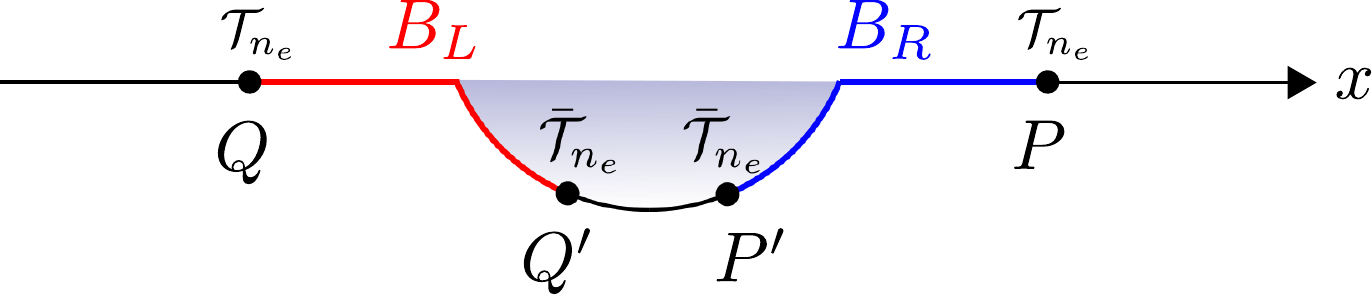}
	\caption{Schematics of the quantum extremal surface for the entanglement negativity between black hole interiors in the disconnected phase}
	\label{fig:BH-BH-disc}
\end{figure}

In this subsection, we concentrate on the disconnected phase for the extremal surface for $B_L\cup B_R$, depicted in \cref{fig:BH-BH-disc}. In this case, there are entanglement islands corresponding to the radiation bath on the EOW brane, and a part of the entanglement wedge for the radiation bath is subtended on the brane. This splits the black hole regions into two disjoint subsystems, namely $B_R\equiv|PP'|$ and $B_L\equiv|QQ'|$, where the points $P'$ and $Q'$ are determined by the extremal surface for $B_L\cup B_R$.

\subsubsection*{Boundary description}
In the two-dimensional effective boundary description, the area term for the generalized entanglement negativity vanishes since there is no non-trivial island cross-section, $\partial B_L\cap \partial B_R=\emptyset$. The effective semi-classical entanglement negativity between  $B_L$ and $B_R$ may be computed through the following four-point correlator of twist operators placed at the endpoints of the intervals,
\begin{align}
\mathcal{E}^{\text{eff}}\left(B_L:B_R\right)=\lim_{n_e \to 1}\log \left[\left(\epsilon_y\Omega_{P'}\right)^{\Delta_{n_e}}\left(\epsilon_y\Omega_{Q'}\right)^{\Delta_{n_e}}\left<\mathcal{T}_{n_e}(Q)\bar{\mathcal{T}}_{n_e}(Q')\bar{\mathcal{T}}_{n_e}(P')\mathcal{T}_{n_e}(P)\right>\right]
\end{align}
As indicated by the disconnected extremal surfaces shown in \cref{fig:BH-BH-disc}, the above four-point correlator factorizes into the product of two 2-point correlators as follows
\begin{align}
	\left<\mathcal{T}_{n_e}(Q)\bar{\mathcal{T}}_{n_e}(Q')\bar{\mathcal{T}}_{n_e}(P')\mathcal{T}_{n_e}(P)\right>\approx 	\left<\mathcal{T}_{n_e}(Q)\bar{\mathcal{T}}_{n_e}(Q')\right>\, \left<\mathcal{T}_{n_e}(P)\bar{\mathcal{T}}_{n_e}(P')\right>\,.
\end{align}
Now utilizing \cref{Conformal-dim}, we may observe that, in the replica limit $n_e\to 1$, the above correlation function vanishes identically. Hence, in this phase the total entanglement negativity between the black hole interiors is also vanishing.

\subsubsection*{Bulk description}
As depicted in \cref{fig:BH-BH-disc}, the entanglement wedges corresponding to the subsystems ${B}_L$ and ${B}_R$ are naturally disconnected and hence, the configuration corresponds to two disjoint intervals on the boundary which are far away from each other. In this case, the area contribution to the bulk entanglement negativity vanishes \cite{Malvimat:2018txq,Malvimat:2018ood}.
The effective entanglement negativity between portions of bulk quantum matter on the EOW brane is given by the BCFT correlation function of twist fields inserted at $P'$ and $Q'$ as follows
\begin{align}
	\mathcal{E}^{\text{eff}}=\lim_{n_e \to 1}\log\left<\mathcal{T}_{n_e}(P')\bar{\mathcal{T}}_{n_e}(Q')\right>_{\mathrm{BCFT}^{\bigotimes n_e}}\,.\label{effEN-BH-BH-disc}
\end{align}
The coordinates of $P'$ and $Q'$ are obtained via extremizing the generalized entropy functional for $B_L\cup B_R$ which, in the $(y,x)$ coordinates, are given by
$(\tau_0,x_0)$ and $(\tau_0,-x_0)$ respectively \cite{Chu:2021gdb}. Now employing the doubling trick \cite{Cardy:2004hm,Sully:2020pza}, the correlation function in \cref{effEN-BH-BH-disc} may be expressed as a chiral four-point function on the full complex plane as 
\begin{align}
	\left<\mathcal{T}_{n_e}(P')\bar{\mathcal{T}}_{n_e}(Q')\right>_{\mathrm{BCFT}^{\bigotimes n_e}}=\left<\mathcal{T}_{n_e}(P')\bar{\mathcal{T}}_{n_e}(Q')\bar{\mathcal{T}}_{n_e}(Q'')\mathcal{T}_{n_e}(P'')\right>_{\mathrm{CFT}^{\bigotimes n_e}}
\end{align}
where $P'':(-\tau_0,x_0)$ and $Q'': (-\tau_0,-x_0)$ are the image points of $P'$ and $Q'$ upon reflection through the boundary at $\tau=0$. The above four point correlator is again factorized into two two-point functions in the dominant channel and similar to the previous subsection, the effective semi-classical entanglement negativity vanishes. Hence, the boundary QES result is reproduced through the bulk computations.
\begin{figure}[H]
	\centering
	\includegraphics[scale=.7]{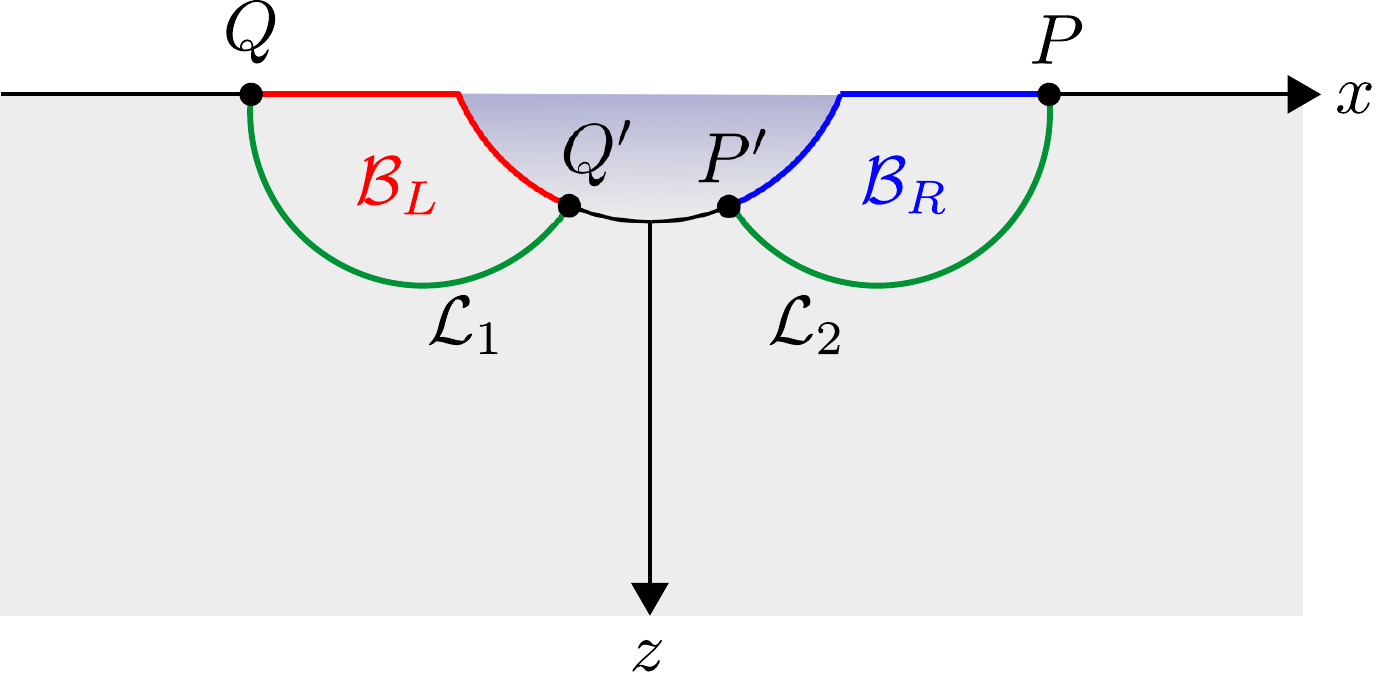}
	\caption{Schematics of the defect extremal surface for the entanglement negativity between black hole interiors in the disconnected phase}
	\label{fig:BH-BH-disc-des}
\end{figure}

\subsubsection{Page curve}
\begin{figure}[ht]
	\centering
	\includegraphics[scale=0.6]{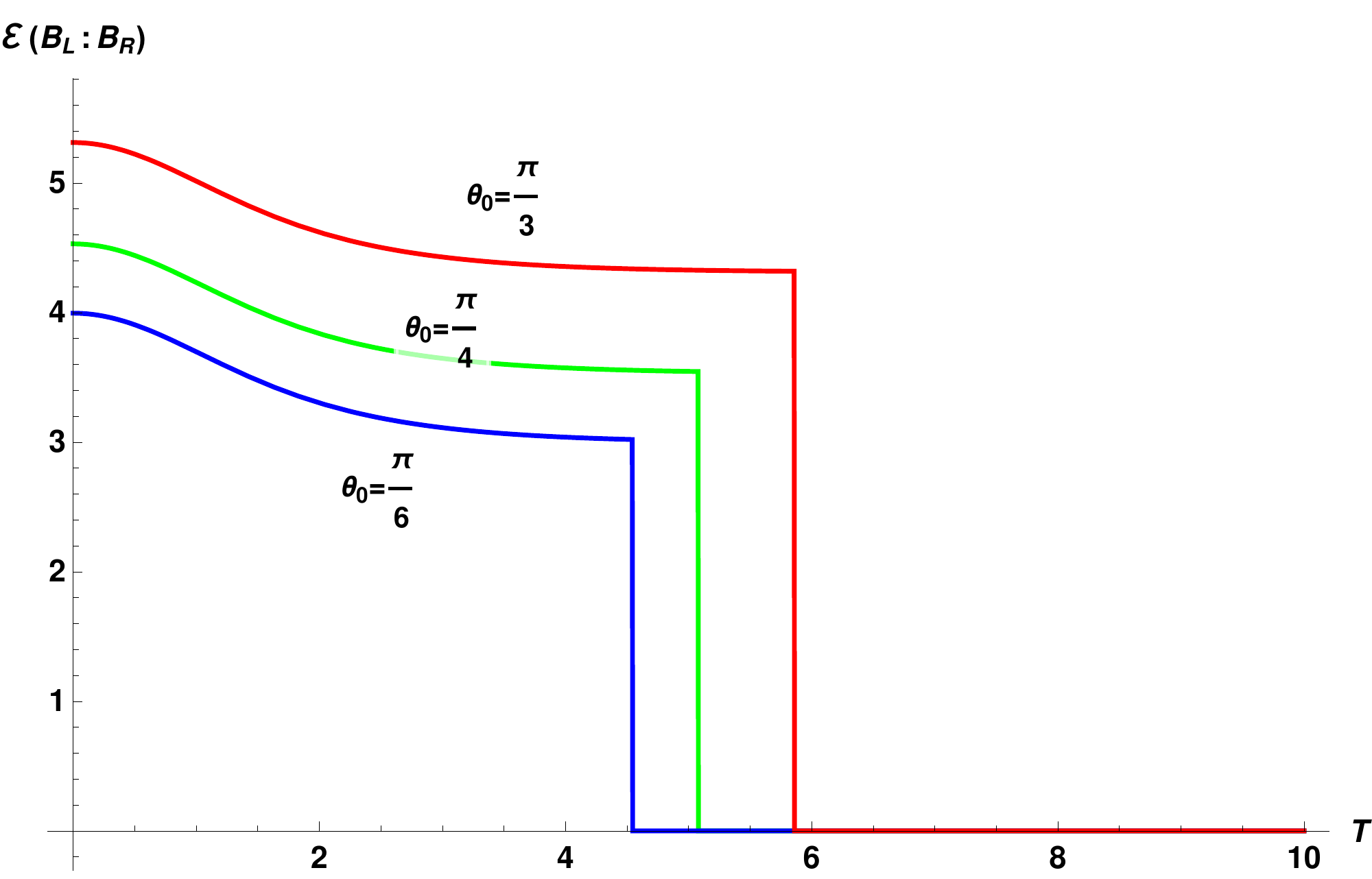}
	\caption{The Page curve for entanglement negativity between black hole interiors for three different values of the EOW brane angle $\theta_0$. Here the variation of the entanglement negativity with respect to the Rindler time $T$ is shown in units of $\frac{c}{4}$ with $X_0 = 1$, $\epsilon_y = 0.1$, $\ell = 1$ and $\theta_0 = \frac{\pi}{3}, \frac{\pi}{4}, \frac{\pi}{6}$.}
	\label{fig:PageCurve-BH-BH}
\end{figure}

From the results of the last two subsections, we may infer that the time evolution of the entanglement negativity between the black hole interiors is governed by the two phases of the extremal surfaces corresponding to the entanglement entropy of $B_L\cup B_R$.
It is well known that the unitary time evolution of the entanglement entropy for a subsystem in the Hawking radiation flux from a black hole is governed by the \textit{Page curve} \cite{Page:1993df,Page:1993wv,Page:2013dx}. Hence the transition between the two different phases of the entanglement negativity between $B_L$ and $B_R$ occurs precisely at the Page time $T_P$, 
given by \cite{Li:2021dmf,Chu:2021gdb}
\begin{align}
	T_P=\cosh^{-1}\left(\sinh X_0\, e^{\tanh^{-1}(\sin\theta_0)}\,\frac{2\ell}{\epsilon_y \cos\theta_0}\right)\,. \label{Page-time}
\end{align}
In the first phase the entanglement negativity is a decreasing function of the Rindler time given by \cref{BH-BH-bdy-fin}. At the Page time $T_P$ the extremal surface for the entanglement entropy transits to the disconnected phase and an entanglement entropy island corresponding to the radiation bath appears inside the gravitational regions on the EOW brane $\mathbb{Q}$. At this time, the entanglement negativity also transits to the corresponding disconnected phase and vanishes identically. The variation of the entanglement negativity between black hole interiors with the Rindler time $T$ is plotted in \cref{fig:PageCurve-BH-BH}. 

\subsection{Entanglement negativity between the black hole and the radiation}
In this subsection, we now proceed to the computation of the entanglement negativity between the black hole region and the radiation region in the time-dependent defect AdS$_3$/BCFT$_2$ scenario. To this end, we consider the black hole region to be described by a space-like interval $B_L$ on the left-half of the two sided eternal black hole and the radiation region to be described by a semi-infinite interval $R_L$ adjacent to $B_L$ as shown in \cref{fig:BH-rad-connected-qes}. Similar to the previous case, there are two phases possible in this case which are investigated below.

\subsubsection{Connected phase}
\begin{figure}[ht]
	\centering
	\includegraphics[scale=.7]{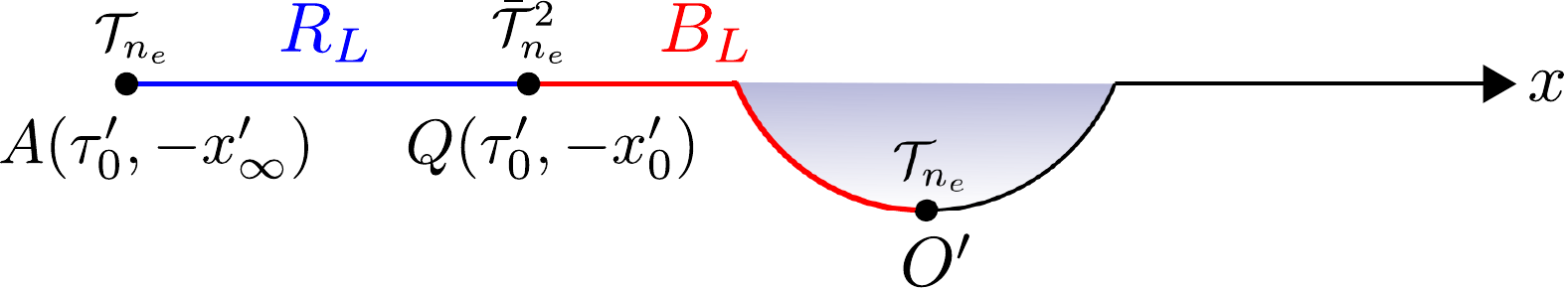}
	\caption{Schematics of the defect extremal surface for the entanglement negativity between the black hole and the radiation in the connected phase}
	\label{fig:BH-rad-connected-qes}
\end{figure}

The connected phase corresponds to the case where $R_L$ does not posses an entanglement island and thus $B_L$ covers the complete left black hole region on the EOW brane as shown in \cref{fig:BH-rad-connected-qes}. From the doubly holographic perspective, this corresponds to an extremal surface for $R_L \cup B_L$ extending from the dynamical endpoint $O'$ of $B_L$ on the EOW brane to spatial infinity. We compute the entanglement negativity between the two adjacent intervals $B_L \equiv |O'Q|$ and $R_L \equiv |QA|$ in this phase where we have regularized the semi-infinite interval $R_L$ to end at some point $A : (\tau'_0 , - x'_\infty)$ which is later taken to infinity.

\subsubsection*{Boundary description}
For the connected phase, the absence of the entanglement island for the radiation region $R_L$ implies that the generalized entanglement negativity doesn't receive any area contribution in the $2d$ effective boundary description. The remaining effective semi-classical entanglement negativity between $B_L$ and $R_L$ may be computed through the following three-point twist correlator
\begin{equation} \label{BH-rad-twist-con}
	\mathcal{E}^{\text{eff}} \left( B_L : R_L \right) = \lim_{n_e \to 1} \log \left[ \left( \epsilon_y \Omega_{O'} \right)^{\Delta_{n_e}} \left< \mathcal{T}_{n_e} (O') \bar{\mathcal{T}}^2_{n_e} (Q) \mathcal{T}_{n_e} (A) \right> \right],
\end{equation}
where $\epsilon_y$ is a UV cut off on the dynamical EOW $\mathbb{Q}$ and $\Omega_{O'}$ is the warp factor as given in \cref{warp-factor-BH}. In the un-primed coordinates, the points $O'$ and $Q$ are located at $(-y, 0)$ and $(\tau_0, - x_0)$ respectively. Using \cref{Banados}, we may locate the spatial infinity $A$ in the un-primed coordinates at $(\tau, x) = (2, 0)$. Now, by utilizing the usual form of a CFT$_2$ three-point twist correlator given in \cref{3-point}, we may obtain the total generalized entanglement negativity in the boundary description for this case to be
\begin{equation} \label{E-gen-qes-BH-rad-con}
	\begin{aligned} 
		\mathcal{E}^\text{bdy}_\text{gen} = \frac{c}{8} \left[ \log \frac{((\tau_0 + y)^2 + x_0^2) (x_0^2 + (2 - \tau_0)^2)}{(y + 2)^2} - 2 \log \frac{4\epsilon}{(\tau'_0 + 1)^2 + x^{\prime 2}_0} \right] \, ,
	\end{aligned}
\end{equation}
where $\epsilon$ is the UV cut-off in the primed coordinates. The above expression is then extremized over the position $y$ of the dynamical point $O'$ on the EOW brane to obtain
\begin{equation}
	y = \frac{x_0^2}{2 - \tau_0} - \tau_0 \, .
\end{equation}
It may be checked through \cref{Banados,Rindler} that $\tau < 2$ for the Rindler time $T > 0$ which guarantees the non-negativity of $y$ for large $x_0$. We may now compute the entanglement negativity by substituting the above value of $y$ in \cref{E-gen-qes-BH-rad-con} to be
\begin{equation} \label{E-qes-BH-rad-con-unprimed}
	\begin{aligned}
		\mathcal{E}^\text{bdy} = \frac{c}{4} \log x_0 - \frac{c}{4} \log \frac{4\epsilon}{(\tau'_0 + 1)^2 + x^{\prime 2}_0}\, .
	\end{aligned}
\end{equation}
Transforming this result to the Rindler coordinates $(X, T)$ through \cref{Banados,Rindler}, we may obtain the final expression for the entanglement negativity between the black hole region $B_L$ and the radiation region $R_L$ in the boundary description to be
\begin{equation} \label{E-qes-BH-rad-con}
	\begin{aligned}
		\mathcal{E}^\text{bdy} (B_L : R_L) = \frac{c}{4} \log \frac{\cosh T}{\epsilon} + X_0 ,
	\end{aligned}
\end{equation}
where $X_0$ corresponds to the endpoint $Q$ of the black hole region $B_L$ at the fixed Rindler time $T$. We note here that the entanglement negativity in the above expression is an increasing function of the Rindler time $T$ in this connected phase.

\subsubsection*{Bulk description}
\begin{figure}[H]
	\centering
	\includegraphics[scale=.7]{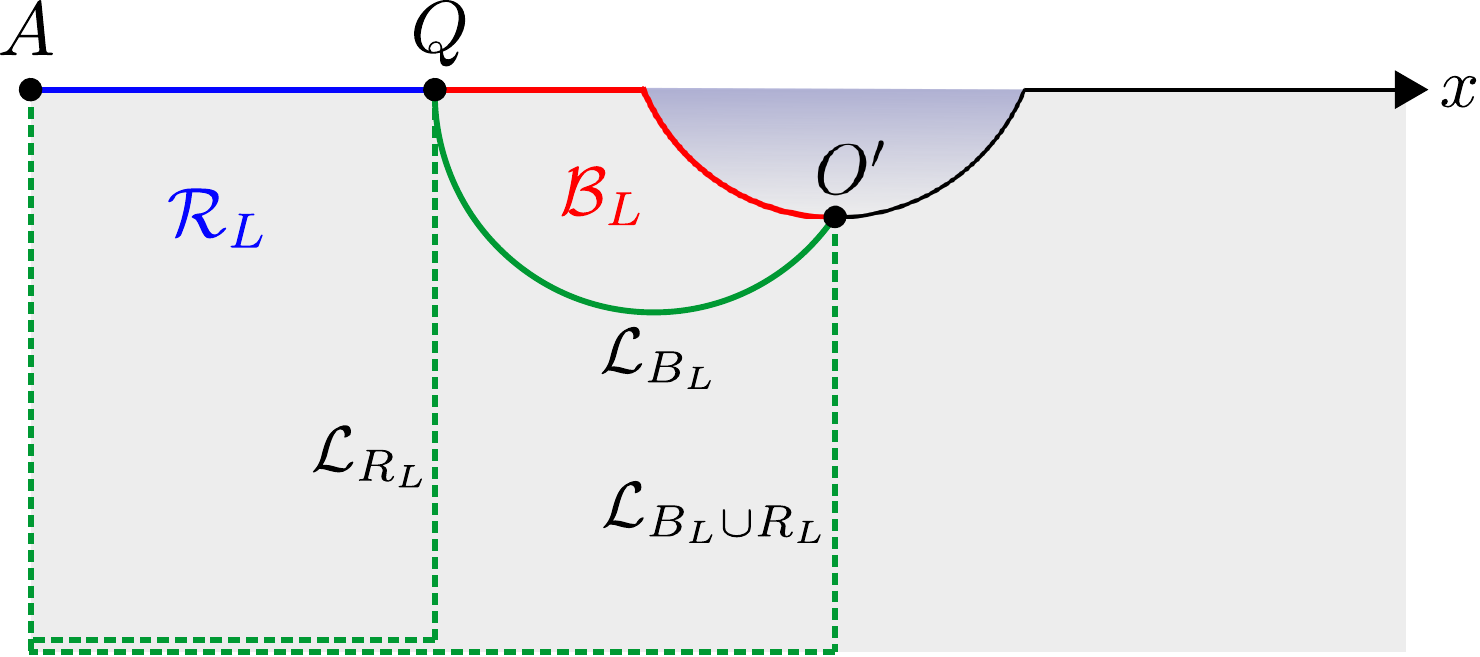}
	\caption{Schematics of the defect extremal surface for the entanglement negativity between the black hole and the radiation in the connected phase}
	\label{fig:BH-rad-connected-des}
\end{figure}

In the bulk description for this phase as depicted in \cref{fig:BH-rad-connected-des}, the generalized entanglement negativity between the black hole region $B_L\equiv|O'Q|$ and the radiation region $R_L\equiv|QA|$ is computed by employing the following formula
\begin{equation} \label{E-BH-rad-con-des-formula}
\begin{aligned}
\mathcal{E}^\text{bulk}_\text{gen} (\mathcal{B}_L : \mathcal{R}_L) & = \frac{3}{16 G_N} \big(\mathcal{L}_{R_L} + \mathcal{L}_{B_L} - \mathcal{L}_{B_L \cup R_L} \big) \\
& = \frac{c}{8} \Bigg( \log \frac{ \big[ (\tau_0 + z \tan \theta_0)^2 + x_0^2 +z^2 \big] \big(x_0^2 + (2 - t_0)^2 \big)}{ \big[ (2 + z \tan \theta_0)^2 + z^2 \big]} - 2 \log \frac{4\epsilon}{(\tau'_0 + 1)^2 + x^{\prime 2}_0} \Bigg) ,
\end{aligned}
\end{equation}
where $O':(-z\tan \theta_0, 0, z)$ and $Q:(\tau_0, - x_0, 0)$ are the endpoints of $B_L$, and $A:(2, 0, 0)$ is the regularized endpoint of the semi infinite radiation region $R_L$ in the un-primed coordinates. We have also used the Brown-Henneaux formula \cite{Brown:1986nw} in the second equality of the above expression. Note that the semi-classical effective entanglement negativity appearing as the second term in \cref{DES-adj} vanishes in this case as $R_L$ does not posses an entanglement island. We may extremize \cref{E-BH-rad-con-des-formula} over the position of the dynamical point $O'$ to obtain
\begin{equation}\label{extremization-bh-rad-con-des}
\partial_z \mathcal{E}^\text{bulk}_\text{gen} = 0 ~~~~ \implies ~~~~z = \frac{x_0^2 \cot \theta_0}{2 - \tau_0} \, ,
\end{equation}
where we have used the approximation that $x_0$ is large. The total entanglement negativity between the black hole region $B_L$ and the radiation region $R_L$ in the Rindler coordinates $(X,T)$ for this phase may now be obtained by utilizing \cref{extremization-bh-rad-con-des,E-BH-rad-con-des-formula,Banados,Rindler} to be
\begin{equation} \label{E-bh-rad-con-des}
\mathcal{E}^\text{bulk} (\mathcal{B}_L : \mathcal{R}_L) = \frac{c}{4} \log \frac{\cosh T}{\epsilon} + X_0 \, ,
\end{equation}
where $X_0$ corresponds to the point $Q$ at the fixed Rindler time $T$. Remarkably, the above expression for the entanglement negativity matches exactly with the boundary description result in \cref{E-qes-BH-rad-con}.

\subsubsection{Disconnected phase}
\begin{figure}[H]
	\centering
	\includegraphics[scale=.7]{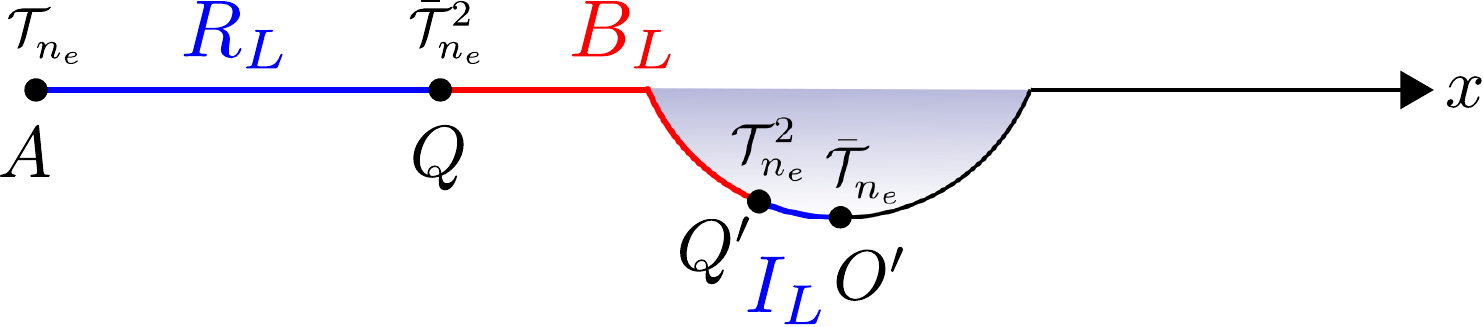}
	\caption{Schematics of the defect extremal surface for the entanglement negativity between the black hole and the radiation in the disconnected phase}
	\label{fig:BH-rad-disconnected-qes}
\end{figure}

The disconnected phase is described by the case where the semi-infinite radiation region $R_L$ has an entanglement island labelled as $I_L \equiv |O'Q'|$ as depicted in \cref{fig:BH-rad-disconnected-qes}. From the bulk perspective, this corresponds to an extremal surface for $R_L$ to end on some point $Q'$ on the EOW brane. The entanglement negativity between the black hole region $B_L$ and the radiation region $R_L$ for this case will thus receive contribution from the island region $I_L$ on the EOW brane.

\subsubsection*{Boundary description}
In the boundary description, the area contribution to the entanglement negativity corresponding to the point $Q' = \partial B_L \cap \partial I_L$ is as given in \cref{area-term}. The remaining effective semi-classical entanglement negativity between $B_L$ and $R_L$ may be computed through the following four-point twist correlator,
\begin{equation} \label{BH-rad-twist-discon}
	\mathcal{E}^{\text{eff}} \left( B_L : R_L \cup I_L\right) = \lim_{n_e \to 1} \log \left[ \left( \epsilon_y \Omega_{Q'} \right)^{\Delta_{n_e}^{(2)}}\left( \epsilon_y \Omega_{O'} \right)^{\Delta_{n_e}} \left< \mathcal{T}_{n_e} (A) \bar{\mathcal{T}}^2_{n_e} (Q) \mathcal{T}^2_{n_e} (Q') \bar{\mathcal{T}}_{n_e} (O') \right> \right],
\end{equation}
where $\Omega$ are the warp factors as given in \cref{warp-factor-BH}, $A$ is the regularized endpoint of the semi-infinite interval $R_L$ and the point $Q'$ and $Q$ are at position $(-y, - x)$ and $(\tau_0, - x_0)$, respectively in the un-primed coordinates. For the bipartite configuration under consideration, the above four-point twist correlator factorizes into two two-point twist correlators in the following way,
\begin{equation}\label{bh-rad-disc-qes-factorization}
\left< \mathcal{T}_{n_e} (A) \bar{\mathcal{T}}^2_{n_e} (Q) \mathcal{T}^2_{n_e} (Q') \bar{\mathcal{T}}_{n_e} (O') \right> \approx \left< \mathcal{T}_{n_e} (A) \bar{\mathcal{T}}_{n_e} (O')) \right> \left< \bar{\mathcal{T}}^2_{n_e} (Q) \mathcal{T}^2_{n_e} (Q') \right>.
\end{equation}
Utilizing the above factorization in \cref{BH-rad-twist-discon} along with the area term in \cref{area-term}, the generalized entanglement negativity for this case may be expressed as
\begin{equation} \label{E-qes-gen-BH-rad-disc}
	\begin{aligned}
		\mathcal{E}^\text{bdy}_\text{gen} = \frac{c}{4} \left( \tanh^{-1} (\sin \theta_0) + \log \frac{\ell}{\epsilon_y y \cos \theta_0} + \log \left( (y + \tau_0)^2 + (x-x_0)^2 \right) - \log \frac{4\epsilon}{(\tau'_0 + 1)^2 + x^{\prime 2}_0} \right),
	\end{aligned}
\end{equation}
where again $\epsilon$ is the UV cut-off in the primed coordinates. Interestingly, the regularized point $A$ does not enter the computation in this case. We may now extremize the above generalized entanglement negativity over the position of the dynamical point $Q'$ {\it i.e.}, $\partial_y \, \mathcal{E}^\text{bdy}_\text{gen} = 0$ and $\partial_{x} \, \mathcal{E}^\text{bdy}_\text{gen} = 0$ to obtain 
\begin{equation} \label{ext-parameters-bh-rad-qes}
y = \tau_0 ~, \qquad x = x_0 \, .
\end{equation}
Using the above values of the coordinates $y$ and $x$ in \cref{E-qes-gen-BH-rad-disc} and transforming the result to the primed coordinates \cref{Banados} and subsequently to the Rindler coordinates \eqref{Rindler} via the analytic continuation $\tau = i t'$, we may obtain the total entanglement negativity between the black hole region $B_L$ and the radiation region $R_L$ to be
\begin{equation} \label{E-qes-BH-rad-disc}
	\mathcal{E}^\text{bdy} (B_L : R_L) = \frac{c}{4} \left( \tanh^{-1} (\sin \theta_0) + \log \frac{\text{e}^{2 X_0} - 1}{\epsilon} + \log \frac{2\ell}{\epsilon_y \cos \theta_0} \right).
\end{equation}
Here $X_0$ corresponds to the endpoint $Q$ of the black hole region $B_L$. Note that the above expression for the entanglement negativity is independent of the Rindler time $T$ and only depends on the position of the point $Q$.

\subsubsection*{Bulk description}
\begin{figure}[H]
	\centering
	\includegraphics[scale=.6]{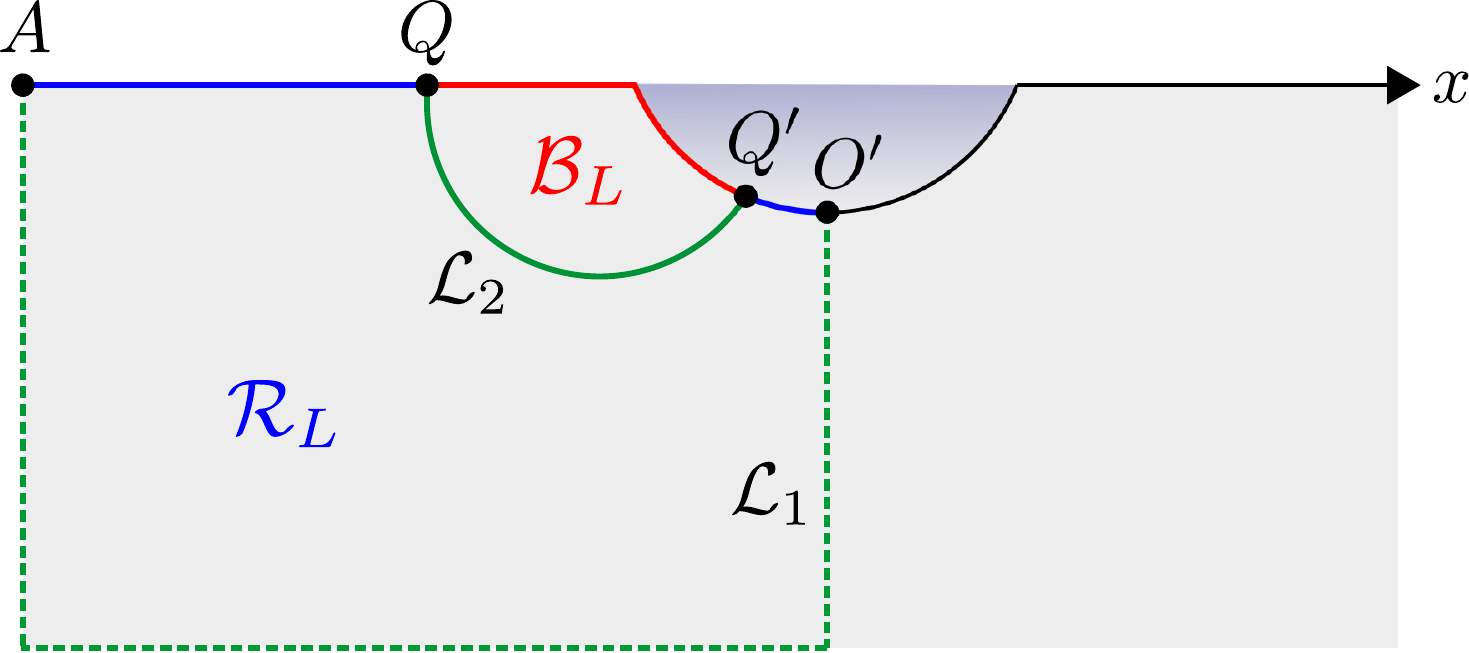}
	\caption{Schematics of the defect extremal surface for the entanglement negativity between the black hole and the radiation in the disconnected phase}
	\label{fig:BH-rad-disconnected-des}
\end{figure}

In the $3d$ bulk description for the disconnected phase, the entanglement negativity between the black hole region $B_L$ and the radiation region $R_L$ is computed by employing the DES formula in \cref{DES-adj} as follows
\begin{equation} \label{E-BH-rad-disc-des-formula}
	\begin{aligned}
		\mathcal{E}^\text{bulk}_\text{gen} (\mathcal{B}_L : \mathcal{R}_L) & = \frac{3}{16 G_N} \left(\mathcal{L}_{B_L} + \mathcal{L}_{R_L} - \mathcal{L}_{B_L \cup R_L}\right) + \mathcal{E}^\text{eff} (\mathcal{B}_L : \mathcal{R}_L) \\
		& = \frac{3}{8 G_N} \mathcal{L}_2 + \mathcal{E}^\text{eff} (B_L : I_L),
	\end{aligned}
\end{equation}
where $\mathcal{L}_2$ is the extremal curve between points $Q' : (-z \tan \theta_0, - x_1, z)$ and $Q : (\tau_0, - x_0, 0)$ and $I_L$ is the island region corresponding to the radiation region $R_L$ as depicted in \cref{fig:BH-rad-disconnected-des}. In the second term of the above expression we have also utilized the fact that bulk matter fields are only localized on the EOW brane $\mathbb{Q}$. We note here that, consistent with the boundary description, the regularized point $A$ does not enter the computation in this phase. The length of the extremal curve $\mathcal{L}_2$ in the un-primed coordinates may be expressed as \cite{Ryu:2006bv, Hubeny:2007xt}
\begin{equation} \label{L2-bh-rad-disc}
	\mathcal{L}_2 \equiv \mathcal{L}_{QQ'} = \ell\,\log \left[\frac{(\tau_0 + z \tan \theta_0)^2 + (x_0 - x_1)^2 + z^2}{z}\right] - \ell\,\log \left(\frac{4\epsilon}{(\tau'_0 + 1)^2 + x^{\prime 2}_0}\right).
\end{equation}
The semi-classical effective entanglement negativity appearing as the last term in \cref{E-BH-rad-disc-des-formula} may be obtained to be
\begin{equation} \label{E-eff-des-bh-rad-disc}
	\begin{aligned}
		\mathcal{E}^\text{eff} (B_L : I_L) = \frac{c}{4} \log \frac{2\ell}{\epsilon_y \cos \theta_0},
	\end{aligned}
\end{equation}
where $\epsilon_y$ is the UV cut-off on the dynamical EOW brane. Extremizing the generalized entanglement negativity obtained by substituting \cref{L2-bh-rad-disc,E-eff-des-bh-rad-disc} in \cref{E-BH-rad-disc-des-formula}, with respect to the position of $Q'$ {\it i.e.}, $\partial_z \mathcal{E}^\text{bulk}_\text{gen} = 0$ and $\partial_{x_1} \mathcal{E}^\text{bulk}_\text{gen} = 0$, we may obtain 
\begin{equation} \label{ext-parameters-bh-rad-des}
	z = \tau_0 \cos \theta_0 ~, \qquad x_1 = x_0 \, .
\end{equation}
The total entanglement negativity between the black hole region $B_L$ and the radiation region $R_L$ in the primed coordinates may then be obtained through \cref{ext-parameters-bh-rad-des,Banados} to be
\begin{equation} \label{E-des-BH-rad-disc}
	\begin{aligned}
		\mathcal{E}^\text{bulk} (\mathcal{B}_L : \mathcal{R}_L) = \frac{c}{4} \left[\log \left(\frac{x'^2_0 + \tau'^2_0 - 1}{\epsilon}\right) + \tanh^{-1} (\sin \theta_0)+  \log \left(\frac{2\ell}{\epsilon_y \cos \theta_0}\right) \right],
	\end{aligned}
\end{equation}
where the Brown-Henneaux formula \cite{Brown:1986nw} has been used. On transformation to the Rindler coordinates \cref{Rindler} via the analytic continuation $\tau' = i t'$, the above expression matches exactly with the boundary perspective result in \cref{E-qes-BH-rad-disc} which serves as a strong consistency check for our proposals.

\subsubsection{Page curve}

\begin{figure}[H]
	\centering
	\includegraphics[scale=.7]{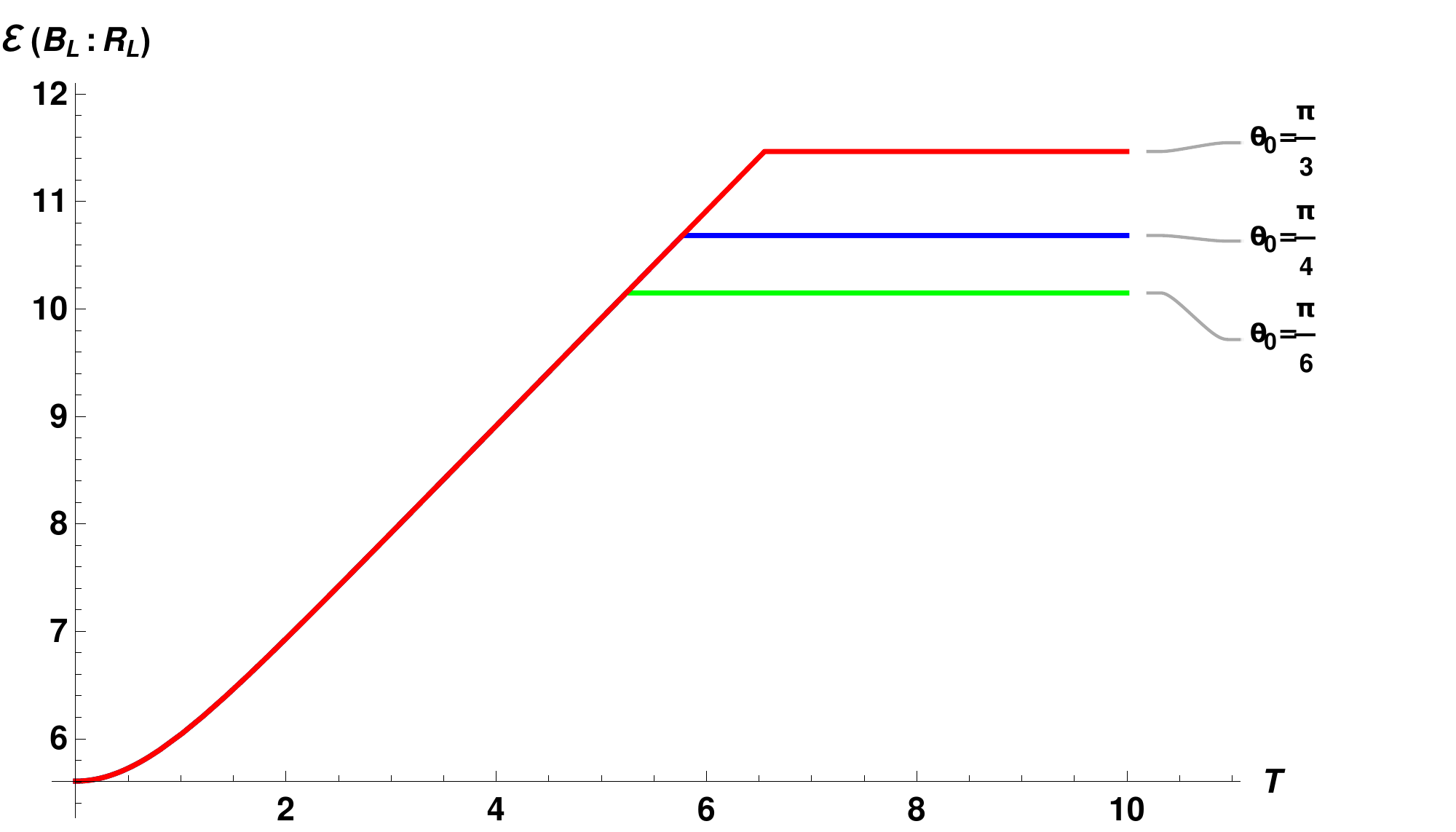}
	\caption{The Page curve for the entanglement negativity between the black hole region and the radiation region for three different values of the EOW brane angle $\theta_0$. Here the variation of the entanglement negativity with respect to the Rindler time $T$ is shown in units of $\frac{c}{4}$ with $X_0 = 1$, $\epsilon = 0.01$, $\epsilon_y = 0.1$, $\ell = 1$ and $\theta_0 = \frac{\pi}{3}, \frac{\pi}{4}, \frac{\pi}{6}$.}
	\label{fig:BH-Rad-Page curve}
\end{figure}

We now analyse the results of the last two subsections where we have computed the entanglement negativity between the black hole region $B_L$ and the radiation region $R_L$ for the two possible phases. In the connected phase, the entanglement negativity computed in \cref{E-qes-BH-rad-con} is an increasing function of the Rindler time $T$. In contrast, for the disconnected phase, the entanglement negativity given in \cref{E-qes-BH-rad-disc} is independent of $T$ and only depends on the size of the black hole region $B_L$. A transition from the connected phase to the disconnected phase is observed at the Page time for the entanglement entropy given in \cref{Page-time}. In \cref{fig:BH-Rad-Page curve}, we show the variation of the time dependent entanglement negativity between the black hole region $B_L$ and the radiation $R_L$ with the Rindler time $T$ for three different values of the EOW brane angle $\theta_0$.

\subsection{Entanglement negativity between subsystems in the radiation bath}
In this subsection, we compute the entanglement negativity between the right subsystem $R_R$ and the left subsystem $R_L$ in the radiation region as shown in \cref{fig:RD-RD-conn-qes}. In the Rindler coordinates $(X,T)$, the right radiation subsystem $R_R$ extends from $(X_0,T)$ to $(X_1,T)$ and the left radiation subsystem $R_L$ extends from $(X_0,-T+i \pi)$ to $(X_1,-T+i \pi)$. In the primed coordinates, the subsystems $R_R \equiv |NQ|$ and $R_L \equiv |PM|$ are mapped to the intervals $[(\tau'_0, x'_0), (\tau'_1, x'_1)]$ and $[(\tau'_1, -x'_1), (\tau'_0, -x'_0)]$ respectively. Similar to the earlier subsections, we perform the computation in the Euclidean signature and subsequently transform the results to Rindler coordinates in the Lorentzian signature. Depending on the phase transition of the extremal surfaces corresponding to $R_L \cup R_R$, the DES corresponding to the entanglement negativity between them crosses from a connected phase to a disconnected phase. In the following, we investigate the time evolution of the entanglement negativity between $R_L$ and $R_R$ from both the bulk and the boundary perspective.

\subsubsection{Connected phase}
\begin{figure}[H]
	\centering
	\includegraphics[scale=0.7]{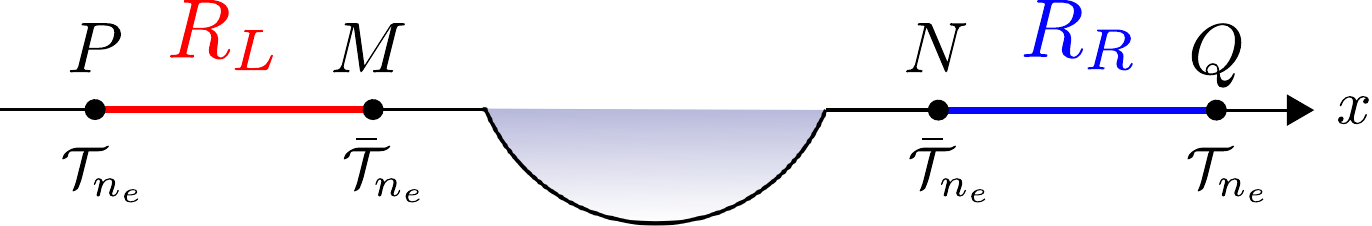}
	\caption{Schematics of the quantum extremal surface for the entanglement negativity between intervals in the radiation region in the connected phase.}
	\label{fig:RD-RD-conn-qes}
\end{figure}

In the connected phase, there are no entanglement entropy islands corresponding to $R_L$ and $R_R$ in the effective boundary description as illustrated in \cref{fig:RD-RD-conn-qes}. In this phase, we compute the entanglement negativity between the disjoint radiation subsystems $R_L$ and $R_R$.

\subsubsection*{Boundary description}
As there are no island contributions in this phase, we observe from \cref{EN_QES} that the entanglement negativity between the radiation subsystems in $2d$ effective boundary description reduces to the effective entanglement negativity between two disjoint intervals as follows
\begin{equation}
\begin{aligned}
\mathcal{E}^{\text{bdy}}(R_L:R_R)=\mathcal{E}^{\text{eff}}(R_L:R_R)=\lim_{n_e \to 1} \log \left[\left<\mathcal{T}_{n_e}(P) \bar{\mathcal{T}}_{n_e}(M) \bar{\mathcal{T}}_{n_e}(N)\mathcal{T}_{n_e}(Q)\right>_{\mathrm{CFT}^{\bigotimes n_e}}\right].
\end{aligned}
\end{equation}
As described in \cref{sec:disj-2}, for the two disjoint subsystems in the $t$-channel, the above four point twist correlator may be computed in the large central charge limit as follows \cite{Malvimat:2018txq}
\begin{equation}\label{neg-con-rdrd-bdy}
\begin{aligned}
\mathcal{E}^{\text{bdy}}(R_L:R_R)&=\frac{c}{4}\log \left( \frac{|PN| |MQ|}{|MN| |PQ|}\right)\\
&=\frac{c}{4}\log \frac{(e^{X_0}+e^{X_1})^2-(e^{X_0}-e^{X_1})^2 \tanh^2 T}{4 e^{X_0+X_1}}.
\end{aligned}
\end{equation}
Note that the entanglement negativity between the radiation subsystems $R_L$ and $R_R$ is a monotonically decreasing function of the Rindler time $T$ in this phase.

\subsubsection*{Bulk description}
In the $3d$ bulk description, the effective entanglement negativity in \cref{DES-disj} vanishes as the corresponding entanglement wedges contain no quantum matter fields. The entanglement negativity between $R_L$ and $R_R$ is then given entirely by the combination of the lengths of the defect extremal surfaces as follows
\begin{equation}\label{neg-con-rdrd-bulk}
\begin{aligned}
\mathcal{E}^{\text{bulk}}\left(\mathcal{R}_L:\mathcal{R}_R\right)&=\frac{3}{16G_N}\left(\mathcal{L}_{PN}+\mathcal{L}_{MQ}-\mathcal{L}_{MN}
-\mathcal{L}_{PQ}\right)\\
&=\frac{3\ell}{8G_N}\left[\log\left(\frac{(x'_1+x'_0)^2+(\tau'_1-\tau'_0)^2}{\epsilon^2}\right)-\log\left(\frac{2x'_0}{\epsilon}\right)-\log\left(\frac{2x'_1}{\epsilon}\right)\right]\,,
\end{aligned}
\end{equation}
where we have used the fact that the length of an extremal curve $\mathcal{L}_{ab}$ connecting two points $(\tau'_a,x'_a)$ and $(\tau'_b, x'_b)$ on the boundary is given by \cite{Hubeny:2007xt}
\begin{equation}
\mathcal{L}_{ab}=\ell\,\log\left[\frac{{(x'_a - x'_b)^2 + (\tau'_a - \tau'_b)^2}}{\epsilon^2}\right].
\end{equation}
Now analytically continuing to the Lorentzian signature and transforming to the Rindler coordinates in \cref{Rindler}, we obtain the entanglement negativity between $R_L$ and $R_R$ in the bulk description to be
\begin{align}
\mathcal{E}^{\text{bulk}}\left(\mathcal{R}_L:\mathcal{R}_R\right)=\frac{c}{4}\log \frac{(e^{X_0}+e^{X_1})^2-(e^{X_0}-e^{X_1})^2 \tanh^2 T}{4 e^{X_0+X_1}}\,,
\end{align}
which matches exactly with the result from the boundary description, given in \cref{neg-con-rdrd-bdy}.

\begin{figure}[H]
	\centering
	\includegraphics[scale=0.7]{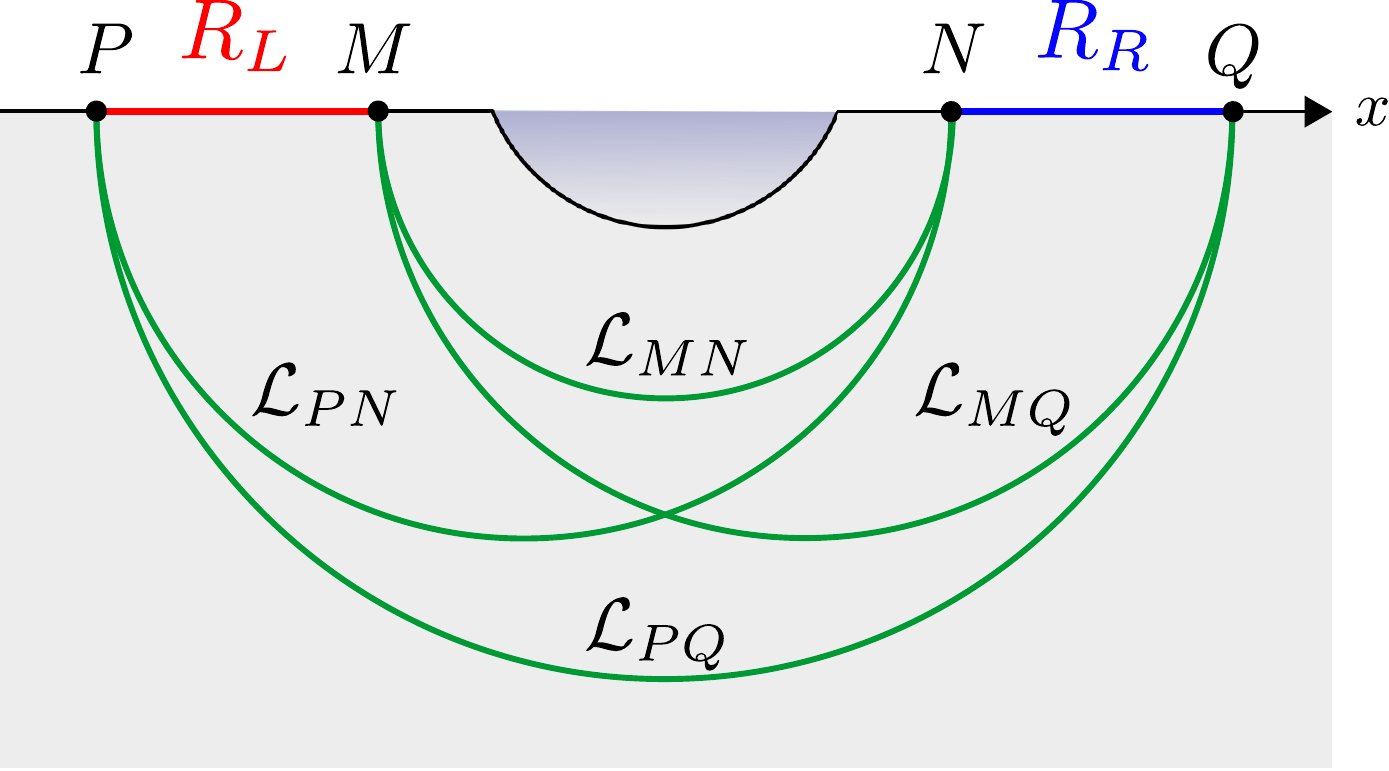}
	\caption{Schematics of the defect extremal surface for the entanglement negativity between intervals in the radiation region in the connected phase.}
	\label{fig:RD-RD-conn}
\end{figure}
\subsubsection{Disconnected phase}
\begin{figure}[H]
	\centering
	\includegraphics[scale=0.7]{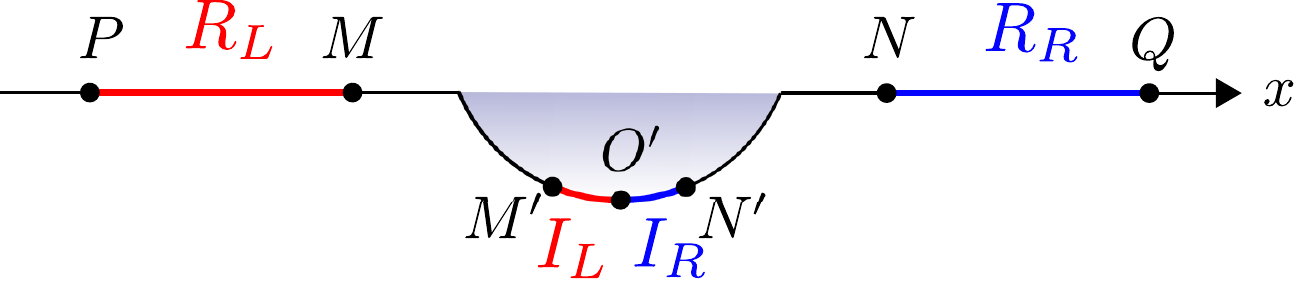}
	\caption{Schematics of the quantum extremal surface for the entanglement negativity between intervals in the radiation region in the disconnected phase. $I_L$ and $I_R$ denote the entanglement negativity islands corresponding to $R_L$ and $R_R$ respectively.}
	\label{fig:RD-RD-disconn}
\end{figure}

For the disconnected phase, the entanglement entropy corresponding to the radiation subsystems receives island contributions as depicted in \cref{fig:RD-RD-disconn}. The entanglement negativity islands corresponding to the radiation subsystems $R_L$ and $R_R$, located on the EOW brane, are denoted as $I_L\equiv |M'O'|$ and $I_R\equiv |O'N'|$ respectively\footnote{Note that the entanglement negativity islands together constitute the entanglement entropy island for $R_L \cup R_R$.}. We now proceed to compute the entanglement negativity between the radiation subsystems in the boundary and bulk descriptions in this phase.

\subsubsection*{Boundary description}
In the $2d$ boundary perspective, the area term corresponding to the point $O'= \partial I_L \cap \partial I_R$ is a constant given by \cref{area-term}. The remaining effective semi-classical entanglement negativity in \cref{EN_QES} may be expressed as
\begin{align}
\mathcal{E}^{\text{eff}}\left(R_L\cup I_L:R_R\cup I_R\right)
=&\lim_{n_e\to 1}\log \Bigg[\left(\epsilon_y\Omega_{M'}\right)^{\Delta_{n_e}}\left(\epsilon_y\Omega_{N'}\right)^{\Delta_{n_e}}\left(\epsilon_y\Omega_{O'}\right)^{\Delta_{n_e}^{(2)}}
\notag\\
&\quad\quad\quad\times\left<\mathcal{T}_{n_e}(P)\bar{\mathcal{T}}_{n_e}(M)\mathcal{T}_{n_e}(M')\bar{\mathcal{T}}^2_{n_e}(O')\mathcal{T}_{n_e}(N')\bar{\mathcal{T}}_{n_e}(N)\mathcal{T}_{n_e}(Q)\right>_{n_e}\Bigg].\label{Discon-eff-rdrd1}
\end{align}
In the large central charge limit, the above twist correlator may be factorized in the dominant channel as follows\footnote{The correlators are factorized into their respective contractions as depicted by the choice of the extremal surfaces in \cref{fig:RD-RD-disconn-des}.}
\begin{align}
	\left<\bar{\mathcal{T}}_{n_e}(M)\mathcal{T}_{n_e}(M')\right>\left<\mathcal{T}_{n_e}(P)\bar{\mathcal{T}}^2_{n_e}(O')\mathcal{T}_{n_e}(Q)\right>\left<\mathcal{T}_{n_e}(N')\bar{\mathcal{T}}_{n_e}(N)\right>\label{RD-RD-disc-factorization}
\end{align}
Now, employing the replica limit $n_e\to 1$, the effective semi-classical entanglement negativity \cref{Discon-eff-rdrd1} in the disconnected phase reduces to
\begin{align}
\mathcal{E}^{\text{eff}}\left(R_L\cup I_L:R_R\cup I_R\right)&\approx\lim_{n_e\to 1}\log\left[\left(\epsilon_y\Omega_{O'}\right)^{\Delta_{n_e}^{(2)}}
\left<\mathcal{T}_{n_e}(P)\bar{\mathcal{T}}^2_{n_e}(O')\mathcal{T}_{n_e}(Q)\right>\right]\notag\\
&=\frac{c}{4}\log\left(\frac{\ell}{\epsilon_y y\cos\theta_0}\right)+\frac{c}{4}\log \left[\frac{(y+\tau_1)^2+x_1^2}{2x_1}\right],\label{Discon-eff-rdrd}
\end{align}
where the coordinates for $P$, $Q$ and $O'$ are given by $(\tau_1,-x_1)$, $ (\tau_1,x_1)$ and $(-y,0)$ respectively.
The generalized entanglement negativity between $R_L$ and $R_R$ in the $2d$ boundary description may now be obtained using \cref{Discon-eff-rdrd,EN_QES} as follows

\begin{equation}\label{Discon-qes-rdrd}
\mathcal{E}^{\text{bdy}}_{\text{gen}}\left(R_L:R_R\right)=\frac{c}{4}\log\left(\frac{\ell}{\epsilon_y y\cos\theta_0}\right)+\frac{c}{4}\log \left[\frac{(y+\tau_1)^2+x_1^2}{2x_1}\right]+\frac{c}{4}\tanh^{-1}(\sin\theta_0).
\end{equation}
Extremizing the above generalized entanglement negativity with respect to $y$ we obtain
\begin{align}
\partial_y\mathcal{E}^{\text{bdy}}_{\text{gen}}\left(R_L:R_R\right)=0\implies y=\sqrt{\tau_1^2+x_1^2}.
\end{align}
Now, substituting the value of $y$ in eq. (\ref{Discon-qes-rdrd}), the entanglement negativity between the radiation subsystems for the disconnected phase in the boundary description is given by
\begin{equation}
\mathcal{E}^{\text{bdy}}\left(R_L:R_R\right)=\frac{c}{4}\left[\log \frac{\sqrt{\tau_1^2+x_1^2}+\tau_1}{x_1}+\log\left(\frac{\ell}{\epsilon_y \cos\theta_0}\right)+
\tanh^{-1}(\sin\theta_0)\right].
\end{equation}
Finally, transforming to the primed coordinates in \cref{Banados}, performing the Lorentzian continuation and utilizing \cref{Rindler} we obtain the entanglement negativity  between the radiation subsystems in terms of the Rindler coordinates $(X,T)$ in the $2d$ effective boundary description as follows
\begin{align}\label{neg-bdy-rdrd}
\mathcal{E}^{\text{bdy}}\left(R_L:R_R\right)=\frac{c}{4}\Bigg[\log\frac{e^{2X_1}-1+\sqrt{4 e^{2X_1}\cosh^2 T+\left(e^{2X_1}-1\right)^2}}{2 e^{X_1}\cosh T}&+\log\left(\frac{\ell}{\epsilon_y\cos\theta_0}\right)\notag\\
&+\log\left(\frac{\cos\theta_0}{1-\sin\theta_0}\right)\Bigg].
\end{align}

\subsubsection*{Bulk description}
\begin{figure}[H]
	\centering
	\includegraphics[scale=0.7]{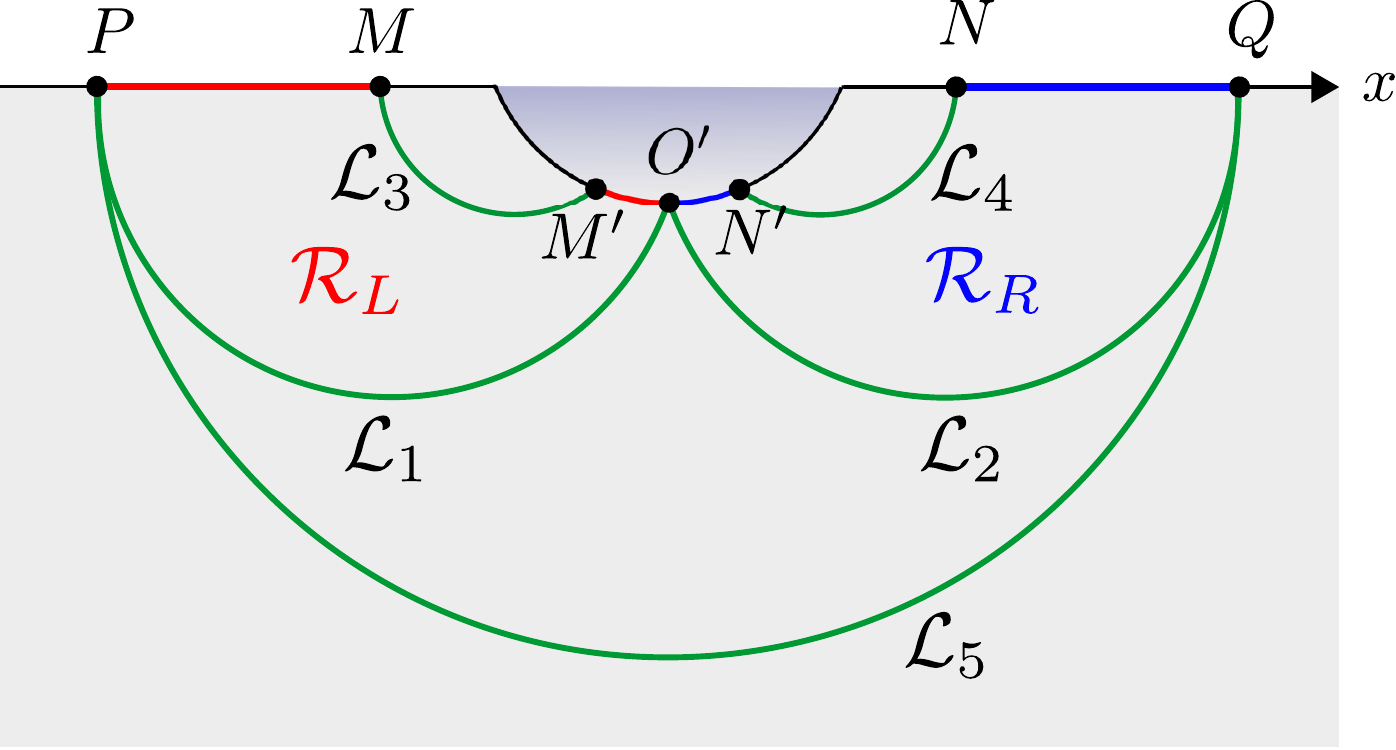}
	\caption{Schematics of the defect extremal surface for the entanglement negativity between intervals in the radiation region in the disconnected phase.}
	\label{fig:RD-RD-disconn-des}
\end{figure}

In the disconnected phase, due to the presence of entanglement islands, a portion of the EOW brane $\mathbb{Q}$ is contained within the entanglement wedge of the radiation in the $3d$ bulk description. As depicted in \cref{fig:RD-RD-disconn-des}, $|M'N'|$ denotes the entanglement entropy island corresponding to $R_L\cup R_R$. The bulk EWCS ends on the EOW brane at the point $O'$ and splits the entanglement wedge corresponding to $R_L\cup R_R$ into two codimension one regions $\mathcal{R}_L$ and $\mathcal{R}_R$ respectively. For this phase, the entanglement negativity between $R_L$ and $R_R$ corresponds to the configuration of disjoint subsystems $|PM|\cup |O'M'|$ and $|NQ|\cup |O'N'|$, sandwiching the region $|MM'|\cup |NN'|$ in between. Therefore, we may employ the DES formula in \cref{DES-disj} to obtain
\begin{align}
\mathcal{E}^{\text{bulk}}_{\text{gen}}\left(\mathcal{R}_L:\mathcal{R}_R\right)&=\frac{3}{16G_N}\Big[\left(\mathcal{L}_1+\mathcal{L}_4\right)+\left(\mathcal{L}_2+\mathcal{L}_3\right)-\left(\mathcal{L}_3+\mathcal{L}_4\right)-\mathcal{L}_5\Big]+\mathcal{E}^{\text{eff}}\left(\mathcal{R}_L:\mathcal{R}_R\right)\notag\\
&=\frac{3}{16G_N}\Big(\mathcal{L}_1+\mathcal{L}_2-\mathcal{L}_5\Big)+\mathcal{E}^{\text{eff}}\left({I}_L:{I}_R\right)\,,\label{Discon-bulk-gen-rdrd}
\end{align}
where as earlier, the effective entanglement negativity between bulk matter fields reduces to that between the adjacent intervals $I_L\equiv |O'M'|$ and $I_R\equiv |O'N'|$ on the EOW brane. The lengths of the extremal surfaces in \cref{Discon-bulk-gen-rdrd} are given by \cite{Ryu:2006bv, Hubeny:2007xt,Chu:2021gdb}
\begin{equation}\label{geods-rdrd}
\begin{aligned}
\mathcal{L}_1&=\ell\,\log\left[\frac{(\tau_1+z\tan\theta_0)^2+x_1^2+z^2}{z}\right]-\ell\,\log\left[\frac{4\epsilon}{(\tau'_1+1)^2+x^{\prime 2}_1}\right]=\mathcal{L}_2\,,\\
\mathcal{L}_5&=2\ell\,\log (2x_1)-2\ell\,\log\left[\frac{4\epsilon}{(\tau'_1+1)^2+x^{\prime 2}_1}\right].
\end{aligned}
\end{equation}
where the second logarithmic term corresponds to the UV cut-off in the unprimed coordinates \cite{Chu:2021gdb, Li:2021dmf}.
\noindent The effective entanglement negativity in \cref{Discon-bulk-gen-rdrd} between the island regions $I_L$ and $I_R$ may be computed through a three-point correlator of twist fields inserted at the endpoints of the intervals as follows
\begin{align}
 \mathcal{E}^\text{eff} (I_L : I_R) = \lim_{n_e \to 1} \log \left[\left(\epsilon_y^2\Omega_{M'}\Omega_{N'}\right)^{\Delta_{n_e}}
 \left(\epsilon_y \, \Omega_{O'}\right)^{\Delta_{n_e}^{(2)}}\left<\mathcal{T}_{n_e}(M')\bar{\mathcal{T}}^2_{n_e}(O')\mathcal{T}_{n_e}(N')\right>_{\mathrm{BCFT}^{\bigotimes n_e}}\right]\,.
\end{align}
The above three-point twist correlator on the half plane describing the BCFT may be expressed as a six-point correlator of chiral twist fields on the whole complex plane using the doubling trick \cite{Cardy:2004hm, Sully:2020pza} as follows 
\begin{align}
	\left<\mathcal{T}_{n_e}(M')\bar{\mathcal{T}}^2_{n_e}(O')\mathcal{T}_{n_e}(N')\right>_{\mathrm{BCFT}^{\bigotimes n_e}}=\left<\mathcal{T}_{n_e}(M')\bar{\mathcal{T}}_{n_e}(M)\bar{\mathcal{T}}^2_{n_e}(O')\mathcal{T}^2_{n_e}(O)\mathcal{T}_{n_e}(N')\bar{\mathcal{T}}_{n_e}(N')\right>_{\mathrm{CFT}^{\bigotimes n_e}}\,,
\end{align}
where $M$, $N$ and $O$ are the image of the points $M'$, $N'$ and $O'$ on the EOW brane respectively. In the large central charge limit, the six-point correlator may be further factorized in the dominant channel similar to \cref{RD-RD-disc-factorization} as follows
\begin{align}
	\left<\bar{\mathcal{T}}_{n_e}(M)\mathcal{T}_{n_e}(M')\right>\left<\bar{\mathcal{T}}^2_{n_e}(O')\mathcal{T}^2_{n_e}(A)\right>\left<\mathcal{T}_{n_e}(N')\bar{\mathcal{T}}_{n_e}(N)\right>\label{RD-RD-disc-factorization1}.
\end{align}
Now, reversing the doubling trick and subsequently employing the replica limit $n_e\to 1$, we may obtain the effective entanglement negativity between $I_L$ and $I_R$ as
\begin{align}
\mathcal{E}^\text{eff} (I_L : I_R)=\lim_{n_e \to 1} \log\left[\left(\epsilon_y \, \Omega_{O'}\right)^{\Delta_{n_e}^{(2)}}\left<\bar{\mathcal{T}}^2_{n_e}(O')\right>_{\mathrm{BCFT}^{\bigotimes n_e}}\right]= \frac{c}{4} \log \frac{2\ell}{\epsilon_y \cos \theta_0} ,\label{eff-rdrd-neg}
\end{align}
We may obtain the generalized entanglement negativity by substituting \cref{geods-rdrd,eff-rdrd-neg} in \cref{Discon-bulk-gen-rdrd} to be
\begin{equation}\label{Discon-des-rdrd}
\mathcal{E}^{\text{bulk}}_{\text{gen}}\left(\mathcal{R}_L:\mathcal{R}_R\right)=\frac{c}{4}\bigg[\log \left(\frac{(\tau_1+z\tan\theta_0)^2+x_1^2+z^2}{2 x_1 z}\right)+\log\left(\frac{2\ell}{\epsilon_y \cos\theta_0}\right)\bigg],
\end{equation}
where we have used the Brown-Henneaux formula \cite{Brown:1986nw} in the first term. The location of the dynamical point $O'$ in the above expression is fixed through extremization at
\begin{equation}
z=\sqrt{\tau_1^2+x_1^2}\cos\theta_0.
\end{equation}
Substituting the above value of $z$ in \cref{Discon-des-rdrd} and subsequently transforming the result to the Rindler coordinates using \cref{Banados,Rindler}, we may obtain the entanglement negativity between the radiation subsystems $R_L$ and $R_R$ to be
\begin{align}\label{neg-bulk-rdrd}
\mathcal{E}^{\text{bulk}}\left(\mathcal{R}_L:\mathcal{R}_R\right)=\frac{c}{4}\Bigg[\log\frac{e^{2X_1}-1+\sqrt{4 e^{2X_1}\cosh^2 T+\left(e^{2X_1}-1\right)^2}}{2 e^{X_1}\cosh T}&+\log\left(\frac{\ell}{\epsilon_y\cos\theta_0}\right)\notag\\
&+\log\left(\frac{\cos\theta_0}{1-\sin\theta_0}\right)\Bigg].
\end{align}
We again observe that the above result in the bulk description matches exactly with the entanglement negativity from the boundary description in \cref{neg-bdy-rdrd} for the disconnected phase.

\subsubsection{Page curve}
\begin{figure}[H]
	\centering
	\includegraphics[scale=.6]{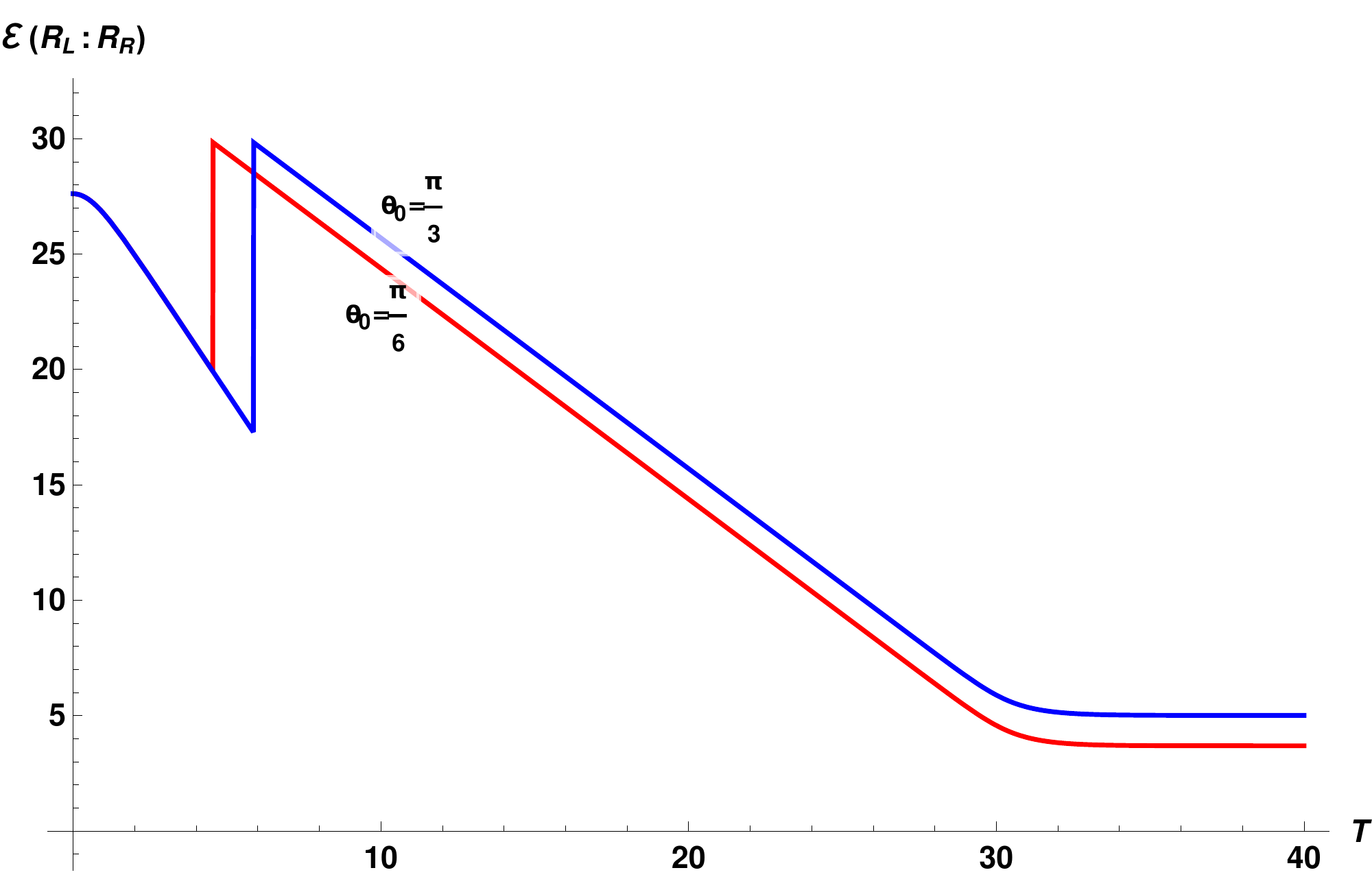}
	\caption{The Page curve for entanglement negativity between the radiation and the radiation with respect to the Rindler time $T$ in units of $\frac{c}{4}$. Here we choose $X_0 = 1$, $X_1 = 30$, $\epsilon_y = 0.1$, $\ell = 1$ and $\theta_0 = \frac{\pi}{3}, \frac{\pi}{6}$.}
	\label{fig:Rad-Rad-Page curve}
\end{figure}

We now analyse the behaviour of the time dependent entanglement negativity between radiation subsystems as discussed in last two subsections. We observe that the entanglement negativity decreases with the Rindler time $T$ in both phases and eventually plateaus out. In the limit when both the radiation subsystems $R_L$ and $R_R$ extends to spatial infinity, namely $X_1\to\infty$, the asymptotic behaviour of the entanglement negativity is given as
\begin{equation}
\mathcal{E}^{\text{bulk}}\left(\mathcal{R}_L:\mathcal{R}_R\right)=
	\begin{cases}
		\frac{c}{4}\left(X_1-X_0-2\log( 2\cosh T)\right) &~~ T<T_P\\
			\frac{c}{4}\left(X_1-\log\cosh T+\log\frac{\cos\theta_0}{1-\sin\theta_0}+\log\frac{\ell}{\epsilon_y\cos\theta_0}\right) & ~~T>T_P\,,
	\end{cases}       
\end{equation}
where $T_P$ is the Page time for the entanglement entropy as given in \cref{Page-time}. Hence, the entanglement negativity shows a sudden jump at $T_P$, and in the limit $X_1\to\infty$ this jump is given by
\begin{align}
	\Delta\mathcal{E}=\frac{c}{4}\left[2\log\left(\frac{2\ell}{\epsilon_y\cos\theta_0}\right)+2\log\left(\frac{\cos\theta_0}{1-\sin\theta_0}\right)+\log\left(e^{2X_0}-1\right)\right]\,.
\end{align}
The analogue of the Page curve for the entanglement negativity for this bipartite configuration is shown in \cref{fig:Rad-Rad-Page curve}.

\section{Summary}\label{sec:summary}

To summarize, in this article, we have proposed a defect extremal surface (DES) prescription for the entanglement negativity of bipartite mixed state configurations in the AdS$_3$/BCFT$_2$ scenarios which include defect conformal matter on the EOW brane. Furthermore, we have extended the island formula for the entanglement negativity to the framework of the defect AdS$_3$/BCFT$_2$, utilizing the lower dimensional effective description involving a CFT$_2$ coupled to semiclassical gravity. Interestingly, the bulk DES formula may be understood as the doubly holographic counterpart of the island formula for the entanglement negativity.

To begin with, we computed the entanglement negativity in the time independent scenarios involving adjacent and disjoint intervals on a static time slice of the conformal boundary of the $3d$ braneworld. To this end, we have demonstrated that the entanglement negativity for various bipartite states obtained through the DES formula matches exactly with the results from the corresponding QES prescription involving entanglement negativity islands. Subsequently we obtained the entanglement negativity in various time-dependent scenarios involving an eternal black hole coupled to a radiation bath in the effective lower dimensional picture. In such time-dependent scenarios, we have obtained the entanglement negativity between subsystems in the black hole interior, between subsystems involving black hole and the radiation bath, and between subsystems in the radiation bath utilizing the island as well as the bulk DES formulae. In this connection, we have studied the time evolution of the entanglement negativity for the above configurations and obtained the analogues of the Page curves. Interestingly, the transitions between different phases of the defect extremal surfaces corresponding to the entanglement negativity for the above configurations occur precisely at the Page time for the corresponding entanglement entropy. Remarkably, it was observed that the entanglement negativity from the boundary and bulk proposals are in perfect agreement for the time-dependent cases, thus demonstrating the equivalence of both formulations. This serves as a strong consistency check for our proposals. We would like to emphasize that our results might lead to several insights about the structure of quantum information about the black hole interior encoded in the Hawking radiation. 

There are several possible future directions to investigate. One such issue would be the extension of our proposals to higher dimensional defect AdS/BCFT scenarios. One may also generalize our doubly holographic formulation for the entanglement negativity with the defect brane at a constant tension to arbitrary embeddings of the brane in the $3d$ bulk geometry. Furthermore, it would also be interesting to derive the bulk DES formula for the entanglement negativity through the gravitational path integral techniques utilizing the replica symmetry breaking wormhole saddles. We leave these open issues for future investigations.

\section{Acknowledgement}
The work of GS is partially supported by the Jag Mohan Chair Professor position at the Indian Institute of Technology, Kanpur.

\bibliographystyle{utphys}
\bibliography{reference}

\end{document}